\documentclass[twocolumn,superscriptaddress,amsmath, amssymb, amsfonts,preprintnumbers,aps,prd,longbibliography,
nofootinbib
]{revtex4-1}

\allowdisplaybreaks

\usepackage{graphicx}
\usepackage{dcolumn}
\usepackage{bm,amstext,amssymb,amsmath}

\usepackage{graphicx,color}
\usepackage{delimset} 
\usepackage[colorlinks=true]{hyperref}
\usepackage{orcidlink}
\usepackage[dvipsnames]{xcolor}
\usepackage{tikz}
\usetikzlibrary{decorations.pathmorphing}
\usetikzlibrary{decorations.markings}
\tikzset{
	compobj/.style={draw=black},
	binary/.style={draw=black, double distance=0.03cm},
	potgrav/.style={draw=black},
	phigrav/.style={draw=blue},
	Agrav/.style={draw=red},
	sigmagrav/.style={draw=LimeGreen},
	radgrav/.style={decorate, draw=black, segment length=5pt, decoration={snake,amplitude=2pt} },
	radgravlarge/.style={decorate,segment length=20pt, decoration={snake,amplitude=2pt}, draw},
	partial ellipse/.style args={#1:#2:#3}{
        insert path={+ (#1:#3) arc (#1:#2:#3)}},
    blackdot/.style={
		circle,
		draw=black,
		fill=black,
		inner sep=0pt,
		minimum size=0.3ex
	}
}
\tikzstyle directed=[postaction={decorate,decoration={markings,
		mark=at position .25 with {\arrow{stealth}},
		mark=at position .75 with {\arrow{stealth}}}}]

\def\ii{ {\bf i}}

\newcommand{\emphh}[1]{\textit{#1}}

\DeclareFontFamily{OT1}{pzc}{}
\DeclareFontShape{OT1}{pzc}{m}{it}{<-> s * [1.350] pzcmi7t}{}
\DeclareMathAlphabet{\mathpzc}{OT1}{pzc}{m}{it}

\definecolor{darkred}{rgb}{0.5,0.0,0.0}
\hypersetup{
  colorlinks=true,
  linkcolor=darkred,   
  citecolor=darkred,   
  urlcolor=darkred     
}

\begin{document}

\preprint{HU-EP-25/43-RTG}

\title{Six-loop gravitational interactions at the sixth post-Newtonian order}

\newcommand{\liverpool}{Department of Mathematical Sciences, University of Liverpool, Liverpool L69 3BX, 
U.K.
}
\newcommand{\aei}{Max Planck Institute for Gravitational Physics (Albert Einstein Institute), D-14476 Potsdam, Germany}
\newcommand{\hu}{Institut f{\"u}r Physik und IRIS Adlershof, Humboldt-Universit {\"a}t zu Berlin, Zum Großen Windkanal 2, D-12489 Berlin, Germany}
\newcommand{\sns}{Scuola Normale Superiore, Piazza dei Cavalieri 7, 56126, Pisa, Italy and INFN Sezione di Pisa, Largo
Pontecorvo 3, 56127 Pisa, Italy}
\newcommand{\edi}{Higgs Centre for Theoretical Physics, University of Edinburgh, James Clerk Maxwell Building,Peter Guthrie Tait Road, Edinburgh, EH9 3FD, United Kingdom}
\newcommand{\unipd}{Dipartimento di Fisica e Astronomia, Universita di Padova, Via Marzolo 8, 35131 Padova, Italy}
\newcommand{\infnpd}{INFN, Sezione di Padova,
Via Marzolo 8, I-35131 Padova, Italy.}

\author{Giacomo Brunello\,\orcidlink{0009-0004-4788-738X}}
\affiliation{\sns}

\author{Manoj K. Mandal\, 
\orcidlink{0000-0003-0850-7685}}
\affiliation{\unipd}
\affiliation{\infnpd}

\author{Pierpaolo Mastrolia\,\orcidlink{0000-0001-9711-7798}}
\affiliation{\unipd}
\affiliation{\infnpd}

\author{Raj Patil\,\orcidlink{0000-0002-7055-0345}}
\affiliation{\aei}
\affiliation{\hu}

\author{\\Matteo Pegorin\,~\orcidlink{0009-0003-1248-871X}}
\affiliation{\unipd}
\affiliation{\infnpd}
\affiliation{\aei}

\author{Jonathan Ronca\,
\orcidlink{0000-0001-9968-311X}}
\affiliation{\unipd}
\affiliation{\infnpd}

\author{Sid Smith\,\orcidlink{0009-0007-7799-0136}}
\affiliation{\unipd}
\affiliation{\infnpd}
\affiliation{\edi}

\author{Jan Steinhoff\,\orcidlink{0000-0002-1614-0214}}
\affiliation{\aei}

\author{William J. Torres~Bobadilla\,\orcidlink{0000-0001-6797-7607}\,}
\affiliation{\liverpool}

\begin{abstract}

We compute the gravitational interaction of two coalescing compact objects at sixth post-Newtonian order in the static limit, employing the diagrammatic approach within the effective field theory framework of General Relativity. The calculation requires the evaluation of six-loop Feynman diagrams that are mapped onto two-point integrals with a gauge-theory-like structure, which are computed here for the first time. The resulting seventh-order contribution in Newton's constant is finite in three space dimensions. This result provides the most technically demanding missing ingredient for the determination of the conservative dynamics of the gravitational two-body system at sixth post-Newtonian order. 

\end{abstract}

\maketitle

\section{Introduction} 
Progress in gravitational-wave (GW) astronomy is driven by observations of gravitational radiation emitted from coalescing compact-object binaries. To date, the LIGO-Virgo-KAGRA collaboration has reported over 200 GW events from such mergers~\cite{KAGRA:2021vkt, LIGOScientific:2025slb}. Future observatories --- including next-generation ground-based detectors such as the Einstein Telescope~\cite{Punturo:2010zz} and Cosmic Explorer~\cite{Reitze:2019iox}, and space missions like LISA~\cite{LISA:2017pwj} --- will enhance our detection reach and sensitivity. Thus, they will offer unprecedented probes of gravity and the astrophysics of compact objects.
Realizing the full scientific potential of these facilities, like the distributions of black hole (BH) masses and spins~\cite{LIGOScientific:2025pvj}, constraints on the neutron-star equation-of-state \cite{LIGOScientific:2018cki}, measurements of the Hubble-Lema\^itre parameter \cite{LIGOScientific:2017adf,LIGOScientific:2021aug,LIGOScientific:2025jau}, and constraints on the theory of General Relativity (GR) \cite{LIGOScientific:2016lio,LIGOScientific:2020tif,LIGOScientific:2021sio,LIGOScientific:2025rid}, requires highly accurate theoretical models for the GW signals. Limited-accuracy models may lead to systematic biases \cite{Owen:2023mid,Dhani:2024jja} that are already evident in the current detectors~\cite{LIGOScientific:2025rsn}, a problem that will only escalate as the signal-to-noise ratio increases for future detectors. 

Currently the modeling of the two-body system combines multiple theoretical and numerical frameworks.
Numerical relativity~\cite{Pretorius:2005gq,Campanelli:2005dd,Baker:2005vv} is used in the highly nonlinear merger regime, where Einstein's equations are solved numerically on supercomputers. This method is highly accurate but computationally very intensive. On the other hand perturbative methods, particularly the post-Newtonian (PN)~\cite{lorentzanddroste,Einstein:1938yz,
Futamase:2007zz,Blanchet:2013haa,Porto:2016pyg,Schafer:2018kuf,Levi:2018nxp,Jaranowski:1997ky,Damour:2014jta,Jaranowski:2015lha,Bernard:2015njp,Bernard:2016wrg,Damour:2016abl, Blanchet:2023sbv,Blanchet:2023bwj, Damour:2017ced,
Goldberger:2004jt,Kol:2013ega,Foffa:2012rn,Foffa:2019rdf,Foffa:2016rgu,Foffa:2019yfl,Blumlein:2020pog,Blumlein:2020znm,Goldberger:2012kf,Galley:2015kus,
Foffa:2019hrb,Blumlein:2019zku,Foffa:2020nqe,Blumlein:2020pyo,Blumlein:2021txe,Porto:2024cwd,
Blumlein:2021txj,
Porto:2005ac,Levi:2015msa,Kim:2021rfj,Levi:2020uwu,Levi:2019kgk,Kim:2022pou,Kim:2022bwv,Levi:2022dqm,Levi:2022rrq,Mandal:2022nty,Mandal:2022ufb,
Goldberger:2009qd,Ross:2012fc,
Cho:2021mqw,Cho:2022syn,Amalberti:2023ohj,Amalberti:2024jaa,Mandal:2024iug} methods are used to approximate the solution to Einstein's equations analytically in the inspiral regime, where the compact objects move with slow velocities and are widely separated. Other methods, such as the post-Minkowskian approximation~\cite{Westpfahl:1979gu,Westpfahl:1980mk,Bel:1981be,Westpfahl:1985tsl,schafer1986adm,Ledvinka:2008tk,Buonanno:2022pgc,Travaglini:2022uwo,Bjerrum-Bohr:2022blt,Kosower:2022yvp,Damour:2016gwp,
Cheung:2018wkq,Bern:2019nnu,Bern:2022kto,Bern:2024adl,Kosower:2018adc,Bjerrum-Bohr:2018xdl,Cristofoli:2019neg,Damgaard:2019lfh,Brandhuber:2021eyq,Vines:2017hyw,Kalin:2020mvi,Kalin:2020fhe,Mogull:2020sak,Jakobsen:2021zvh,Jakobsen:2022psy,Driesse:2024xad,Kalin:2022hph,Driesse:2024feo,Bern:2025zno,Dlapa:2025biy,Kalin:2019rwq,Dlapa:2024cje,
Bern:2025wyd}
and the gravitational self-force (SF)~\cite{Mino:1996nk,Quinn:1996am,Barack:2001gx,Barack:2002mh,Gralla:2008fg,Detweiler:2008ft,Keidl:2010pm,vandeMeent:2017bcc,Pound:2012nt,Pound:2019lzj,Gralla:2021qaf,Pound:2021qin,Warburton:2021kwk,Wardell:2021fyy} are also used for systems in the weak-field regime and for small mass-ratios, respectively.
The most commonly used approaches to build complete inspiral-merger-ringdown  waveform models
of compact binaries are the phenomenological methods~\cite{Ajith:2007qp,Pratten:2020ceb,Estelles:2021gvs} and effective-one-body formalism~\cite{Buonanno:1998gg,Buonanno:2000ef,Damour:2000we,Damour:2001tu,Buonanno:2005xu,Damour:2008gu}. 

\begin{figure}[t]
\begin{tikzpicture}[scale=0.45,line width=1 pt]
\begin{scope}[shift={(0,0)}]
\draw[compobj] (-2,1.5)--(2,1.5);
\draw[compobj] (-2,-1.5)--(2,-1.5);
\draw[phigrav] (0,1.5)--(0,0.5);
\draw[phigrav] (0,0.5)--(-1.5,-1.5);
\draw[sigmagrav] (0,0.5)--(-0.3,-0.7);
\draw[sigmagrav] (0,0.5)--(0.4,-0.7);
\draw[sigmagrav] (0,0.5)--(1.2,-0.7);
\draw[phigrav] (-0.3,-0.7)--(-0.9,-1.5);
\draw[phigrav] (-0.3,-0.7)--(-0.3,-1.5);
\draw[phigrav] (0.4,-0.7)--(0.7,-1.5);
\draw[phigrav] (0.4,-0.7)--(0.1,-1.5);
\draw[phigrav] (1.2,-0.7)--(1.6,-1.5);
\draw[phigrav] (1.2,-0.7)--(1.1,-1.5);
\node[blackdot] at (0,1.5) {};
\node[blackdot] at (0,0.5) {};
\node[blackdot] at (-0.3,-0.7) {};
\node[blackdot] at (0.4,-0.7) {};
\node[blackdot] at (1.2,-0.7) {};
\node[blackdot] at (-1.5,-1.5) {};
\node[blackdot] at (-0.9,-1.5) {};
\node[blackdot] at (-0.3,-1.5) {};
\node[blackdot] at (0.7,-1.5) {};
\node[blackdot] at (0.1,-1.5) {};
\node[blackdot] at (1.6,-1.5) {};
\node[blackdot] at (1.1,-1.5) {};
\node at (0,2.2) {\scalebox{0.9}{(0SF)}};	
\node at (0,-2.4) {\scalebox{0.9}{+35 diag.}};	
\end{scope}
\begin{scope}[shift={(5,0)}]
\draw[compobj] (-2,1.5)--(2,1.5);
\draw[compobj] (-2,-1.5)--(2,-1.5);
\draw[phigrav] (-1,1.5)--(-1,0.5);
\draw[phigrav] (1,1.5)--(1,0.5);
\draw[phigrav] (-1,0.5)--(-1.5,-1.5);
\draw[phigrav] (1,0.5)--(1.5,-1.5);
\draw[sigmagrav] (1,0.5)--(-1,0.5);
\draw[sigmagrav] (0,0.5)--(-0.5,-0.5);
\draw[phigrav] (-0.5,-0.5)--(-1,-1.5);
\draw[phigrav] (-0.5,-0.5)--(-0.2,-1.5);
\draw[sigmagrav] (-0.5,-0.5)--(0.5,-0.5);
\draw[phigrav] (0.5,-0.5)--(1,-1.5);
\draw[phigrav] (0.5,-0.5)--(0.2,-1.5);
\node[blackdot] at (-1,1.5) {};
\node[blackdot] at (1,1.5) {};
\node[blackdot] at (-1,0.5) {};
\node[blackdot] at (1,0.5) {};
\node[blackdot] at (0,0.5) {};
\node[blackdot] at (-0.5,-0.5) {};
\node[blackdot] at (0.5,-0.5) {};
\node[blackdot] at (-1.5,-1.5) {};
\node[blackdot] at (-1,-1.5) {};
\node[blackdot] at (-0.2,-1.5) {};
\node[blackdot] at (0.2,-1.5) {};
\node[blackdot] at (1,-1.5) {};
\node[blackdot] at (1.5,-1.5) {};
\node at (0,2.2) {\scalebox{0.9}{(1SF)}};	
\node at (0,-2.4) {\scalebox{0.9}{+160 diag.}};	
\end{scope}
\begin{scope}[shift={(10,0)}]
\draw[compobj] (-2,1.5)--(2,1.5);
\draw[compobj] (-2,-1.5)--(2,-1.5);
\draw[phigrav] (-1,1.5)--(-0.5,0.5);
\draw[phigrav] (0,1.5)--(-0.5,0.5);
\draw[sigmagrav] (-0.5,0.5)--(-1,-0.5);
\draw[phigrav] (-1,-0.5)--(-1.5,-1.5);
\draw[phigrav] (-1,-0.5)--(-0.5,-1.5);
\draw[phigrav] (1,1.5)--(1,0);
\draw[phigrav] (1,0)--(1.5,-1.5);
\draw[phigrav] (-0.2,-0.5)--(0.2,-1.5);
\draw[phigrav] (0.5,-0.5)--(0.9,-1.5);
\draw[sigmagrav] (-1,-0.5)--(-0.2,-0.5);
\draw[phigrav] (-0.2,-0.5)--(0.5,-0.5);
\draw[sigmagrav] (0.5,-0.5)--(1,0);
\node[blackdot] at (-1,1.5) {};
\node[blackdot] at (0,1.5) {};
\node[blackdot] at (-0.5,0.5) {};
\node[blackdot] at (-1,-0.5) {};
\node[blackdot] at (-0.2,-0.5) {};
\node[blackdot] at (0.5,-0.5) {};
\node[blackdot] at (1,0) {};
\node[blackdot] at (1,1.5) {};
\node[blackdot] at (-1.5,-1.5) {};
\node[blackdot] at (-0.5,-1.5) {};
\node[blackdot] at (0.2,-1.5) {};
\node[blackdot] at (0.9,-1.5) {};
\node[blackdot] at (1.5,-1.5) {};
\node at (0,2.2) {\scalebox{0.9}{(2SF)}};	
\node at (0,-2.4) {\scalebox{0.9}{+386 diag.}};	
\end{scope}
\begin{scope}[shift={(15,0)}]
\draw[compobj] (-2,1.5)--(2,1.5);
\draw[compobj] (-2,-1.5)--(2,-1.5);
\draw[sigmagrav] (0,0.5)--(0,-0.5);
\draw[sigmagrav] (0,0.5)--(0.45,1);
\draw[sigmagrav] (0,0.5)--(1.3,0);
\draw[sigmagrav] (0,-0.5)--(-0.45,-1);
\draw[sigmagrav] (0,-0.5)--(-1.3,0);
\draw[white,fill=white] (-0.5,-0.2) rectangle (-0.4,-0.4);
\draw[white,fill=white] (0.5,0.2) rectangle (0.4,0.4);
\draw[phigrav] (-1.3,1.5)--(-1.3,-1.5);
\draw[phigrav] (0.45,1.5)--(0.45,-1.5);
\draw[phigrav] (-0.45,1.5)--(-0.45,-1.5);
\draw[phigrav] (1.3,1.5)--(1.3,-1.5);
\node[blackdot] at (0,0.5) {};
\node[blackdot] at (0,-0.5) {};
\node[blackdot] at (0.45,1) {};
\node[blackdot] at (1.3,0) {};
\node[blackdot] at (-0.45,-1) {};
\node[blackdot] at (-1.3,0) {};
\node[blackdot] at (-1.3,1.5) {};
\node[blackdot] at (-1.3,-1.5) {};
\node[blackdot] at (0.45,1.5) {};
\node[blackdot] at (0.45,-1.5) {};
\node[blackdot] at (-0.45,1.5) {};
\node[blackdot] at (-0.45,-1.5) {};
\node[blackdot] at (1.3,1.5) {};
\node[blackdot] at (1.3,-1.5) {};
\node at (0,2.2) {\scalebox{0.9}{(3SF)}};	
\node at (0,-2.4) {\scalebox{0.9}{+532 diag.}};	
\end{scope}
\end{tikzpicture}
\label{fig:sixloopgrphs}
\vspace*{-5mm}
\caption{Representative Feynman diagrams at six-loop contributing to the effective potential at 6PN, in the static limit 
${\cal O}(G_N^7)$: massive compact objects, $\phi$ and $\sigma$ fields are represented by black, blue and green lines, respectively. The counting does not include diagrams that can be obtained by the exchange of the massive objects.} 
\end{figure}
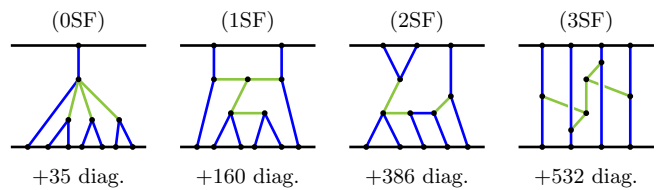

All binary systems observed thus far consist of two compact objects on bound trajectories that spiral toward each other until undergoing a nonlinear merger process. In the inspiral regime, the binary's components move at non-relativistic velocities, and their orbital separation slowly decays. The non-relativistic regime of motion at this stage allows for a perturbative treatment of the problem, which can be studied in powers of $v/c < 1$, where $v$ is the orbital velocity of the compact binary and $c$ is the speed of light. Consequently, physical quantities admit a series expansion, and the $n^{\rm th}$-order coefficient ($n$PN) receives contributions from terms proportional to $G_N^k \, (v^2)^{n+1-k}$, with $0 \le k \le n+1$. 
Here, $G_N$ is the Newton's constant and $v^2$ is the squared relative velocity, a parameter that is virial-related to $G_N$ in the case of bound orbits ($v^2\simeq G_N m/r$, where $m$ and $r$ are the typical mass and size of the system, respectively).

The systematic PN expansion of the two-body problem began with the pioneering 1PN analysis in Refs.~\cite{lorentzanddroste,Einstein:1938yz}. 
Since that foundational result, computing successive higher-order corrections has required sustained efforts and increasingly sophisticated analytical and computational techniques. The current state-of-the-art is the 4PN conservative effective interaction between the two objects first derived in Refs.~\cite{Damour:2014jta,Jaranowski:2015lha,Bernard:2015njp,Bernard:2016wrg,Damour:2016abl} and the 4.5PN radiation emitted by the binary~\cite{Blanchet:2023sbv,Blanchet:2023bwj}.
However, to avoid systematic bias (like those seen in current observations~\cite{LIGOScientific:2025rsn}), 
it is urgent to complete the 5PN order or higher while current facilities reach their design sensitivity~\cite{Owen:2023mid}. Next-generation detectors require further improvement in waveform accuracy of more than two orders of magnitude~\cite{Purrer:2019jcp,Hu:2022rjq}, hence the PN program should aim to reach 7PN accuracy within a decade. 

Over the past decade, substantial progress has been made in pushing the conservative two-body dynamics \cite{Goldberger:2004jt,Foffa:2012rn,Foffa:2019rdf,Foffa:2016rgu,Foffa:2019yfl,Blumlein:2020pog,Foffa:2019hrb,Blumlein:2021txe,Porto:2024cwd,Blumlein:2021txj,Blumlein:2020znm,Blumlein:2019zku,Foffa:2020nqe,Blumlein:2020pyo,
Porto:2005ac,Levi:2015msa,Kim:2021rfj,Levi:2020uwu,Levi:2019kgk,Kim:2022pou,Kim:2022bwv,Levi:2022dqm,Levi:2022rrq,Mandal:2022nty,Mandal:2022ufb} and emitted radiation \cite{Goldberger:2009qd,Ross:2012fc,
Cho:2021mqw,Cho:2022syn,Amalberti:2023ohj,Amalberti:2024jaa,Mandal:2024iug} to increasingly high PN orders within the effective field theory (EFT) approach. In particular, this approach has enabled a systematic reformulation of the PN expansion in terms of classical Feynman diagrams~\cite{Goldberger:2004jt,Goldberger:2009qd}. When combined with modern multi-loop techniques~\cite{Kol:2013ega,Foffa:2016rgu}, this approach has led to the complete determination of the conservative dynamics at 4PN order~\cite{Foffa:2016rgu, Foffa:2019rdf, Foffa:2019yfl}, and partial determination at 5PN order~\cite{Foffa:2019hrb,Blumlein:2019zku,Foffa:2020nqe,Blumlein:2020pyo,Blumlein:2021txe,Porto:2024cwd}. 
The current state-of-the-art is represented by the ${\cal O}(G_N^4)$ 6PN corrections \cite{Blumlein:2021txj}, a subset of the complete 6PN corrections that requires the evaluation of the ${\cal O}(G_N^k)$ terms up to $k=7$. On the other hand, the emitted radiation from a binary has been computed up to 3PN order\cite{Amalberti:2024jaa}.

As observed in Refs.~\cite{Kol:2013ega,Foffa:2016rgu}, PN calculations are technically equivalent to evaluating multi-loop massless two-point Feynman integrals. This brings classical gravity into direct contact with modern multi-loop methods developed in quantum field theory (QFT). 
In the diagrammatic approach, the $n$PN correction at order 
${\cal O}$($G_N^k \, (v^2)^{n+1-k}$) receives contributions from multi-loop diagrams, with the number of loops decreasing as the power of $v^2$ increases. Particularly, the static sector (independent of $v^2$ and corresponding to $k=n+1$) requires evaluating diagrams with the maximal number of loops, namely $n$.  In contrast, the $G_N(v^2)^{n}$ term (corresponding to $k=1$) receives contributions from tree-level diagrams.

It is the main goal of this communication to present the first 
evaluation of the static two-body effective interaction potential at the ${\cal O}(G_N^7)$ 6PN order.

\vspace{-5pt}
\section{Diagrammatic Structure of Gravitational Two-Body Dynamics} 
\label{sec_comp_meth}

The ${\cal O}(G_N^7)$ static term\footnote{Here by `static' we mean $k=n+1$ contribution to the effective Lagrangian, where the $k<n+1$ terms usually contain higher-order-time derivatives and removing them could also lead to contributions to the $k=n+1$ sector.} is the most challenging component of 6PN conservative dynamics and includes, for the first time, a third-order SF term that is absent from lower PN orders. The corresponding amplitude gets contributions from 1117 six-loop Feynman diagrams.
Their integrands stem from a combinatorial construction of classical interaction graphs dressed by PN-expanded vertices, and involve thousands of distinct dimensionally regulated multi-loop  scalar integrals \cite{panther:inprep}.
Their evaluation is made possible by the optimized use of state-of-the art algorithms and software, partly developed to achieve this goal. 
The phase of the diagram generation is followed by 
the application of integration-by-parts identities (IBPs)~\cite{Tkachov:1981wb,Chetyrkin:1981qh,Laporta:2000dsw}, 
implemented in a novel system solving strategy that combines the standard decomposition algorithm 
with modern concepts including spanning cuts~\cite{Larsen:2015ped}, syzygy-based methods~\cite{Gluza:2010ws,Wu:2023upw,Wu:2025aeg,Smith:2025xes}, and improved seeding strategies~\cite{Lange:2025fba,Bern:2024adl,Driesse:2024feo,Driesse:2024xad,vonHippel:2025okr,Song:2025pwy,Zeng:2025xbh,Bern:2025zno}, to reduce the several thousands of integrals contributing to the amplitude onto a set of 39 master integrals (MIs), out of which, 
after taking the Fourier transform and the $d=3$ limit, only 21 turn out to give a non-vanishing contribution to the potential. 
While the evaluation of the static contribution at lower PN orders involved MIs associated to planar diagrams, 
at 6PN we observe the first contributions from two non-planar graphs.
The MIs are evaluated using both analytic methods and numerical techniques. For those cases where a direct analytic result cannot be obtained, we reconstruct the analytic expression from their numerical evaluation. In these cases, we opt for 
a combination of the auxiliary mass flow method~\cite{Liu:2017jxz,Liu:2022chg}  
with the integer-relation finding algorithm \cite{PSLQ} for the analytic reconstruction of high-precision numerical results.

\vspace{-10pt}
\subsection{Post-Newtonian Action}
\label{sec:PN_action}
We consider the Einstein-Hilbert action for the dynamics of the gravitational degrees of freedom given by 
\begin{align}\label{eq_action_EH}
S_{\text{EH}} &= -\frac{c^4}{16 \pi G_N} \int d^4x \sqrt{|g|} \Big(R-\frac{1}{2} g_{\mu\nu}\Gamma^\mu \Gamma^\nu\Big) \, , 
\end{align}
where $G_N$ is Newton's gravitational constant
\footnote{Within the dimensional regularisation scheme, the gravitational coupling constant in $d$ dimensions is defined $G_d = G_N \Big(\sqrt{4\pi e^{\gamma_E}} R_0\Big)^{d-3}$ where, $R_0$ is an arbitrary length scale.}
, $g_{\mu\nu}$ is the spacetime metric and $g$ its determinant, $R$ is the Ricci scalar and $\Gamma^{\mu}_{\rho\sigma}$ is the metric compatible Christoffel connection. Here, the second term enforces harmonic gauge, $\Gamma^\mu = \Gamma^{\mu}_{\rho\sigma}g^{\rho\sigma} = 0$.
The dynamics of each compact object in the binary is described by a point-particle worldline action given as,
\begin{align}\label{eq_action_pp}
S_{\text{pp}} = \sum_{a=1,2} \int d\tau \left( - m_{(a)}c \sqrt{u_{(a)}^2}  \right)\, ,
\end{align}
where $u_{(a)}^\mu=dx_{(a)}^\mu/d\tau$ is the four-velocity, such that $u_{(a)}^2 = u_{(a)}^\mu u_{(a)}^\nu g_{\mu\nu}$, $m_{(a)}$ denotes the mass of each object, the proper time $\tau$ is related to the co-ordinate time $t$ as $d\tau=c~dt$, and we work with mostly negative metric signature.

In the inspiral regime, the dynamics are characterized by three well separated length scales: the Schwarzschild radius $R_s$ of each compact object, the orbital separation $r$, and the wavelength of gravitational radiation $\lambda$, with $\lambda \gg r \gg R_s$.  In this regime, due to small velocities and weak fields, we expand the metric around flat spacetime, $g_{\mu\nu} = \eta_{\mu\nu} + h_{\mu\nu}$,
where the gravitational interaction is mediated by the gravitons $h_{\mu\nu}$. 
The separation of scales allows us to decompose the graviton field in short-distance instantaneous potential modes $H_{\mu\nu}$ scaling as $(k_0,\textbf{k})\sim(v/r,1/r)$, and long-distance radiation modes $\bar{h}_{\mu\nu}$, scaling as $(k_0,\textbf{k})\sim(v/r,v/r)$ \cite{Goldberger:2004jt}.
Since here we are only interested in the conservative binding potential, we discard the radiation modes and decompose the potential modes using a Kaluza-Klein (KK) decomposition~\cite{Kol:2007bc,Kol:2007rx}.\footnote{In the KK decomposition, the 3-dimensional vector field $\textbf{A}_i$ does not contribute to the static potential, which is crucial to the factorization theorem \cite{Foffa:2019hrb}.} 
The metric components $g_{\mu\nu}$ ($=\eta_{\mu\nu}+H_{\mu\nu}$) are encoded in a scalar field $\bm{\phi}$,
a 3-dimensional vector $\bm{A}_i$, and a 3-dimensional symmetric tensor $\bm{\sigma}_{ij}$ as,
\begin{equation}
g_{\mu\nu} = e^{2\bm{\phi} / c^2}
\begin{pmatrix}
 1 & - {\bm{A}_j / c^2}\\
- {\bm{A}_i / c^2} \,\,\,\,\,\, & -e^{-c_d\bm{\phi}/c^2}\bm{\gamma}_{ij}
+ {\bm{A}_i\bm{A}_j / c^4}  
\end{pmatrix}\, ,
\end{equation}
with $\bm{\gamma}_{ij}=\bm{\delta}_{ij}+\bm{\sigma}_{ij}/c^2$ and $c_d=2(d-1)/(d-2)$. 
Then, integrating out the gravitational fields gives the two-body effective action,
\begin{align}
\text{exp}\Big[{\ii \int dt ~ \mathcal{L}_{\text{eff}}}\Big] = 
\int D\bm{\phi} \, D\bm{A}_i \, D\bm{\sigma}_{ij} ~e^{\ii(S_{\text{EH}}+S_{\text{pp}})} \, ,
\end{align}
where the effective Lagrangian $\mathcal{L}_{\rm eff}$ is decomposed as 
$\mathcal{L}_{\rm eff} = \mathcal{K}_{\rm eff} - \mathcal{V}_{\rm eff}$
with $\mathcal{K}_{\rm eff}$ the kinetic term and $\mathcal{V}_{\rm eff}$ the effective interaction potential. 
The kinetic term is independent of graviton exchange and, at any PN order, follows from expanding $\sqrt{1-v^2/c^2}$. By contrast, the potential encodes the gravitational interaction between the two bodies and is the primary quantity computed in this work.

\vspace{-10pt}
\subsection{Integrand generation}

The effective two-body potential $\mathcal{V}_{\text{eff}}$ can be expressed in terms of the connected, classical, one-particle-irreducible Feynman diagram as, 
\begin{align} \label{eq:effective_potential_diagrammatic}
\mathcal{V}_{\text{eff}}=\ii \lim_{d\rightarrow 3} \int \frac{d^d \textbf{p}}{(2\pi)^d} ~e^{\ii \textbf{p}\cdot (\textbf{x}_{(1)}-\textbf{x}_{(2)})} \ \times  
\parbox{20mm}{
\begin{tikzpicture}[line width=0.8 pt, scale=0.25]
    \begin{scope}[shift={(0,0)}]
        \filldraw[color=gray!40, fill=gray!40, thick](0,0) rectangle (3,3);
        \draw (-1,0)--(4,0);
        \draw (-1,3)--(4,3);
    \end{scope}
\end{tikzpicture}
} 
\end{align}
where, $\textbf{p}$ is the momentum transferred between the two massive objects. The four-point amplitude consists of a sum of 1117 six-loop diagrams\footnote{This counting does not include diagrams that can be obtained by the exchange of $m_1\leftrightarrow m_2$.}, 
as depicted in Fig.~\ref{fig:sixloopgrphs},
grouped according to the power of $m_1^{7-n} m_2^{1+n}$, respectively indicating the $n^{\rm th}$-order self force ($n$SF) contribution: 
the 0SF component corresponds to the simplest set of diagrams, encoding the test-body motion; whereas the 1SF, 2SF, and 3SF include contributions with an increasing level of complexity.

The code \texttt{PNTHR}: Post-Newtonian Toolkit for Hamiltonian and Radiation \cite{panther:inprep} was extended\footnote{The code \texttt{PNTHR} \cite{panther:inprep} has been previously used for the analyses 
up to 4 loops
of spinning compact objects in ref.\ \cite{Mandal:2022nty,Mandal:2022ufb} and tidally deformed compact objects in ref.\ \cite{Mandal:2023lgy,Mandal:2023hqa,Mandal:2024iug}.} to generate the required 6 loop graphs and the corresponding integrands.
\texttt{PNTHR} uses \texttt{QGRAF}~\cite{Nogueira:1991ex} to generate the skeleton of the diagrams, and after dressing them with the KK field and Feynman rules derived from the actions of PN expansion of GR given in eqs.~\eqref{eq_action_EH} and \eqref{eq_action_pp}, prepares the integrands by carrying out the tensor algebra using 
\texttt{xTensor}~\cite{xAct}.

\renewcommand{\arraystretch}{1.2}
\begin{table}[t]
\centering
\begin{tabular}{ccc}
\hline
\textbf{~~~SF order~~~} & \textbf{~~~Diagrams~~~} &  \textbf{~~~MIs~~~} \\ \hline
0 & 36  &1 \\ 
1 & 161 &3\\ 
2 & 387 &13\\ 
3 & 533 &16\\ 
\hline
Total & 1117  &21\\
\hline
\end{tabular}
\caption{Contributions to different SF sectors at 6PN. Multiple MIs are common to different SF orders.
}
\label{tbl_no_diag}
\end{table}

As observed in \cite{Kol:2013ega,Foffa:2016rgu}, the static integrands of the 6PN four-point diagrams within the EFT approach to GR can be remarkably mapped onto those of six-loop two-point functions with massless internal lines, typical of higher order corrections in QFT, whose external momentum is the momentum $\textbf{p}$ transferred of the massive bodies as, 
\begin{align}
\parbox{20mm}{ \  }
\parbox{15mm}{
\begin{tikzpicture}[line width=0.8 pt, scale=0.25]
    \begin{scope}[shift={(0,0)}]
        \filldraw[color=gray!40, fill=gray!40, thick](0,0) rectangle (3,3);
        \draw (-1,0)--(4,0);
        \draw (-1,3)--(4,3);
        \node at (1.5,1.5) {\scalebox{1}{{\tiny EFT}}};	
    \end{scope}
\end{tikzpicture}
} 
\longleftrightarrow \quad
\parbox{30mm}{
\begin{tikzpicture}[line width=0.8 pt, scale=0.20]
    \begin{scope}[shift={(0,0)}]
	\filldraw[color=gray!40, fill=gray!40, thick](0,1.5) circle (1.5);
	\draw (0,3)--(0,4);
	\draw (0,0)--(0,-1);	
    \node at (0.0,1.5) {{\tiny QFT}};	
    \end{scope}
\end{tikzpicture} 
}    
\label{eq:relation_nrgreft_qft}
\end{align}
We employ an in-house topology-mapping algorithm to group the contributing diagrams into 97 equivalence classes identified by their graphs (Symanzik) polynomials.  

Each topology represents a scalar integral, 
whose generic form reads as:
\begin{equation}
    I_{n_{1},\dots,n_{27}} = 
    \int_{k}\prod_{i=1}^{27}D_{i}^{-n_{i}}\,, \  {\rm with} \,\ \int_{k} := \int\prod_{i=1}^{6}\frac{d^{d}k_{i}}{(2\pi)^{d}} \, .
    \label{eq:genericintegral}
\end{equation}
in terms of 27 generalised denominators $D_i$, 
out of which 13 are genuine propagators' denominators  
($n_i \ge 1, \ i=1,\ldots, 13$), 
while the left-over 14 corresponds to irreducible scalar products ($n_{i}\leq0$ for $i=14,\dots,27$).

\vspace{-10pt}
\subsection{Integrals' decomposition}

Symanzik polynomials are also known as the $\mathcal{U}-$ and $\mathcal{F}-$ polynomials, and arise from considering the Feynman integrals in the Feynman parameter or Lee-Pomeransky representations.

IBP relations for the integrals of the considered problems stem from a system of symbolic equations reading as:
\begin{equation}
    \int_k\frac{\partial}{\partial k_{\ell}^{\mu}}\left(q^{\mu}\prod_{i=1}^{27}D_{i}^{-n_{i}}\right)= 0  \, ,
\end{equation}
where $q\in\{k_1,k_2,k_3,k_4,k_5,k_6,p\}$. 
By {\it seeding} the equations, namely 
by inserting different values of the indices $n_{i}$, and including symmetry relations (linear relations originating from shifting the loop momenta), we can generate a large system of linear equations for the integrals $I_{n_{1},\dots,n_{27}}$, whose solutions gives the decomposition relations, exactly in $d$, in terms of 39 MIs.

Let us remark that the decomposition of the integrals arising in this problem constitutes a highly demanding computational task. The relevant integrals are scalar six-loop integrals of rank up to six, corresponding to at most six powers of scalar products in the numerator, or equivalently to numerators with mass dimension not exceeding twelve. 
Their reduction to MIs cannot be achieved with commonly used public codes and instead required the development of a novel algorithm implemented within the code \texttt{PRISM} \cite{PRISM}: Prime-field Reduction of Integrals using Syzygy Methods. This algorithm combines multiple optimization strategies to efficiently generate and solve a minimal IBP system.

\emphh{Spanning Cuts:} 
Starting from the Symanzik polynomials of each topology, graph-theoretic considerations allow one to identify minimal edge cuts whose removal separates the diagram into two connected subgraphs, each containing at least one external leg.
Loosely speaking, a cut can be considered by replacing propagators with Dirac delta functions $D_{i}^{-1}\rightarrow\delta(D_{i})$ \cite{Larsen:2015ped}. For higher denominator powers it is not so simple, however due to the application of syzygies we do encounter integrals with these higher powers.
Applying IBP reduction to integrals associated with spanning cuts decomposes the original IBP system into smaller cut subsystems that are significantly easier to solve. The solution of the full system is then obtained by consistently combining the results from all spanning cuts that collectively cover the original graph \cite{Lee:2013mka,Maierhofer:2017gsa}.

\emphh{Syzygies:} Recent progress in syzygy-based approaches \cite{Gluza:2010ws,Wu:2023upw,Wu:2025aeg,Smith:2025xes} has significantly enhanced the algorithmic efficiency of IBP reductions. When integrated with spanning cuts, these methods substantially reduce the size and complexity of the IBP systems to be generated and solved. In \texttt{PRISM}, syzygy relations are computed using \texttt{Singular} \cite{DGPS} and are embedded directly into the IBP generation step, thereby constraining the system at the outset and improving the overall performance of the reduction algorithm.

\emphh{Seeding strategy:}  
\texttt{PRISM} employs an improved seeding strategy, inspired by techniques used in Kira and NeatIBP \cite{Lange:2025fba,Wu:2023upw,Wu:2025aeg}.

In the top sector, integrals are seeded up to rank six, with the maximal rank reduced according to the degree of the corresponding syzygy solution. As the algorithm descends to lower sectors, the maximum seed rank is decreased by one at each step, while integrals with squared denominators' exponents are excluded. Additional constraints inspired by rectangular seeding are imposed and tailored to the specific integrals appearing for each topology. In several instances, these combined strategies lead to a substantial reduction in both the size of the IBP systems and the overall computational cost.

After generating the IBP systems for each spanning cut and topology, the resulting linear systems are solved using \texttt{FiniteFlow} \cite{Peraro:2019svx}. The solver first applies a mark-and-sweep procedure by solving the system at a random numerical phase-space point modulo some large prime number (sample point). This allows for the elimination of redundant equations and identification of the relevant MIs from the predefined basis obtained through counting. Furthermore, the row reduction operations are remembered and can be applied much faster for subsequent runs. The reduced systems are then solved over finite fields at many more sample points, and the rational coefficients of the MIs are reconstructed from these evaluations.

The IBP reduction of these integrals constituted the most computationally demanding part of the calculation. In particular, the generation of the IBP system, which required solving syzygy equations, together with the initial mark-and-sweep procedure for each spanning cut, represented the most time-consuming steps.

\vspace{-10pt}
\subsection{Master Integrals}
\begin{figure}[ht]
\includegraphics[width=0.08\textwidth]{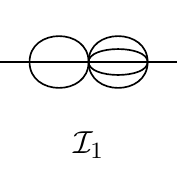}
\includegraphics[width=0.08\textwidth]{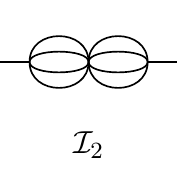}
\includegraphics[width=0.08\textwidth]{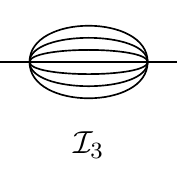}
\includegraphics[width=0.08\textwidth]{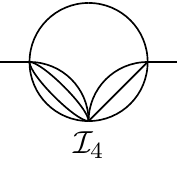}
\includegraphics[width=0.08\textwidth]{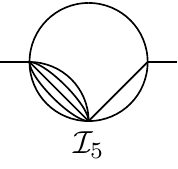}
\vspace{1mm}

\includegraphics[width=0.08\textwidth]{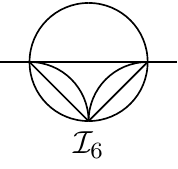}
\includegraphics[width=0.08\textwidth]{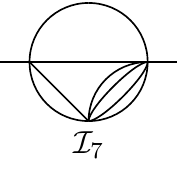}
\includegraphics[width=0.08\textwidth]{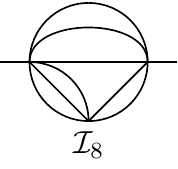}
\includegraphics[width=0.08\textwidth]{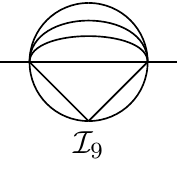}
\includegraphics[width=0.08\textwidth]{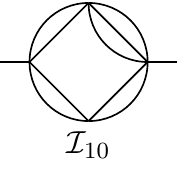}
\vspace{1mm}

\includegraphics[width=0.08\textwidth]{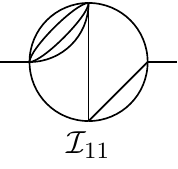}
\includegraphics[width=0.08\textwidth]{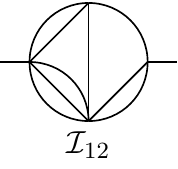}
\includegraphics[width=0.08\textwidth]{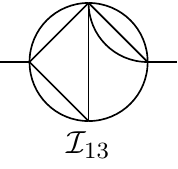}
\includegraphics[width=0.08\textwidth]{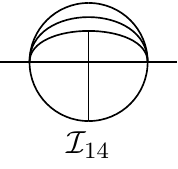}
\includegraphics[width=0.08\textwidth]{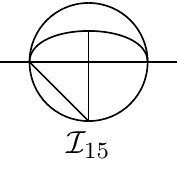}
\vspace{1mm}

\includegraphics[width=0.08\textwidth]{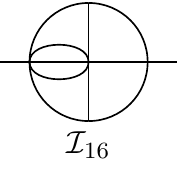}
\includegraphics[width=0.08\textwidth]{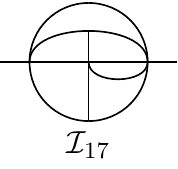}
\includegraphics[width=0.08\textwidth]{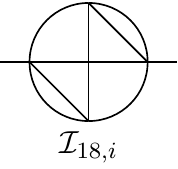}
\includegraphics[width=0.08\textwidth]{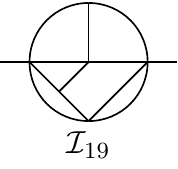}
\vspace{1mm}

\includegraphics[width=0.08\textwidth]{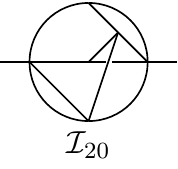}
\includegraphics[width=0.08\textwidth]{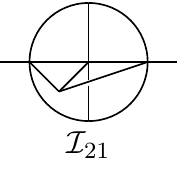}

\caption{The 21 MIs contributing to the 6PN static potential. The topology $\mathcal{I}_{18,i}$ has two MIs, $i\in\{1,2\}$, but only $\mathcal{I}_{18,2}$ appears in the final result. 
}
\label{fig:masterintegrals}
\end{figure}

The IBP reduction yields 39 MIs, out of which
only 21 contribute to the conservative potential. 
They are shown in Fig.~\ref{fig:masterintegrals}.
Let us observe that the 0SF sector gets contribution just from one  MI, namely $\mathcal{I}_3$, as expected \cite{Damgaard:2024fqj,Mougiakakos:2024lif},
whereas the 1SF, 2SF and 3SF sectors get contributions respectively from 3, 13 and 16 MIs.

The MIs $\mathcal{I}_{1}$ through  $\mathcal{I}_{10}$ are directly 
solvable as iterated massless one-loop two-point integrals, and are evaluated analytically in generic dimension $d$ using the standard identity: 
\begin{align}
\label{eq:one_loop_identity}
    &\int \frac{d^d k}{(2 \pi)^d} \frac{1}{(k^2)^{\nu_1} [(k-p)^2]^{\nu_2}}  \\
    &= \frac{(p^2)^{\frac{d}{2} - (\nu_1+\nu_2)}}{(4\pi)^{\frac{d}{2}} } \frac{  \Gamma \left(\frac{d}{2}-\nu_1\right) \Gamma \left(\frac{d}{2}-\nu_2\right) \Gamma \left(\nu_1+\nu_2-\frac{d}{2}\right) }{\Gamma (\nu_1) \Gamma (\nu_2) \Gamma (d-\nu_1-\nu_2)}\ .\nonumber
\end{align}
Remarkably, this subset is sufficient for the determination of the 0SF and 1SF terms of the complete result.
Using the identity in Eq.~\eqref{eq:one_loop_identity}, MIs  $\mathcal{I}_{11},\, \mathcal{I}_{12},\, \mathcal{I}_{13}$ are solvable as two-loop kite integrals with $d$-dependent denominator exponents, and are evaluated analytically, using the techniques proposed in Refs.~\cite{Chetyrkin:1980pr,Kotikov:1995cw,Moch:2001zr,Bierenbaum:2003ud,Grozin:2012xi}.
The MIs $\mathcal{I}_{14}$ through $\mathcal{I}_{21}$ are evaluated numerically around $d = 3 + \epsilon$ dimensions, using the auxiliary mass flow method implemented in \texttt{AMFlow}~\cite{Liu:2017jxz,Liu:2022chg}. 
Their analytic Laurent coefficients are reconstructed via the \texttt{PSLQ} algorithm~\cite{PSLQ}, employing 
a tailored ans\"atze for the expected transcendental constants, 
built from 
Harmonic Polylogarithm
constants~\cite{Remiddi:1999ew} evaluated at unit argument with transcendental weight up to three.~\footnote{
The \texttt{AMFlow} precision was at least 50 digits. The last eight digits were withheld from \texttt{PSLQ} and used a posteriori to validate the reconstruction.
}
We note that this is the first PN order at which the effective potential receives contributions from non-planar master integrals, namely, $\mathcal{I}_{20}$ and $\mathcal{I}_{21}$.

The definitions and analytic expressions of the 21 MIs of the problem can be found in the Appendix \ref{app:master_integrals} and \ref{app_expofMI}.

\emphh{Checks.} 
The 13 MIs that have been obtained analytically in $d$ dimensions have been checked by verifying that they satisfy dimensional recurrence relations~\cite{Lee:2009dh,Lee:2010wea,Lee:2015eva,Lee:2017ftw}, and via an independent numerical evaluation using \texttt{AMFlow}.
Independent numerical checks of all the six-loop MIs were performed using \texttt{feyntrop}~\cite{Borinsky:2023jdv}, which evaluates the integrals via Monte Carlo integration within a tropical-geometry framework. Such a comparison requires the determination of a quasi-finite basis of MIs in $d=5$, the construction of dimensional recurrence relations, and additional IBP reductions to relate the two bases. The resulting agreement between these computational strategies can be considered a stringent validation test on the MIs as well as on the IBP decomposition.\\

\section{Results}\label{sec_results}

Following Eq.~\eqref{eq:effective_potential_diagrammatic}, the effective two-body potential is obtained by Fourier transforming the amplitude using expression given in Appendix \ref{app_fourier},
expressed as linear combination of MIs
, and performing the Laurent expansion in $\epsilon = d-3$.
Finally, the resulting static two-body effective interaction potential at 6PN order reads
\begin{widetext}
\begin{align}\label{eq_6PN_potential}
    \mathcal{V}_{\rm 6PN}^{\rm G_N^7} = 
    -\frac{G_N^7}{r^7} 
    \Bigg( &\frac{5}{16}m_1^7 m_2
           +\frac{190}{9}m_1^6 m_2^2 +\frac{37651}{144}m_1^5 m_2^3
           +\frac{5852}{9}\frac{m_1^4 m_2^4}{2}
           +(1\leftrightarrow 2)\Bigg) \ .
\end{align}
\end{widetext}
This remarkably simple expression constitutes the main result of this letter.
The robustness of this result has been assessed in several ways.

\emphh{Finite and rational potential.} 
The final potential~\eqref{eq_6PN_potential} is finite in $d = 3$ spatial dimensions and purely rational, due to the exact cancellation of the $\epsilon$ poles, appearing at 2SF and 3SF orders, as well as the cancellation of the transcendental constant $\pi^2$ \cite{Damour:2017ced} which is present in the contributions of the individual 1SF-, 2SF- and 3SF-order terms.
The purely rational expression of the 6PN potential at 
${\cal O}(G_N^7)$ is in line with the known result of the static contribution at the lower PN orders, and the cancellation of the divergent terms along with the transcendental constant appearing in the intermediate expressions can be considered a non-trivial validation of our result.

\emphh{The $m_1^7 m_2$ term.} 
We further verified the result against the known test body limit. The static contribution to the interaction potential, linear in $m_2$, is given by
\begin{align}
    \mathcal{V}^{m_2\ll m_1} = m_2 \sqrt{-g_{00}} = m_2 \sqrt{\frac{1-\frac{G_Nm_1}{r}}{1+\frac{G_Nm_1}{r}}} \ ,
\end{align}
where the metric component $g_{00}$ is expressed in harmonic coordinates.
The Taylor expansion in $G_N$ yields the numerical coefficient $-5/16$ at $\mathcal{O}(G_N^7)$, which exactly agrees with the coefficient of the $m_1^7 m_2$ term in Eq.~\eqref{eq_6PN_potential}.

\emphh{The static contribution at 5PN.} 
Let us remark that our complete computational pipeline, 
from the automatic diagrams and integrands generation to the master integral decomposition, and the Fourier transform of the amplitude from momentum to position space in $d=3$ dimensions
has been validated by computing the 5PN static contribution, 
reproducing the result of Refs.~\cite{Foffa:2019hrb,Blumlein:2019zku}. 

\vspace{-5pt}
\section{Conclusion}\label{sec_conc}

In this Letter, we computed the static contribution to the conservative two-body potential at 6PN order within the EFT framework. This term constitutes the most technically demanding component of the 6PN conservative dynamics and, for the first time, incorporates a third-order SF contribution. The calculation required the evaluation of six-loop massless two-point Feynman integrals, never addressed before.
We achieved this by combining state-of-the-art multi-loop techniques, including automatic diagrams generation, IBP identities supplemented by spanning cuts, syzygy-based methods, and optimized seeding strategies. The analytic expressions of the MIs are obtained through a combination of analytic and numerical evaluations, together with reconstruction algorithms. 
Our result, together with the methods developed to obtain it, has significant implications for several future research directions. On the one hand, it represents a major milestone toward the completion of the full 6PN conservative dynamics, which is essential for achieving accurate theoretical predictions of the inspiral phase of compact binary coalescences for current and future gravitational-wave observatories. In particular, velocity-dependent contributions at 6PN order—requiring the evaluation of lower-loop integrals with higher-rank numerators—as well as radiation-reaction effects at this order, can be addressed using the automated framework developed in this work. This approach naturally extends and complements existing partial results obtained up to three-loop order using EFT methods \cite{Blumlein:2021txj}, as well as complementary information on local-in-time 6PN contributions derived from alternative approaches \cite{Bini:2020nsb,Bini:2020rzn,Bini:2020uiq}. Our result thus fills a crucial missing gap and completes the currently known local contributions to the 6PN dynamics, and can be considered the state-of-the art for post-Newtonian correction in General Relativity.

On the other hand, important missing ingredients remain in the form of hereditary, or non-local-in-time, effects.  
These contributions have been partially determined at 5PN and 6PN order in Refs.~\cite{Almeida:2021xwn,Bini:2020hmy,Edison:2023qvg,Porto:2024cwd}. 
A dedicated and systematic analysis will be required to complete these terms at 6PN.
With regard to higher-order PN corrections, the static contribution at 6PN order provides a key input for the application of the factorization theorem \cite{Foffa:2019hrb} to the derivation of the 7PN static term, which follows from purely combinatorial properties together with all lower-order static contributions.

Beyond its implications for General Relativity, this work illustrates a fruitful cross-fertilization between gravitational and gauge-theory computations. The diagrammatic techniques, IBP strategies, and reduction algorithms developed for high-order post-Newtonian gravity closely mirror—and in several respects extend—those employed in multi-loop gauge-theory calculations. As such, they pave the way for previously inaccessible high-precision results in gauge theories, including the six-loop 
$\beta$-functions in quantum electrodynamics and quantum chromodynamics and related observables, while at the same time benefiting from conceptual and technical advances originally developed in the context of perturbative QFT.

\vspace{-10pt}
\section*{Acknowledgments}

We thank Mao Zeng for granting access to his private Mathematica–Singular interface and to his implementation of the algorithm for identifying non-vanishing sectors.
We wish to thank Stefano Foffa, for stimulating discussions and collaboration at early stages.
We acknowledge Donato Bini and Thibault Damour for comments on the manuscript, and 
Nicola Bartolo, 
Jungwon Lim,
Angelo Ricciardone, 
Benjamin Sauer,
and
Trevor Scheopner
for useful discussions.
We gratefully acknowledge the stimulating scientific environments of ``Domoschool 2025" and ``Loop-The-Loop 2025", which provided valuable opportunities for discussion and for the development of the novel aspects of this work.
The computations for this work were performed on the Slarti and Hypatia clusters at the Max Planck Institute for Gravitational Physics in Potsdam, with additional computing and storage resources provided by the CloudVeneto initiative at the University of Padova and INFN.
G.B.'s research is supported by the Italian MIUR under contract 20223ANFHR (PRIN2022) and by the ERC (NOTIMEFORCOSMO, 101126304).
R.P.’s research is funded by the Deutsche Forschungsgemeinschaft (DFG, German Research Foundation), Projektnummer 417533893/GRK2575 ``Rethinking Quantum Field Theory''.
M.P.'s research is supported by the European Union under the Next Generation EU programme. M.P. wishes to thank the support of Fondazione Ing. Aldo Gini, and the hospitality and support of the Albert Einstein Institute.
W.J.T.'s research is supported by the Leverhulme Trust, LIP-2021-014.
This work has been supported by the INFN initiatives \textit{Amplitudes} and \textit{InDark}.

\appendix

\vspace{-10pt}
\section{Master integrals}
\label{app:master_integrals}
The list of denominators appearing in the MIs are
\begin{align}
&
D_{i\in\{1,\dots,6\}} = k_{i}^{2}\,, 
\quad 
D_7 = (k_1-k_3)^2\,, 
\nonumber \\ &
D_8 = (k_1-k_4)^2\,,
\quad
D_9 = (k_2-k_4)^2\,,
\nonumber \\ &
D_{10} = (k_3-k_4)^2\,, 
\quad 
D_{11} = (k_2-k_5)^2\,, 
\nonumber \\ &
D_{12} = (k_3-k_5)^2\,, 
\quad 
D_{13} = (k_4-k_5)^2\,, 
\nonumber \\ &
D_{14} = (k_3-k_6)^2\,, 
\quad
D_{15} = (k_4-k_6)^2\,, 
\nonumber \\ &
D_{16} = (k_5-k_6)^2\,, 
\quad 
D_{17} = (k_2+k_6)^2\,,
\nonumber \\ &
D_{18} = (k_3+k_6)^2\,, 
\quad 
D_{19} = (k_4+k_6)^2\,, 
\nonumber \\ & 
D_{20}= (k_5+k_6)^2\,,
\quad
D_{21} = (k_1+k_3-k_6)^2\,, 
\nonumber \\ &  
D_{22} = (k_2+k_3-k_6)^2\,, 
\quad
D_{23} = (k_2+k_4-k_6)^2\,,
\nonumber \\ &  
D_{24} = (k_3+k_4-k_6)^2\,,
\quad
D_{25} = (k_1+k_2+k_6)^2\,, 
\nonumber \\ &  
D_{26} = (k_2+k_3+k_6)^2\,, 
\quad
D_{27} = (k_2+k_4+k_6)^2\,, 
\nonumber \\ & 
D_{28} = (k_1+k_5+k_6)^2\,, 
\quad
D_{29} =  (k_2+k_5+k_6)^2\,, 
\nonumber \\ &
D_{30} = (k_3+k_5+k_6)^2\,,
\quad
D_{31} = (k_2+k_3-p)^2\,, 
\nonumber \\ &
D_{32} = (k_2+k_4-p)^2\,,
\quad
D_{33} = (k_2-k_6-p)^2\,, 
\nonumber \\ &
D_{34} = (k_2+k_6-p)^2\,, 
\quad
D_{35} = (k_4-k_6+p)^2\,, 
\nonumber \\ &
D_{36} = (k_2+k_3+k_4-k_6)^2\,,
\quad
D_{37} = (k_1+k_2+k_3-p)^2\,,
\nonumber \\ &
D_{38} = (k_1+k_3+k_5-p)^2\,,  
\quad 
D_{39} = (k_2+k_3+k_5-p)^2\,, 
\nonumber \\ &
D_{40} = (k_2-k_4-k_6-p)^2\,, 
\quad
D_{41} = (k_3-k_5-k_6-p)^2\,, 
\nonumber \\ &
D_{42} = (k_2+k_3-k_5-k_6-p)^2\,, 
\nonumber \\ &
D_{43} = (k_1+k_2+k_3-k_4+k_6-p)^2\,,
\nonumber \\ & 
D_{44} = (k_6+p)^2\,.
\label{eq:denominators_definitions}
\end{align}

The MIs appearing in Fig.~\ref{fig:masterintegrals} are defined as:
\begin{align}
&\mathcal{I}_{1} = \int_{k}\frac{1}{D_{1}D_{2}D_{3}D_{8}D_{11}D_{24}D_{31}D_{42}}\, , \nonumber\\
&\mathcal{I}_{2} = \int_{k}\frac{1}{D_{1}D_{2}D_{3}D_{9}D_{13}D_{16}D_{33}D_{37}}\, , \nonumber\\
&\mathcal{I}_{3} = \int_{k}\frac{1}{D_{1}D_{2}D_{4}D_{7}D_{12}D_{16}D_{40}} 
\, ,\nonumber\\ 
&\mathcal{I}_{4} = \int_{k}\frac{1}{D_{1}D_{2}D_{3}D_{8}D_{11}D_{20}D_{24}D_{42}}\, , \nonumber\\
&\mathcal{I}_{5} = \int_{k}\frac{1}{D_{1}D_{2}D_{7}D_{9}D_{12}D_{16}D_{19}D_{40}}\, ,
\nonumber\\ &
\mathcal{I}_{6} = \int_{k}\frac{1}{D_{1}D_{2}D_{3}D_{8}D_{18}D_{23}D_{30}D_{39}} \, ,
\nonumber\\ &
\mathcal{I}_{7} = \int_{k}\frac{1}{D_{1}D_{2}D_{7}D_{12}D_{16}D_{17}D_{27}D_{32}}\, , \nonumber\\ &
\mathcal{I}_{8} = \int_{k}\frac{1}{D_{1}D_{2}D_{6}D_{7}D_{13}D_{16}D_{26}D_{32}} \, ,\nonumber\\
&
\mathcal{I}_{9} = \int_{k}\frac{1}{D_{1}D_{2}D_{3}D_{6}D_{8}D_{13}D_{22}D_{41}} \, ,
\nonumber\\ &
\mathcal{I}_{10} = \int_{k}\frac{1}{D_{1}D_{2}D_{3}D_{4}D_{5}D_{20}D_{21}D_{34}D_{35}} \, ,
\nonumber\\
&
\mathcal{I}_{11} = \int_{k}\frac{1}{D_{1}D_{2}D_{3}D_{6}D_{8}D_{13}D_{22}D_{31}D_{42}} \, ,
\nonumber\\ 
&
\mathcal{I}_{12} = \int_{k}\frac{1}{D_{1}D_{2}D_{3}D_{8}D_{18}D_{23}D_{30}D_{31}D_{39}} \, ,
\nonumber\\
&
\mathcal{I}_{13} = \int_{k}\frac{1}{D_{1}D_{2}D_{3}D_{5}D_{9}D_{13}D_{16}D_{33}D_{37}} \, ,
\nonumber\\ &
\mathcal{I}_{14} = \int_{k}\frac{1}{D_{1}D_{2}D_{4}D_{5}D_{10}D_{13}D_{16}D_{17}D_{25}D_{37}} \, ,
\nonumber\\
&
\mathcal{I}_{15} = \int_{k}\frac{1}{D_{1}D_{2}D_{3}D_{5}D_{10}D_{13}D_{16}D_{17}D_{25}D_{37}}\, ,
\nonumber\\
&
\mathcal{I}_{16} = \int_{k}\frac{1}{D_{1}D_{2}D_{3}D_{8}D_{13}D_{14}D_{20}D_{29}D_{31}D_{34}}\, ,
\nonumber\\
&
\mathcal{I}_{17} = \int_{k}\frac{1}{D_{1}D_{2}D_{3}D_{4}D_{10}D_{13}D_{16}D_{17}D_{25}D_{37}} \, ,
\nonumber\\
&
\mathcal{I}_{18,1} = \int_{k}\frac{1}{D_{1}D_{2}D_{3}D_{4}D_{5}D_{14}D_{20}D_{36}D_{37}D_{39}}\, ,
\nonumber\\
&
\mathcal{I}_{18,2} = \int_{k}\frac{D_{44}}{D_{1}D_{2}D_{3}D_{4}D_{5}D_{14}D_{20}D_{36}D_{37}D_{39}}\, ,
\nonumber\\
&
\mathcal{I}_{19} = \int_{k}\frac{1}{D_{1}D_{2}D_{3}D_{6}D_{8}D_{11}D_{14}D_{20}D_{24}D_{37}D_{38}} \, ,
\nonumber\\
&
\mathcal{I}_{20} = \int_{k}\frac{1}{D_{1}D_{2}D_{3}D_{4}D_{6}D_{11}D_{20}D_{24}D_{28}D_{37}D_{38}} \, ,
\nonumber\\
&
\mathcal{I}_{21} = \int_{k}\frac{1}{D_{1}D_{2}D_{3}D_{5}D_{8}D_{12}D_{15}D_{20}D_{29}D_{37}D_{43}}\, .
\end{align}

\section{Analytic expression of the MIs}
\label{app_expofMI}
The analytic expression of the MIs, given as a Laurent series in $\epsilon$ up to the needed order, with $\epsilon= d-3$, and with overall normalization
\begin{equation}
 c_\epsilon=  \frac{(p^2)^{2+3\epsilon}e^{3 \gamma_E  \epsilon}}{(4 \pi) ^{3( \epsilon+2)}} \, ,
\end{equation}
reads as:
\begin{align}
\mathcal{I}_{1}
 &
 =
 \frac{c_\epsilon}{(p^2)}\biggl[
 -\frac{1}{48 \epsilon ^2}+\frac{35}{144 \epsilon }-\frac{199}{108}+\frac{29 \pi ^2}{1152}+\mathcal{O}(\epsilon)
 \biggr] \, , 
\nonumber\\ 
 \mathcal{I}_{2}
 &
 =
 \frac{c_\epsilon}{(p^2)}\biggl[
 \frac{\pi ^2}{16}+\mathcal{O}(\epsilon)
 \biggr]\, ,
 \nonumber\\  
 \mathcal{I}_{3}
 &
 =
 c_\epsilon\biggl[
 -\frac{1}{720\epsilon}+\frac{457}{21600}+\mathcal{O}(\epsilon)
 \biggr]\, ,
 \nonumber\\ 
 \mathcal{I}_{4}
 &
 =
 \frac{c_\epsilon}{(p^2)}\biggl[
 -\frac{1}{72 \epsilon ^2}+\frac{37}{216 \epsilon }-\frac{229}{162}+\frac{23 \pi ^2}{576}+\mathcal{O}(\epsilon)
 \biggr]\, ,
 \nonumber\\ 
 \mathcal{I}_{5}
 &
 =
 \frac{c_\epsilon}{(p^2)}\biggl[
 -\frac{1}{432 \epsilon ^2}+\frac{13}{432 \epsilon }-\frac{259}{972}+\frac{31 \pi ^2}{3456}+\mathcal{O}(\epsilon)
 \biggr]\, ,
 \nonumber\\  
 \mathcal{I}_{6}
 &
 =
 \frac{c_\epsilon}{(p^2)}\biggl[
 -\frac{1}{36 \epsilon ^2}+\frac{10}{27 \epsilon }-\frac{527}{162}+\frac{37 \pi ^2}{864}+\mathcal{O}(\epsilon)
 \biggr]\, ,
 \nonumber\\  
 \mathcal{I}_{7}
 &
 =
 \frac{c_\epsilon}{(p^2)}\biggl[
 \frac{\pi ^2}{72}+\mathcal{O}(\epsilon)
 \biggr]\, , 
 \nonumber\\  
 \mathcal{I}_{8}
 &
 =
 \frac{c_\epsilon}{(p^2)}\biggl[
 -\frac{1}{72 \epsilon ^2}+\frac{23}{108 \epsilon }-\frac{173}{81}+\frac{61 \pi ^2}{1728}+\mathcal{O}(\epsilon)
 \biggr]\, ,
 \nonumber\\  
 \mathcal{I}_{9}
 &
 =
 \frac{c_\epsilon}{(p^2)}\biggl[
 \frac{\pi ^2}{144 \epsilon }-\frac{37 \pi ^2}{432}-\frac{1}{72} \pi ^2 \log (2)+\mathcal{O}(\epsilon)
 \biggr]\, ,
 \nonumber\\  
 \mathcal{I}_{10}
 &
 =
 \frac{c_\epsilon}{(p^2)^2}
 \biggl[
 \frac{\pi ^2}{48 \epsilon ^2}-\frac{\pi ^2 }{24 \epsilon }(5+\log (2))+\mathcal{O}(\epsilon^0)
 \biggr]\, , 
 \nonumber\\  
 \mathcal{I}_{11}
 &
 =
 \frac{c_\epsilon}{(p^2)^{2}}\biggl[
 -\frac{1}{48 \epsilon^2}+\frac{17}{48 \epsilon}-\frac{47}{12}-\frac{\pi ^2}{384}+\mathcal{O}(\epsilon)
 \biggr]\, , 
 \nonumber\\  
 \mathcal{I}_{12}
 &
 =
\frac{c_\epsilon}{(p^2)^{2}}\biggl[
 -\frac{1}{48 \epsilon ^3}+\frac{5}{16 \epsilon ^2}-\frac{1}{24 \varepsilon}\left(77-\frac{61 \pi ^2}{48}\right)+\mathcal{O}(\epsilon^0)
 \biggr]\, , 
 \nonumber\\  
 \mathcal{I}_{13}
 &
 =
 \frac{c_\epsilon}{(p^2)^{2}}\biggl[
 \frac{\pi ^2}{48 \epsilon ^2}-\frac{\pi^2}{24\epsilon}\left(5+\log(2)\right)+\mathcal{O}(\epsilon^0)
 \biggr]\, ,
 \nonumber\\  
 \mathcal{I}_{14}
 &
 =
\frac{c_\epsilon}{(p^2)^{3}}\biggl[
-\frac{1}{4 \epsilon ^2}-\frac{3}{2 \epsilon }+\mathcal{O}(\epsilon^0)
\biggr]\, ,
 \nonumber\\  
 \mathcal{I}_{15}
 &
 =
 \frac{c_\epsilon}{(p^2)^{3}}\biggl[
 \frac{1}{48 \epsilon ^3}+\frac{3}{16 \epsilon ^2}+\frac{1}{24\epsilon}\left(29-\frac{47 \pi ^2}{16 }\right)+\mathcal{O}(\epsilon^0)
 \biggr]\, ,
 \nonumber\\ 
 \mathcal{I}_{16}
 &
 =
 \frac{c_\epsilon}{(p^2)^{3}}\biggl[
 -\frac{1}{8 \epsilon ^3}+\frac{1}{8 \epsilon ^2}-\frac{1}{4\epsilon}\left(3-\frac{13 \pi ^2}{48 }\right)+\mathcal{O}(\epsilon^0)
 \biggr]\, ,
 \nonumber\\  
 \mathcal{I}_{17}
 &
 =
  \frac{c_\epsilon}{(p^2)^{3}}\biggl[
 \frac{1}{24 \epsilon ^3}+\frac{7}{24 \epsilon ^2}+\frac{1}{12\epsilon}\left(19-\frac{47 \pi ^2}{16  }\right)+\mathcal{O}(\epsilon^0)
 \biggr]\, ,
 \nonumber\\  
 \mathcal{I}_{18,2}
 &
 =
\frac{c_\epsilon}{(p^2)^{2}}\biggl[
 \frac{\pi ^2}{24 \epsilon ^2}-\frac{\pi ^2}{3 \epsilon }\left(1+\frac{\log (2)}{4}\right)+\mathcal{O}(\epsilon^0)
 \biggr]\, ,
 \nonumber\\  
 \mathcal{I}_{19}
 &
 =
\frac{c_\epsilon}{(p^2)^{4}}\biggl[
 \frac{5}{48 \epsilon ^3}-\frac{7}{24 \epsilon ^2}-\frac{1}{48\epsilon}\left(139+\frac{95 \pi ^2}{8}\right)+\mathcal{O}(\epsilon^0)
 \biggr]\, ,
 \nonumber\\  
 \mathcal{I}_{20}
 &
 =  
\frac{c_\epsilon}{(p^2)^{4}}\biggl[\frac{1}{24 \epsilon^3}-\frac{1}{24 \epsilon}\left(41+\frac{59}{8}\pi^2\right)+\mathcal{O}(\epsilon^0)\biggr] \, ,
\nonumber\\ 
\mathcal{I}_{21} 
&
=   
\frac{c_\epsilon}{(p^2)^{4}}\biggl[ \frac{1}{4 \epsilon^3}-\frac{1}{4 \epsilon^2} -\frac{1}{
    \epsilon}\left(\frac{7}{2}+\frac{3 \pi^2}{32}\right)+\mathcal{O}(\epsilon^0)\biggr]\, .
    \nonumber\\
\end{align}
We observe that, up to the given order in $\epsilon$, the following relations are valid:
\begin{align}
    \mathcal{I}_{2} -\frac{9}{2}  \, \mathcal{I}_{7} = \mathcal{O}(\epsilon)\, ,\quad \nonumber \rm{and}\,\quad
      \mathcal{I}_{10} - \mathcal{I}_{13} = \mathcal{O}(\epsilon^0)\,  . 
\end{align}
Such relations could originate from Gram determinant or Schouten identities in the $d=3$ case \cite{Remiddi:2013joa,Gluza:2010ws}. This will be a subject for future studies.

\section{Analytic expression of the Fourier integral}
\label{app_fourier}
The expression for the scalar Fourier integral is given by 
\begin{align}
\int_k  e^{\ii \textbf{k}^i  \textbf{x}_i} \left( \textbf{k}^i \textbf{k}_i \right)^{-\alpha} = \frac{2^{-2 \alpha } \pi ^{-\frac{d}{2}} \Gamma \left(\frac{d}{2}-\alpha \right) \left(\textbf{x}^i \textbf{x}_i \right)^{\alpha -\frac{d}{2}}}{\Gamma (\alpha )}\, .
\label{eq_fourier}
\end{align}
which is for generic dimensions but only in Euclidean signature. The required tensor Fourier integrals are then obtained by taking derivatives on both sides with respect to $x^i$.

\bibliography{biblio}

\begin{thebibliography}{177}%
\makeatletter
\providecommand \@ifxundefined [1]{%
 \@ifx{#1\undefined}
}%
\providecommand \@ifnum [1]{%
 \ifnum #1\expandafter \@firstoftwo
 \else \expandafter \@secondoftwo
 \fi
}%
\providecommand \@ifx [1]{%
 \ifx #1\expandafter \@firstoftwo
 \else \expandafter \@secondoftwo
 \fi
}%
\providecommand \natexlab [1]{#1}%
\providecommand \enquote  [1]{``#1''}%
\providecommand \bibnamefont  [1]{#1}%
\providecommand \bibfnamefont [1]{#1}%
\providecommand \citenamefont [1]{#1}%
\providecommand \href@noop [0]{\@secondoftwo}%
\providecommand \href [0]{\begingroup \@sanitize@url \@href}%
\providecommand \@href[1]{\@@startlink{#1}\@@href}%
\providecommand \@@href[1]{\endgroup#1\@@endlink}%
\providecommand \@sanitize@url [0]{\catcode `\\12\catcode `\$12\catcode `\&12\catcode `\#12\catcode `\^12\catcode `\_12\catcode `\%12\relax}%
\providecommand \@@startlink[1]{}%
\providecommand \@@endlink[0]{}%
\providecommand \url  [0]{\begingroup\@sanitize@url \@url }%
\providecommand \@url [1]{\endgroup\@href {#1}{\urlprefix }}%
\providecommand \urlprefix  [0]{URL }%
\providecommand \Eprint [0]{\href }%
\providecommand \doibase [0]{http://dx.doi.org/}%
\providecommand \selectlanguage [0]{\@gobble}%
\providecommand \bibinfo  [0]{\@secondoftwo}%
\providecommand \bibfield  [0]{\@secondoftwo}%
\providecommand \translation [1]{[#1]}%
\providecommand \BibitemOpen [0]{}%
\providecommand \bibitemStop [0]{}%
\providecommand \bibitemNoStop [0]{.\EOS\space}%
\providecommand \EOS [0]{\spacefactor3000\relax}%
\providecommand \BibitemShut  [1]{\csname bibitem#1\endcsname}%
\let\auto@bib@innerbib\@empty
\bibitem [{\citenamefont {Abbott}\ \emph {et~al.}(2023{\natexlab{a}})\citenamefont {Abbott} \emph {et~al.}}]{KAGRA:2021vkt}%
  \BibitemOpen
  \bibfield  {author} {\bibinfo {author} {\bibfnamefont {R.}~\bibnamefont {Abbott}} \emph {et~al.} (\bibinfo {collaboration} {KAGRA, VIRGO, LIGO Scientific}),\ }\bibfield  {title} {\enquote {\bibinfo {title} {{GWTC-3: Compact Binary Coalescences Observed by LIGO and Virgo during the Second Part of the Third Observing Run}},}\ }\href {\doibase 10.1103/PhysRevX.13.041039} {\bibfield  {journal} {\bibinfo  {journal} {Phys. Rev. X}\ }\textbf {\bibinfo {volume} {13}},\ \bibinfo {pages} {041039} (\bibinfo {year} {2023}{\natexlab{a}})},\ \Eprint {http://arxiv.org/abs/2111.03606} {arXiv:2111.03606 [gr-qc]} \BibitemShut {NoStop}%
\bibitem [{\citenamefont {Abac}\ \emph {et~al.}(2025{\natexlab{a}})\citenamefont {Abac} \emph {et~al.}}]{LIGOScientific:2025slb}%
  \BibitemOpen
  \bibfield  {author} {\bibinfo {author} {\bibfnamefont {A.~G.}\ \bibnamefont {Abac}} \emph {et~al.} (\bibinfo {collaboration} {LIGO Scientific, VIRGO, KAGRA}),\ }\bibfield  {title} {\enquote {\bibinfo {title} {{GWTC-4.0: Updating the Gravitational-Wave Transient Catalog with Observations from the First Part of the Fourth LIGO-Virgo-KAGRA Observing Run}},}\ }\href@noop {} {\  (\bibinfo {year} {2025}{\natexlab{a}})},\ \Eprint {http://arxiv.org/abs/2508.18082} {arXiv:2508.18082 [gr-qc]} \BibitemShut {NoStop}%
\bibitem [{\citenamefont {Punturo}\ \emph {et~al.}(2010)\citenamefont {Punturo} \emph {et~al.}}]{Punturo:2010zz}%
  \BibitemOpen
  \bibfield  {author} {\bibinfo {author} {\bibfnamefont {M.}~\bibnamefont {Punturo}} \emph {et~al.},\ }\bibfield  {title} {\enquote {\bibinfo {title} {{The Einstein Telescope: A third-generation gravitational wave observatory}},}\ }\href {\doibase 10.1088/0264-9381/27/19/194002} {\bibfield  {journal} {\bibinfo  {journal} {Class. Quant. Grav.}\ }\textbf {\bibinfo {volume} {27}},\ \bibinfo {pages} {194002} (\bibinfo {year} {2010})}\BibitemShut {NoStop}%
\bibitem [{\citenamefont {Reitze}\ \emph {et~al.}(2019)\citenamefont {Reitze} \emph {et~al.}}]{Reitze:2019iox}%
  \BibitemOpen
  \bibfield  {author} {\bibinfo {author} {\bibfnamefont {David}\ \bibnamefont {Reitze}} \emph {et~al.},\ }\bibfield  {title} {\enquote {\bibinfo {title} {{Cosmic Explorer: The U.S. Contribution to Gravitational-Wave Astronomy beyond LIGO}},}\ }\href@noop {} {\bibfield  {journal} {\bibinfo  {journal} {Bull. Am. Astron. Soc.}\ }\textbf {\bibinfo {volume} {51}},\ \bibinfo {pages} {035} (\bibinfo {year} {2019})},\ \Eprint {http://arxiv.org/abs/1907.04833} {arXiv:1907.04833 [astro-ph.IM]} \BibitemShut {NoStop}%
\bibitem [{\citenamefont {Amaro-Seoane}\ \emph {et~al.}(2017)\citenamefont {Amaro-Seoane} \emph {et~al.}}]{LISA:2017pwj}%
  \BibitemOpen
  \bibfield  {author} {\bibinfo {author} {\bibfnamefont {Pau}\ \bibnamefont {Amaro-Seoane}} \emph {et~al.} (\bibinfo {collaboration} {LISA}),\ }\bibfield  {title} {\enquote {\bibinfo {title} {{Laser Interferometer Space Antenna}},}\ }\href@noop {} {\  (\bibinfo {year} {2017})},\ \Eprint {http://arxiv.org/abs/1702.00786} {arXiv:1702.00786 [astro-ph.IM]} \BibitemShut {NoStop}%
\bibitem [{\citenamefont {Abac}\ \emph {et~al.}(2025{\natexlab{b}})\citenamefont {Abac} \emph {et~al.}}]{LIGOScientific:2025pvj}%
  \BibitemOpen
  \bibfield  {author} {\bibinfo {author} {\bibfnamefont {A.~G.}\ \bibnamefont {Abac}} \emph {et~al.} (\bibinfo {collaboration} {LIGO Scientific, VIRGO, KAGRA}),\ }\bibfield  {title} {\enquote {\bibinfo {title} {{GWTC-4.0: Population Properties of Merging Compact Binaries}},}\ }\href@noop {} {\  (\bibinfo {year} {2025}{\natexlab{b}})},\ \Eprint {http://arxiv.org/abs/2508.18083} {arXiv:2508.18083 [astro-ph.HE]} \BibitemShut {NoStop}%
\bibitem [{\citenamefont {Abbott}\ \emph {et~al.}(2018)\citenamefont {Abbott} \emph {et~al.}}]{LIGOScientific:2018cki}%
  \BibitemOpen
  \bibfield  {author} {\bibinfo {author} {\bibfnamefont {B.~P.}\ \bibnamefont {Abbott}} \emph {et~al.} (\bibinfo {collaboration} {LIGO Scientific, Virgo}),\ }\bibfield  {title} {\enquote {\bibinfo {title} {{GW170817: Measurements of neutron star radii and equation of state}},}\ }\href {\doibase 10.1103/PhysRevLett.121.161101} {\bibfield  {journal} {\bibinfo  {journal} {Phys. Rev. Lett.}\ }\textbf {\bibinfo {volume} {121}},\ \bibinfo {pages} {161101} (\bibinfo {year} {2018})},\ \Eprint {http://arxiv.org/abs/1805.11581} {arXiv:1805.11581 [gr-qc]} \BibitemShut {NoStop}%
\bibitem [{\citenamefont {Abbott}\ \emph {et~al.}(2017)\citenamefont {Abbott} \emph {et~al.}}]{LIGOScientific:2017adf}%
  \BibitemOpen
  \bibfield  {author} {\bibinfo {author} {\bibfnamefont {B.~P.}\ \bibnamefont {Abbott}} \emph {et~al.} (\bibinfo {collaboration} {LIGO Scientific, Virgo, 1M2H, Dark Energy Camera GW-E, DES, DLT40, Las Cumbres Observatory, VINROUGE, MASTER}),\ }\bibfield  {title} {\enquote {\bibinfo {title} {{A gravitational-wave standard siren measurement of the Hubble constant}},}\ }\href {\doibase 10.1038/nature24471} {\bibfield  {journal} {\bibinfo  {journal} {Nature}\ }\textbf {\bibinfo {volume} {551}},\ \bibinfo {pages} {85--88} (\bibinfo {year} {2017})},\ \Eprint {http://arxiv.org/abs/1710.05835} {arXiv:1710.05835 [astro-ph.CO]} \BibitemShut {NoStop}%
\bibitem [{\citenamefont {Abbott}\ \emph {et~al.}(2023{\natexlab{b}})\citenamefont {Abbott} \emph {et~al.}}]{LIGOScientific:2021aug}%
  \BibitemOpen
  \bibfield  {author} {\bibinfo {author} {\bibfnamefont {R.}~\bibnamefont {Abbott}} \emph {et~al.} (\bibinfo {collaboration} {LIGO Scientific, Virgo, KAGRA}),\ }\bibfield  {title} {\enquote {\bibinfo {title} {{Constraints on the Cosmic Expansion History from GWTC{\textendash}3}},}\ }\href {\doibase 10.3847/1538-4357/ac74bb} {\bibfield  {journal} {\bibinfo  {journal} {Astrophys. J.}\ }\textbf {\bibinfo {volume} {949}},\ \bibinfo {pages} {76} (\bibinfo {year} {2023}{\natexlab{b}})},\ \Eprint {http://arxiv.org/abs/2111.03604} {arXiv:2111.03604 [astro-ph.CO]} \BibitemShut {NoStop}%
\bibitem [{\citenamefont {Abac}\ \emph {et~al.}(2025{\natexlab{c}})\citenamefont {Abac} \emph {et~al.}}]{LIGOScientific:2025jau}%
  \BibitemOpen
  \bibfield  {author} {\bibinfo {author} {\bibfnamefont {A.~G.}\ \bibnamefont {Abac}} \emph {et~al.} (\bibinfo {collaboration} {LIGO Scientific, VIRGO, KAGRA}),\ }\bibfield  {title} {\enquote {\bibinfo {title} {{GWTC-4.0: Constraints on the Cosmic Expansion Rate and Modified Gravitational-wave Propagation}},}\ }\href@noop {} {\  (\bibinfo {year} {2025}{\natexlab{c}})},\ \Eprint {http://arxiv.org/abs/2509.04348} {arXiv:2509.04348 [astro-ph.CO]} \BibitemShut {NoStop}%
\bibitem [{\citenamefont {Abbott}\ \emph {et~al.}(2016)\citenamefont {Abbott} \emph {et~al.}}]{LIGOScientific:2016lio}%
  \BibitemOpen
  \bibfield  {author} {\bibinfo {author} {\bibfnamefont {B.~P.}\ \bibnamefont {Abbott}} \emph {et~al.} (\bibinfo {collaboration} {LIGO Scientific, Virgo}),\ }\bibfield  {title} {\enquote {\bibinfo {title} {{Tests of general relativity with GW150914}},}\ }\href {\doibase 10.1103/PhysRevLett.116.221101} {\bibfield  {journal} {\bibinfo  {journal} {Phys. Rev. Lett.}\ }\textbf {\bibinfo {volume} {116}},\ \bibinfo {pages} {221101} (\bibinfo {year} {2016})},\ \bibinfo {note} {[Erratum: Phys.Rev.Lett. 121, 129902 (2018)]},\ \Eprint {http://arxiv.org/abs/1602.03841} {arXiv:1602.03841 [gr-qc]} \BibitemShut {NoStop}%
\bibitem [{\citenamefont {Abbott}\ \emph {et~al.}(2021)\citenamefont {Abbott} \emph {et~al.}}]{LIGOScientific:2020tif}%
  \BibitemOpen
  \bibfield  {author} {\bibinfo {author} {\bibfnamefont {R.}~\bibnamefont {Abbott}} \emph {et~al.} (\bibinfo {collaboration} {LIGO Scientific, Virgo}),\ }\bibfield  {title} {\enquote {\bibinfo {title} {{Tests of general relativity with binary black holes from the second LIGO-Virgo gravitational-wave transient catalog}},}\ }\href {\doibase 10.1103/PhysRevD.103.122002} {\bibfield  {journal} {\bibinfo  {journal} {Phys. Rev. D}\ }\textbf {\bibinfo {volume} {103}},\ \bibinfo {pages} {122002} (\bibinfo {year} {2021})},\ \Eprint {http://arxiv.org/abs/2010.14529} {arXiv:2010.14529 [gr-qc]} \BibitemShut {NoStop}%
\bibitem [{\citenamefont {Abbott}\ \emph {et~al.}(2025)\citenamefont {Abbott} \emph {et~al.}}]{LIGOScientific:2021sio}%
  \BibitemOpen
  \bibfield  {author} {\bibinfo {author} {\bibfnamefont {R.}~\bibnamefont {Abbott}} \emph {et~al.} (\bibinfo {collaboration} {LIGO Scientific, VIRGO, KAGRA}),\ }\bibfield  {title} {\enquote {\bibinfo {title} {{Tests of General Relativity with GWTC-3}},}\ }\href {\doibase 10.1103/PhysRevD.112.084080} {\bibfield  {journal} {\bibinfo  {journal} {Phys. Rev. D}\ }\textbf {\bibinfo {volume} {112}},\ \bibinfo {pages} {084080} (\bibinfo {year} {2025})},\ \Eprint {http://arxiv.org/abs/2112.06861} {arXiv:2112.06861 [gr-qc]} \BibitemShut {NoStop}%
\bibitem [{\citenamefont {Abac}\ \emph {et~al.}(2025{\natexlab{d}})\citenamefont {Abac} \emph {et~al.}}]{LIGOScientific:2025rid}%
  \BibitemOpen
  \bibfield  {author} {\bibinfo {author} {\bibfnamefont {A.~G.}\ \bibnamefont {Abac}} \emph {et~al.} (\bibinfo {collaboration} {LIGO Scientific, Virgo, KAGRA}),\ }\bibfield  {title} {\enquote {\bibinfo {title} {{GW250114: Testing Hawking{\textquoteright}s Area Law and the Kerr Nature of Black Holes}},}\ }\href {\doibase 10.1103/kw5g-d732} {\bibfield  {journal} {\bibinfo  {journal} {Phys. Rev. Lett.}\ }\textbf {\bibinfo {volume} {135}},\ \bibinfo {pages} {111403} (\bibinfo {year} {2025}{\natexlab{d}})},\ \Eprint {http://arxiv.org/abs/2509.08054} {arXiv:2509.08054 [gr-qc]} \BibitemShut {NoStop}%
\bibitem [{\citenamefont {Owen}\ \emph {et~al.}(2023)\citenamefont {Owen}, \citenamefont {Haster}, \citenamefont {Perkins}, \citenamefont {Cornish},\ and\ \citenamefont {Yunes}}]{Owen:2023mid}%
  \BibitemOpen
  \bibfield  {author} {\bibinfo {author} {\bibfnamefont {Caroline~B.}\ \bibnamefont {Owen}}, \bibinfo {author} {\bibfnamefont {Carl-Johan}\ \bibnamefont {Haster}}, \bibinfo {author} {\bibfnamefont {Scott}\ \bibnamefont {Perkins}}, \bibinfo {author} {\bibfnamefont {Neil~J.}\ \bibnamefont {Cornish}}, \ and\ \bibinfo {author} {\bibfnamefont {Nicol{\'a}s}\ \bibnamefont {Yunes}},\ }\bibfield  {title} {\enquote {\bibinfo {title} {{Waveform accuracy and systematic uncertainties in current gravitational wave observations}},}\ }\href {\doibase 10.1103/PhysRevD.108.044018} {\bibfield  {journal} {\bibinfo  {journal} {Phys. Rev. D}\ }\textbf {\bibinfo {volume} {108}},\ \bibinfo {pages} {044018} (\bibinfo {year} {2023})},\ \Eprint {http://arxiv.org/abs/2301.11941} {arXiv:2301.11941 [gr-qc]} \BibitemShut {NoStop}%
\bibitem [{\citenamefont {Dhani}\ \emph {et~al.}(2025)\citenamefont {Dhani}, \citenamefont {V{\"o}lkel}, \citenamefont {Buonanno}, \citenamefont {Estelles}, \citenamefont {Gair}, \citenamefont {Pfeiffer}, \citenamefont {Pompili},\ and\ \citenamefont {Toubiana}}]{Dhani:2024jja}%
  \BibitemOpen
  \bibfield  {author} {\bibinfo {author} {\bibfnamefont {Arnab}\ \bibnamefont {Dhani}}, \bibinfo {author} {\bibfnamefont {Sebastian~H.}\ \bibnamefont {V{\"o}lkel}}, \bibinfo {author} {\bibfnamefont {Alessandra}\ \bibnamefont {Buonanno}}, \bibinfo {author} {\bibfnamefont {Hector}\ \bibnamefont {Estelles}}, \bibinfo {author} {\bibfnamefont {Jonathan}\ \bibnamefont {Gair}}, \bibinfo {author} {\bibfnamefont {Harald~P.}\ \bibnamefont {Pfeiffer}}, \bibinfo {author} {\bibfnamefont {Lorenzo}\ \bibnamefont {Pompili}}, \ and\ \bibinfo {author} {\bibfnamefont {Alexandre}\ \bibnamefont {Toubiana}},\ }\bibfield  {title} {\enquote {\bibinfo {title} {{Systematic Biases in Estimating the Properties of Black Holes Due to Inaccurate Gravitational-Wave Models}},}\ }\href {\doibase 10.1103/5pks-qz6b} {\bibfield  {journal} {\bibinfo  {journal} {Phys. Rev. X}\ }\textbf {\bibinfo {volume} {15}},\ \bibinfo {pages} {031036} (\bibinfo {year} {2025})},\ \Eprint {http://arxiv.org/abs/2404.05811} {arXiv:2404.05811 [gr-qc]} \BibitemShut {NoStop}%
\bibitem [{\citenamefont {Abac}\ \emph {et~al.}(2025{\natexlab{e}})\citenamefont {Abac} \emph {et~al.}}]{LIGOScientific:2025rsn}%
  \BibitemOpen
  \bibfield  {author} {\bibinfo {author} {\bibfnamefont {A.~G.}\ \bibnamefont {Abac}} \emph {et~al.} (\bibinfo {collaboration} {LIGO Scientific, VIRGO, KAGRA}),\ }\bibfield  {title} {\enquote {\bibinfo {title} {{GW231123: A Binary Black Hole Merger with Total Mass 190{\textendash}265 M$_{\odot}$}},}\ }\href {\doibase 10.3847/2041-8213/ae0c9c} {\bibfield  {journal} {\bibinfo  {journal} {Astrophys. J. Lett.}\ }\textbf {\bibinfo {volume} {993}},\ \bibinfo {pages} {L25} (\bibinfo {year} {2025}{\natexlab{e}})},\ \Eprint {http://arxiv.org/abs/2507.08219} {arXiv:2507.08219 [astro-ph.HE]} \BibitemShut {NoStop}%
\bibitem [{\citenamefont {Pretorius}(2005)}]{Pretorius:2005gq}%
  \BibitemOpen
  \bibfield  {author} {\bibinfo {author} {\bibfnamefont {Frans}\ \bibnamefont {Pretorius}},\ }\bibfield  {title} {\enquote {\bibinfo {title} {{Evolution of binary black hole spacetimes}},}\ }\href {\doibase 10.1103/PhysRevLett.95.121101} {\bibfield  {journal} {\bibinfo  {journal} {Phys. Rev. Lett.}\ }\textbf {\bibinfo {volume} {95}},\ \bibinfo {pages} {121101} (\bibinfo {year} {2005})},\ \Eprint {http://arxiv.org/abs/gr-qc/0507014} {arXiv:gr-qc/0507014} \BibitemShut {NoStop}%
\bibitem [{\citenamefont {Campanelli}\ \emph {et~al.}(2006)\citenamefont {Campanelli}, \citenamefont {Lousto}, \citenamefont {Marronetti},\ and\ \citenamefont {Zlochower}}]{Campanelli:2005dd}%
  \BibitemOpen
  \bibfield  {author} {\bibinfo {author} {\bibfnamefont {Manuela}\ \bibnamefont {Campanelli}}, \bibinfo {author} {\bibfnamefont {C.~O.}\ \bibnamefont {Lousto}}, \bibinfo {author} {\bibfnamefont {P.}~\bibnamefont {Marronetti}}, \ and\ \bibinfo {author} {\bibfnamefont {Y.}~\bibnamefont {Zlochower}},\ }\bibfield  {title} {\enquote {\bibinfo {title} {{Accurate evolutions of orbiting black-hole binaries without excision}},}\ }\href {\doibase 10.1103/PhysRevLett.96.111101} {\bibfield  {journal} {\bibinfo  {journal} {Phys. Rev. Lett.}\ }\textbf {\bibinfo {volume} {96}},\ \bibinfo {pages} {111101} (\bibinfo {year} {2006})},\ \Eprint {http://arxiv.org/abs/gr-qc/0511048} {arXiv:gr-qc/0511048} \BibitemShut {NoStop}%
\bibitem [{\citenamefont {Baker}\ \emph {et~al.}(2006)\citenamefont {Baker}, \citenamefont {Centrella}, \citenamefont {Choi}, \citenamefont {Koppitz},\ and\ \citenamefont {van Meter}}]{Baker:2005vv}%
  \BibitemOpen
  \bibfield  {author} {\bibinfo {author} {\bibfnamefont {John~G.}\ \bibnamefont {Baker}}, \bibinfo {author} {\bibfnamefont {Joan}\ \bibnamefont {Centrella}}, \bibinfo {author} {\bibfnamefont {Dae-Il}\ \bibnamefont {Choi}}, \bibinfo {author} {\bibfnamefont {Michael}\ \bibnamefont {Koppitz}}, \ and\ \bibinfo {author} {\bibfnamefont {James}\ \bibnamefont {van Meter}},\ }\bibfield  {title} {\enquote {\bibinfo {title} {{Gravitational wave extraction from an inspiraling configuration of merging black holes}},}\ }\href {\doibase 10.1103/PhysRevLett.96.111102} {\bibfield  {journal} {\bibinfo  {journal} {Phys. Rev. Lett.}\ }\textbf {\bibinfo {volume} {96}},\ \bibinfo {pages} {111102} (\bibinfo {year} {2006})},\ \Eprint {http://arxiv.org/abs/gr-qc/0511103} {arXiv:gr-qc/0511103} \BibitemShut {NoStop}%
\bibitem [{\citenamefont {Lorentz}\ and\ \citenamefont {Droste}()}]{lorentzanddroste}%
  \BibitemOpen
  \bibfield  {author} {\bibinfo {author} {\bibfnamefont {Hendrik~Antoon}\ \bibnamefont {Lorentz}}\ and\ \bibinfo {author} {\bibfnamefont {Johannes}\ \bibnamefont {Droste}},\ }\bibfield  {title} {\enquote {\bibinfo {title} {{The motion of a system of bodies under the influence of their mutual attraction, according to Einstein’s theory}},}\ }\href@noop {} {\ }\bibinfo {note} {Page 330. Nijhoff, The Hague, 1937. Versl. K. Akad. Wet. Amsterdam 26, 392 and 649 (1917)}\BibitemShut {NoStop}%
\bibitem [{\citenamefont {Einstein}\ \emph {et~al.}(1938)\citenamefont {Einstein}, \citenamefont {Infeld},\ and\ \citenamefont {Hoffmann}}]{Einstein:1938yz}%
  \BibitemOpen
  \bibfield  {author} {\bibinfo {author} {\bibfnamefont {Albert}\ \bibnamefont {Einstein}}, \bibinfo {author} {\bibfnamefont {L.}~\bibnamefont {Infeld}}, \ and\ \bibinfo {author} {\bibfnamefont {B.}~\bibnamefont {Hoffmann}},\ }\bibfield  {title} {\enquote {\bibinfo {title} {{The Gravitational equations and the problem of motion}},}\ }\href {\doibase 10.2307/1968714} {\bibfield  {journal} {\bibinfo  {journal} {Annals Math.}\ }\textbf {\bibinfo {volume} {39}},\ \bibinfo {pages} {65--100} (\bibinfo {year} {1938})}\BibitemShut {NoStop}%
\bibitem [{\citenamefont {Futamase}\ and\ \citenamefont {Itoh}(2007)}]{Futamase:2007zz}%
  \BibitemOpen
  \bibfield  {author} {\bibinfo {author} {\bibfnamefont {Toshifumi}\ \bibnamefont {Futamase}}\ and\ \bibinfo {author} {\bibfnamefont {Yousuke}\ \bibnamefont {Itoh}},\ }\bibfield  {title} {\enquote {\bibinfo {title} {{The post-Newtonian approximation for relativistic compact binaries}},}\ }\href {\doibase 10.12942/lrr-2007-2} {\bibfield  {journal} {\bibinfo  {journal} {Living Rev. Rel.}\ }\textbf {\bibinfo {volume} {10}},\ \bibinfo {pages} {2} (\bibinfo {year} {2007})}\BibitemShut {NoStop}%
\bibitem [{\citenamefont {Blanchet}(2014)}]{Blanchet:2013haa}%
  \BibitemOpen
  \bibfield  {author} {\bibinfo {author} {\bibfnamefont {Luc}\ \bibnamefont {Blanchet}},\ }\bibfield  {title} {\enquote {\bibinfo {title} {{Post-Newtonian Theory for Gravitational Waves}},}\ }\href {\doibase 10.12942/lrr-2014-2} {\bibfield  {journal} {\bibinfo  {journal} {Living Rev. Rel.}\ }\textbf {\bibinfo {volume} {17}},\ \bibinfo {pages} {2} (\bibinfo {year} {2014})},\ \Eprint {http://arxiv.org/abs/1310.1528} {arXiv:1310.1528 [gr-qc]} \BibitemShut {NoStop}%
\bibitem [{\citenamefont {Porto}(2016)}]{Porto:2016pyg}%
  \BibitemOpen
  \bibfield  {author} {\bibinfo {author} {\bibfnamefont {Rafael~A.}\ \bibnamefont {Porto}},\ }\bibfield  {title} {\enquote {\bibinfo {title} {{The effective field theorist{\textquoteright}s approach to gravitational dynamics}},}\ }\href {\doibase 10.1016/j.physrep.2016.04.003} {\bibfield  {journal} {\bibinfo  {journal} {Phys. Rept.}\ }\textbf {\bibinfo {volume} {633}},\ \bibinfo {pages} {1--104} (\bibinfo {year} {2016})},\ \Eprint {http://arxiv.org/abs/1601.04914} {arXiv:1601.04914 [hep-th]} \BibitemShut {NoStop}%
\bibitem [{\citenamefont {Sch{\"a}fer}\ and\ \citenamefont {Jaranowski}(2018)}]{Schafer:2018kuf}%
  \BibitemOpen
  \bibfield  {author} {\bibinfo {author} {\bibfnamefont {Gerhard}\ \bibnamefont {Sch{\"a}fer}}\ and\ \bibinfo {author} {\bibfnamefont {Piotr}\ \bibnamefont {Jaranowski}},\ }\bibfield  {title} {\enquote {\bibinfo {title} {{Hamiltonian formulation of general relativity and post-Newtonian dynamics of compact binaries}},}\ }\href {\doibase 10.1007/s41114-024-00048-7} {\bibfield  {journal} {\bibinfo  {journal} {Living Rev. Rel.}\ }\textbf {\bibinfo {volume} {21}},\ \bibinfo {pages} {7} (\bibinfo {year} {2018})},\ \Eprint {http://arxiv.org/abs/1805.07240} {arXiv:1805.07240 [gr-qc]} \BibitemShut {NoStop}%
\bibitem [{\citenamefont {Levi}(2020)}]{Levi:2018nxp}%
  \BibitemOpen
  \bibfield  {author} {\bibinfo {author} {\bibfnamefont {Mich{\`e}le}\ \bibnamefont {Levi}},\ }\bibfield  {title} {\enquote {\bibinfo {title} {{Effective Field Theories of Post-Newtonian Gravity: A comprehensive review}},}\ }\href {\doibase 10.1088/1361-6633/ab12bc} {\bibfield  {journal} {\bibinfo  {journal} {Rept. Prog. Phys.}\ }\textbf {\bibinfo {volume} {83}},\ \bibinfo {pages} {075901} (\bibinfo {year} {2020})},\ \Eprint {http://arxiv.org/abs/1807.01699} {arXiv:1807.01699 [hep-th]} \BibitemShut {NoStop}%
\bibitem [{\citenamefont {Jaranowski}\ and\ \citenamefont {Schaefer}(1998)}]{Jaranowski:1997ky}%
  \BibitemOpen
  \bibfield  {author} {\bibinfo {author} {\bibfnamefont {Piotr}\ \bibnamefont {Jaranowski}}\ and\ \bibinfo {author} {\bibfnamefont {Gerhard}\ \bibnamefont {Schaefer}},\ }\bibfield  {title} {\enquote {\bibinfo {title} {{Third postNewtonian higher order ADM Hamilton dynamics for two-body point mass systems}},}\ }\href {\doibase 10.1103/PhysRevD.57.7274} {\bibfield  {journal} {\bibinfo  {journal} {Phys. Rev. D}\ }\textbf {\bibinfo {volume} {57}},\ \bibinfo {pages} {7274--7291} (\bibinfo {year} {1998})},\ \bibinfo {note} {[Erratum: Phys.Rev.D 63, 029902 (2001)]},\ \Eprint {http://arxiv.org/abs/gr-qc/9712075} {arXiv:gr-qc/9712075} \BibitemShut {NoStop}%
\bibitem [{\citenamefont {Damour}\ \emph {et~al.}(2014)\citenamefont {Damour}, \citenamefont {Jaranowski},\ and\ \citenamefont {Sch{\"a}fer}}]{Damour:2014jta}%
  \BibitemOpen
  \bibfield  {author} {\bibinfo {author} {\bibfnamefont {Thibault}\ \bibnamefont {Damour}}, \bibinfo {author} {\bibfnamefont {Piotr}\ \bibnamefont {Jaranowski}}, \ and\ \bibinfo {author} {\bibfnamefont {Gerhard}\ \bibnamefont {Sch{\"a}fer}},\ }\bibfield  {title} {\enquote {\bibinfo {title} {{Nonlocal-in-time action for the fourth post-Newtonian conservative dynamics of two-body systems}},}\ }\href {\doibase 10.1103/PhysRevD.89.064058} {\bibfield  {journal} {\bibinfo  {journal} {Phys. Rev. D}\ }\textbf {\bibinfo {volume} {89}},\ \bibinfo {pages} {064058} (\bibinfo {year} {2014})},\ \Eprint {http://arxiv.org/abs/1401.4548} {arXiv:1401.4548 [gr-qc]} \BibitemShut {NoStop}%
\bibitem [{\citenamefont {Jaranowski}\ and\ \citenamefont {Sch{\"a}fer}(2015)}]{Jaranowski:2015lha}%
  \BibitemOpen
  \bibfield  {author} {\bibinfo {author} {\bibfnamefont {Piotr}\ \bibnamefont {Jaranowski}}\ and\ \bibinfo {author} {\bibfnamefont {Gerhard}\ \bibnamefont {Sch{\"a}fer}},\ }\bibfield  {title} {\enquote {\bibinfo {title} {{Derivation of local-in-time fourth post-Newtonian ADM Hamiltonian for spinless compact binaries}},}\ }\href {\doibase 10.1103/PhysRevD.92.124043} {\bibfield  {journal} {\bibinfo  {journal} {Phys. Rev. D}\ }\textbf {\bibinfo {volume} {92}},\ \bibinfo {pages} {124043} (\bibinfo {year} {2015})},\ \Eprint {http://arxiv.org/abs/1508.01016} {arXiv:1508.01016 [gr-qc]} \BibitemShut {NoStop}%
\bibitem [{\citenamefont {Bernard}\ \emph {et~al.}(2016)\citenamefont {Bernard}, \citenamefont {Blanchet}, \citenamefont {Boh{\'e}}, \citenamefont {Faye},\ and\ \citenamefont {Marsat}}]{Bernard:2015njp}%
  \BibitemOpen
  \bibfield  {author} {\bibinfo {author} {\bibfnamefont {Laura}\ \bibnamefont {Bernard}}, \bibinfo {author} {\bibfnamefont {Luc}\ \bibnamefont {Blanchet}}, \bibinfo {author} {\bibfnamefont {Alejandro}\ \bibnamefont {Boh{\'e}}}, \bibinfo {author} {\bibfnamefont {Guillaume}\ \bibnamefont {Faye}}, \ and\ \bibinfo {author} {\bibfnamefont {Sylvain}\ \bibnamefont {Marsat}},\ }\bibfield  {title} {\enquote {\bibinfo {title} {{Fokker action of nonspinning compact binaries at the fourth post-Newtonian approximation}},}\ }\href {\doibase 10.1103/PhysRevD.93.084037} {\bibfield  {journal} {\bibinfo  {journal} {Phys. Rev. D}\ }\textbf {\bibinfo {volume} {93}},\ \bibinfo {pages} {084037} (\bibinfo {year} {2016})},\ \Eprint {http://arxiv.org/abs/1512.02876} {arXiv:1512.02876 [gr-qc]} \BibitemShut {NoStop}%
\bibitem [{\citenamefont {Bernard}\ \emph {et~al.}(2017)\citenamefont {Bernard}, \citenamefont {Blanchet}, \citenamefont {Boh{\'e}}, \citenamefont {Faye},\ and\ \citenamefont {Marsat}}]{Bernard:2016wrg}%
  \BibitemOpen
  \bibfield  {author} {\bibinfo {author} {\bibfnamefont {Laura}\ \bibnamefont {Bernard}}, \bibinfo {author} {\bibfnamefont {Luc}\ \bibnamefont {Blanchet}}, \bibinfo {author} {\bibfnamefont {Alejandro}\ \bibnamefont {Boh{\'e}}}, \bibinfo {author} {\bibfnamefont {Guillaume}\ \bibnamefont {Faye}}, \ and\ \bibinfo {author} {\bibfnamefont {Sylvain}\ \bibnamefont {Marsat}},\ }\bibfield  {title} {\enquote {\bibinfo {title} {{Energy and periastron advance of compact binaries on circular orbits at the fourth post-Newtonian order}},}\ }\href {\doibase 10.1103/PhysRevD.95.044026} {\bibfield  {journal} {\bibinfo  {journal} {Phys. Rev. D}\ }\textbf {\bibinfo {volume} {95}},\ \bibinfo {pages} {044026} (\bibinfo {year} {2017})},\ \Eprint {http://arxiv.org/abs/1610.07934} {arXiv:1610.07934 [gr-qc]} \BibitemShut {NoStop}%
\bibitem [{\citenamefont {Damour}\ \emph {et~al.}(2016)\citenamefont {Damour}, \citenamefont {Jaranowski},\ and\ \citenamefont {Sch{\"a}fer}}]{Damour:2016abl}%
  \BibitemOpen
  \bibfield  {author} {\bibinfo {author} {\bibfnamefont {Thibault}\ \bibnamefont {Damour}}, \bibinfo {author} {\bibfnamefont {Piotr}\ \bibnamefont {Jaranowski}}, \ and\ \bibinfo {author} {\bibfnamefont {Gerhard}\ \bibnamefont {Sch{\"a}fer}},\ }\bibfield  {title} {\enquote {\bibinfo {title} {{Conservative dynamics of two-body systems at the fourth post-Newtonian approximation of general relativity}},}\ }\href {\doibase 10.1103/PhysRevD.93.084014} {\bibfield  {journal} {\bibinfo  {journal} {Phys. Rev. D}\ }\textbf {\bibinfo {volume} {93}},\ \bibinfo {pages} {084014} (\bibinfo {year} {2016})},\ \Eprint {http://arxiv.org/abs/1601.01283} {arXiv:1601.01283 [gr-qc]} \BibitemShut {NoStop}%
\bibitem [{\citenamefont {Blanchet}\ \emph {et~al.}(2023{\natexlab{a}})\citenamefont {Blanchet}, \citenamefont {Faye}, \citenamefont {Henry}, \citenamefont {Larrouturou},\ and\ \citenamefont {Trestini}}]{Blanchet:2023sbv}%
  \BibitemOpen
  \bibfield  {author} {\bibinfo {author} {\bibfnamefont {Luc}\ \bibnamefont {Blanchet}}, \bibinfo {author} {\bibfnamefont {Guillaume}\ \bibnamefont {Faye}}, \bibinfo {author} {\bibfnamefont {Quentin}\ \bibnamefont {Henry}}, \bibinfo {author} {\bibfnamefont {Fran{\c{c}}ois}\ \bibnamefont {Larrouturou}}, \ and\ \bibinfo {author} {\bibfnamefont {David}\ \bibnamefont {Trestini}},\ }\bibfield  {title} {\enquote {\bibinfo {title} {{Gravitational-wave flux and quadrupole modes from quasicircular nonspinning compact binaries to the fourth post-Newtonian order}},}\ }\href {\doibase 10.1103/PhysRevD.108.064041} {\bibfield  {journal} {\bibinfo  {journal} {Phys. Rev. D}\ }\textbf {\bibinfo {volume} {108}},\ \bibinfo {pages} {064041} (\bibinfo {year} {2023}{\natexlab{a}})},\ \Eprint {http://arxiv.org/abs/2304.11186} {arXiv:2304.11186 [gr-qc]} \BibitemShut {NoStop}%
\bibitem [{\citenamefont {Blanchet}\ \emph {et~al.}(2023{\natexlab{b}})\citenamefont {Blanchet}, \citenamefont {Faye}, \citenamefont {Henry}, \citenamefont {Larrouturou},\ and\ \citenamefont {Trestini}}]{Blanchet:2023bwj}%
  \BibitemOpen
  \bibfield  {author} {\bibinfo {author} {\bibfnamefont {Luc}\ \bibnamefont {Blanchet}}, \bibinfo {author} {\bibfnamefont {Guillaume}\ \bibnamefont {Faye}}, \bibinfo {author} {\bibfnamefont {Quentin}\ \bibnamefont {Henry}}, \bibinfo {author} {\bibfnamefont {Fran{\c{c}}ois}\ \bibnamefont {Larrouturou}}, \ and\ \bibinfo {author} {\bibfnamefont {David}\ \bibnamefont {Trestini}},\ }\bibfield  {title} {\enquote {\bibinfo {title} {{Gravitational-Wave Phasing of Quasicircular Compact Binary Systems to the Fourth-and-a-Half Post-Newtonian Order}},}\ }\href {\doibase 10.1103/PhysRevLett.131.121402} {\bibfield  {journal} {\bibinfo  {journal} {Phys. Rev. Lett.}\ }\textbf {\bibinfo {volume} {131}},\ \bibinfo {pages} {121402} (\bibinfo {year} {2023}{\natexlab{b}})},\ \Eprint {http://arxiv.org/abs/2304.11185} {arXiv:2304.11185 [gr-qc]} \BibitemShut {NoStop}%
\bibitem [{\citenamefont {Damour}\ and\ \citenamefont {Jaranowski}(2017)}]{Damour:2017ced}%
  \BibitemOpen
  \bibfield  {author} {\bibinfo {author} {\bibfnamefont {Thibault}\ \bibnamefont {Damour}}\ and\ \bibinfo {author} {\bibfnamefont {Piotr}\ \bibnamefont {Jaranowski}},\ }\bibfield  {title} {\enquote {\bibinfo {title} {{Four-loop static contribution to the gravitational interaction potential of two point masses}},}\ }\href {\doibase 10.1103/PhysRevD.95.084005} {\bibfield  {journal} {\bibinfo  {journal} {Phys. Rev. D}\ }\textbf {\bibinfo {volume} {95}},\ \bibinfo {pages} {084005} (\bibinfo {year} {2017})},\ \Eprint {http://arxiv.org/abs/1701.02645} {arXiv:1701.02645 [gr-qc]} \BibitemShut {NoStop}%
\bibitem [{\citenamefont {Goldberger}\ and\ \citenamefont {Rothstein}(2006)}]{Goldberger:2004jt}%
  \BibitemOpen
  \bibfield  {author} {\bibinfo {author} {\bibfnamefont {Walter~D.}\ \bibnamefont {Goldberger}}\ and\ \bibinfo {author} {\bibfnamefont {Ira~Z.}\ \bibnamefont {Rothstein}},\ }\bibfield  {title} {\enquote {\bibinfo {title} {{An Effective field theory of gravity for extended objects}},}\ }\href {\doibase 10.1103/PhysRevD.73.104029} {\bibfield  {journal} {\bibinfo  {journal} {Phys. Rev. D}\ }\textbf {\bibinfo {volume} {73}},\ \bibinfo {pages} {104029} (\bibinfo {year} {2006})},\ \Eprint {http://arxiv.org/abs/hep-th/0409156} {arXiv:hep-th/0409156} \BibitemShut {NoStop}%
\bibitem [{\citenamefont {Kol}\ and\ \citenamefont {Shir}(2013)}]{Kol:2013ega}%
  \BibitemOpen
  \bibfield  {author} {\bibinfo {author} {\bibfnamefont {Barak}\ \bibnamefont {Kol}}\ and\ \bibinfo {author} {\bibfnamefont {Ruth}\ \bibnamefont {Shir}},\ }\bibfield  {title} {\enquote {\bibinfo {title} {{Classical 3-loop 2-body diagrams}},}\ }\href {\doibase 10.1007/JHEP09(2013)069} {\bibfield  {journal} {\bibinfo  {journal} {JHEP}\ }\textbf {\bibinfo {volume} {09}},\ \bibinfo {pages} {069} (\bibinfo {year} {2013})},\ \Eprint {http://arxiv.org/abs/1306.3220} {arXiv:1306.3220 [hep-th]} \BibitemShut {NoStop}%
\bibitem [{\citenamefont {Foffa}\ and\ \citenamefont {Sturani}(2013)}]{Foffa:2012rn}%
  \BibitemOpen
  \bibfield  {author} {\bibinfo {author} {\bibfnamefont {Stefano}\ \bibnamefont {Foffa}}\ and\ \bibinfo {author} {\bibfnamefont {Riccardo}\ \bibnamefont {Sturani}},\ }\bibfield  {title} {\enquote {\bibinfo {title} {{Dynamics of the gravitational two-body problem at fourth post-Newtonian order and at quadratic order in the Newton constant}},}\ }\href {\doibase 10.1103/PhysRevD.87.064011} {\bibfield  {journal} {\bibinfo  {journal} {Phys. Rev. D}\ }\textbf {\bibinfo {volume} {87}},\ \bibinfo {pages} {064011} (\bibinfo {year} {2013})},\ \Eprint {http://arxiv.org/abs/1206.7087} {arXiv:1206.7087 [gr-qc]} \BibitemShut {NoStop}%
\bibitem [{\citenamefont {Foffa}\ and\ \citenamefont {Sturani}(2019)}]{Foffa:2019rdf}%
  \BibitemOpen
  \bibfield  {author} {\bibinfo {author} {\bibfnamefont {Stefano}\ \bibnamefont {Foffa}}\ and\ \bibinfo {author} {\bibfnamefont {Riccardo}\ \bibnamefont {Sturani}},\ }\bibfield  {title} {\enquote {\bibinfo {title} {{Conservative dynamics of binary systems to fourth Post-Newtonian order in the EFT approach I: Regularized Lagrangian}},}\ }\href {\doibase 10.1103/PhysRevD.100.024047} {\bibfield  {journal} {\bibinfo  {journal} {Phys. Rev. D}\ }\textbf {\bibinfo {volume} {100}},\ \bibinfo {pages} {024047} (\bibinfo {year} {2019})},\ \Eprint {http://arxiv.org/abs/1903.05113} {arXiv:1903.05113 [gr-qc]} \BibitemShut {NoStop}%
\bibitem [{\citenamefont {Foffa}\ \emph {et~al.}(2017)\citenamefont {Foffa}, \citenamefont {Mastrolia}, \citenamefont {Sturani},\ and\ \citenamefont {Sturm}}]{Foffa:2016rgu}%
  \BibitemOpen
  \bibfield  {author} {\bibinfo {author} {\bibfnamefont {Stefano}\ \bibnamefont {Foffa}}, \bibinfo {author} {\bibfnamefont {Pierpaolo}\ \bibnamefont {Mastrolia}}, \bibinfo {author} {\bibfnamefont {Riccardo}\ \bibnamefont {Sturani}}, \ and\ \bibinfo {author} {\bibfnamefont {Christian}\ \bibnamefont {Sturm}},\ }\bibfield  {title} {\enquote {\bibinfo {title} {{Effective field theory approach to the gravitational two-body dynamics, at fourth post-Newtonian order and quintic in the Newton constant}},}\ }\href {\doibase 10.1103/PhysRevD.95.104009} {\bibfield  {journal} {\bibinfo  {journal} {Phys. Rev. D}\ }\textbf {\bibinfo {volume} {95}},\ \bibinfo {pages} {104009} (\bibinfo {year} {2017})},\ \Eprint {http://arxiv.org/abs/1612.00482} {arXiv:1612.00482 [gr-qc]} \BibitemShut {NoStop}%
\bibitem [{\citenamefont {Foffa}\ \emph {et~al.}(2019{\natexlab{a}})\citenamefont {Foffa}, \citenamefont {Porto}, \citenamefont {Rothstein},\ and\ \citenamefont {Sturani}}]{Foffa:2019yfl}%
  \BibitemOpen
  \bibfield  {author} {\bibinfo {author} {\bibfnamefont {Stefano}\ \bibnamefont {Foffa}}, \bibinfo {author} {\bibfnamefont {Rafael~A.}\ \bibnamefont {Porto}}, \bibinfo {author} {\bibfnamefont {Ira}\ \bibnamefont {Rothstein}}, \ and\ \bibinfo {author} {\bibfnamefont {Riccardo}\ \bibnamefont {Sturani}},\ }\bibfield  {title} {\enquote {\bibinfo {title} {{Conservative dynamics of binary systems to fourth Post-Newtonian order in the EFT approach II: Renormalized Lagrangian}},}\ }\href {\doibase 10.1103/PhysRevD.100.024048} {\bibfield  {journal} {\bibinfo  {journal} {Phys. Rev. D}\ }\textbf {\bibinfo {volume} {100}},\ \bibinfo {pages} {024048} (\bibinfo {year} {2019}{\natexlab{a}})},\ \Eprint {http://arxiv.org/abs/1903.05118} {arXiv:1903.05118 [gr-qc]} \BibitemShut {NoStop}%
\bibitem [{\citenamefont {Bl{\"u}mlein}\ \emph {et~al.}(2020{\natexlab{a}})\citenamefont {Bl{\"u}mlein}, \citenamefont {Maier}, \citenamefont {Marquard},\ and\ \citenamefont {Sch{\"a}fer}}]{Blumlein:2020pog}%
  \BibitemOpen
  \bibfield  {author} {\bibinfo {author} {\bibfnamefont {J.}~\bibnamefont {Bl{\"u}mlein}}, \bibinfo {author} {\bibfnamefont {A.}~\bibnamefont {Maier}}, \bibinfo {author} {\bibfnamefont {P.}~\bibnamefont {Marquard}}, \ and\ \bibinfo {author} {\bibfnamefont {G.}~\bibnamefont {Sch{\"a}fer}},\ }\bibfield  {title} {\enquote {\bibinfo {title} {{Fourth post-Newtonian Hamiltonian dynamics of two-body systems from an effective field theory approach}},}\ }\href {\doibase 10.1016/j.nuclphysb.2020.115041} {\bibfield  {journal} {\bibinfo  {journal} {Nucl. Phys. B}\ }\textbf {\bibinfo {volume} {955}},\ \bibinfo {pages} {115041} (\bibinfo {year} {2020}{\natexlab{a}})},\ \Eprint {http://arxiv.org/abs/2003.01692} {arXiv:2003.01692 [gr-qc]} \BibitemShut {NoStop}%
\bibitem [{\citenamefont {Bl{\"u}mlein}\ \emph {et~al.}(2020{\natexlab{b}})\citenamefont {Bl{\"u}mlein}, \citenamefont {Maier}, \citenamefont {Marquard},\ and\ \citenamefont {Sch{\"a}fer}}]{Blumlein:2020znm}%
  \BibitemOpen
  \bibfield  {author} {\bibinfo {author} {\bibfnamefont {J.}~\bibnamefont {Bl{\"u}mlein}}, \bibinfo {author} {\bibfnamefont {A.}~\bibnamefont {Maier}}, \bibinfo {author} {\bibfnamefont {P.}~\bibnamefont {Marquard}}, \ and\ \bibinfo {author} {\bibfnamefont {G.}~\bibnamefont {Sch{\"a}fer}},\ }\bibfield  {title} {\enquote {\bibinfo {title} {{Testing binary dynamics in gravity at the sixth post-Newtonian level}},}\ }\href {\doibase 10.1016/j.physletb.2020.135496} {\bibfield  {journal} {\bibinfo  {journal} {Phys. Lett. B}\ }\textbf {\bibinfo {volume} {807}},\ \bibinfo {pages} {135496} (\bibinfo {year} {2020}{\natexlab{b}})},\ \Eprint {http://arxiv.org/abs/2003.07145} {arXiv:2003.07145 [gr-qc]} \BibitemShut {NoStop}%
\bibitem [{\citenamefont {Goldberger}\ \emph {et~al.}(2014)\citenamefont {Goldberger}, \citenamefont {Ross},\ and\ \citenamefont {Rothstein}}]{Goldberger:2012kf}%
  \BibitemOpen
  \bibfield  {author} {\bibinfo {author} {\bibfnamefont {Walter~D.}\ \bibnamefont {Goldberger}}, \bibinfo {author} {\bibfnamefont {Andreas}\ \bibnamefont {Ross}}, \ and\ \bibinfo {author} {\bibfnamefont {Ira~Z.}\ \bibnamefont {Rothstein}},\ }\bibfield  {title} {\enquote {\bibinfo {title} {{Black hole mass dynamics and renormalization group evolution}},}\ }\href {\doibase 10.1103/PhysRevD.89.124033} {\bibfield  {journal} {\bibinfo  {journal} {Phys. Rev. D}\ }\textbf {\bibinfo {volume} {89}},\ \bibinfo {pages} {124033} (\bibinfo {year} {2014})},\ \Eprint {http://arxiv.org/abs/1211.6095} {arXiv:1211.6095 [hep-th]} \BibitemShut {NoStop}%
\bibitem [{\citenamefont {Galley}\ \emph {et~al.}(2016)\citenamefont {Galley}, \citenamefont {Leibovich}, \citenamefont {Porto},\ and\ \citenamefont {Ross}}]{Galley:2015kus}%
  \BibitemOpen
  \bibfield  {author} {\bibinfo {author} {\bibfnamefont {Chad~R.}\ \bibnamefont {Galley}}, \bibinfo {author} {\bibfnamefont {Adam~K.}\ \bibnamefont {Leibovich}}, \bibinfo {author} {\bibfnamefont {Rafael~A.}\ \bibnamefont {Porto}}, \ and\ \bibinfo {author} {\bibfnamefont {Andreas}\ \bibnamefont {Ross}},\ }\bibfield  {title} {\enquote {\bibinfo {title} {{Tail effect in gravitational radiation reaction: Time nonlocality and renormalization group evolution}},}\ }\href {\doibase 10.1103/PhysRevD.93.124010} {\bibfield  {journal} {\bibinfo  {journal} {Phys. Rev. D}\ }\textbf {\bibinfo {volume} {93}},\ \bibinfo {pages} {124010} (\bibinfo {year} {2016})},\ \Eprint {http://arxiv.org/abs/1511.07379} {arXiv:1511.07379 [gr-qc]} \BibitemShut {NoStop}%
\bibitem [{\citenamefont {Foffa}\ \emph {et~al.}(2019{\natexlab{b}})\citenamefont {Foffa}, \citenamefont {Mastrolia}, \citenamefont {Sturani}, \citenamefont {Sturm},\ and\ \citenamefont {Torres~Bobadilla}}]{Foffa:2019hrb}%
  \BibitemOpen
  \bibfield  {author} {\bibinfo {author} {\bibfnamefont {Stefano}\ \bibnamefont {Foffa}}, \bibinfo {author} {\bibfnamefont {Pierpaolo}\ \bibnamefont {Mastrolia}}, \bibinfo {author} {\bibfnamefont {Riccardo}\ \bibnamefont {Sturani}}, \bibinfo {author} {\bibfnamefont {Christian}\ \bibnamefont {Sturm}}, \ and\ \bibinfo {author} {\bibfnamefont {William~J.}\ \bibnamefont {Torres~Bobadilla}},\ }\bibfield  {title} {\enquote {\bibinfo {title} {{Static two-body potential at fifth post-Newtonian order}},}\ }\href {\doibase 10.1103/PhysRevLett.122.241605} {\bibfield  {journal} {\bibinfo  {journal} {Phys. Rev. Lett.}\ }\textbf {\bibinfo {volume} {122}},\ \bibinfo {pages} {241605} (\bibinfo {year} {2019}{\natexlab{b}})},\ \Eprint {http://arxiv.org/abs/1902.10571} {arXiv:1902.10571 [gr-qc]} \BibitemShut {NoStop}%
\bibitem [{\citenamefont {Bl{\"u}mlein}\ \emph {et~al.}(2020{\natexlab{c}})\citenamefont {Bl{\"u}mlein}, \citenamefont {Maier},\ and\ \citenamefont {Marquard}}]{Blumlein:2019zku}%
  \BibitemOpen
  \bibfield  {author} {\bibinfo {author} {\bibfnamefont {J.}~\bibnamefont {Bl{\"u}mlein}}, \bibinfo {author} {\bibfnamefont {A.}~\bibnamefont {Maier}}, \ and\ \bibinfo {author} {\bibfnamefont {P.}~\bibnamefont {Marquard}},\ }\bibfield  {title} {\enquote {\bibinfo {title} {{Five-Loop Static Contribution to the Gravitational Interaction Potential of Two Point Masses}},}\ }\href {\doibase 10.1016/j.physletb.2019.135100} {\bibfield  {journal} {\bibinfo  {journal} {Phys. Lett. B}\ }\textbf {\bibinfo {volume} {800}},\ \bibinfo {pages} {135100} (\bibinfo {year} {2020}{\natexlab{c}})},\ \Eprint {http://arxiv.org/abs/1902.11180} {arXiv:1902.11180 [gr-qc]} \BibitemShut {NoStop}%
\bibitem [{\citenamefont {Foffa}\ \emph {et~al.}(2021)\citenamefont {Foffa}, \citenamefont {Sturani},\ and\ \citenamefont {Torres~Bobadilla}}]{Foffa:2020nqe}%
  \BibitemOpen
  \bibfield  {author} {\bibinfo {author} {\bibfnamefont {Stefano}\ \bibnamefont {Foffa}}, \bibinfo {author} {\bibfnamefont {Riccardo}\ \bibnamefont {Sturani}}, \ and\ \bibinfo {author} {\bibfnamefont {William~J.}\ \bibnamefont {Torres~Bobadilla}},\ }\bibfield  {title} {\enquote {\bibinfo {title} {{Efficient resummation of high post-Newtonian contributions to the binding energy}},}\ }\href {\doibase 10.1007/JHEP02(2021)165} {\bibfield  {journal} {\bibinfo  {journal} {JHEP}\ }\textbf {\bibinfo {volume} {02}},\ \bibinfo {pages} {165} (\bibinfo {year} {2021})},\ \Eprint {http://arxiv.org/abs/2010.13730} {arXiv:2010.13730 [gr-qc]} \BibitemShut {NoStop}%
\bibitem [{\citenamefont {Bl{\"u}mlein}\ \emph {et~al.}(2021{\natexlab{a}})\citenamefont {Bl{\"u}mlein}, \citenamefont {Maier}, \citenamefont {Marquard},\ and\ \citenamefont {Sch{\"a}fer}}]{Blumlein:2020pyo}%
  \BibitemOpen
  \bibfield  {author} {\bibinfo {author} {\bibfnamefont {J.}~\bibnamefont {Bl{\"u}mlein}}, \bibinfo {author} {\bibfnamefont {A.}~\bibnamefont {Maier}}, \bibinfo {author} {\bibfnamefont {P.}~\bibnamefont {Marquard}}, \ and\ \bibinfo {author} {\bibfnamefont {G.}~\bibnamefont {Sch{\"a}fer}},\ }\bibfield  {title} {\enquote {\bibinfo {title} {{The fifth-order post-Newtonian Hamiltonian dynamics of two-body systems from an effective field theory approach: potential contributions}},}\ }\href {\doibase 10.1016/j.nuclphysb.2021.115352} {\bibfield  {journal} {\bibinfo  {journal} {Nucl. Phys. B}\ }\textbf {\bibinfo {volume} {965}},\ \bibinfo {pages} {115352} (\bibinfo {year} {2021}{\natexlab{a}})},\ \Eprint {http://arxiv.org/abs/2010.13672} {arXiv:2010.13672 [gr-qc]} \BibitemShut {NoStop}%
\bibitem [{\citenamefont {Bl{\"u}mlein}\ \emph {et~al.}(2022)\citenamefont {Bl{\"u}mlein}, \citenamefont {Maier}, \citenamefont {Marquard},\ and\ \citenamefont {Sch{\"a}fer}}]{Blumlein:2021txe}%
  \BibitemOpen
  \bibfield  {author} {\bibinfo {author} {\bibfnamefont {J.}~\bibnamefont {Bl{\"u}mlein}}, \bibinfo {author} {\bibfnamefont {A.}~\bibnamefont {Maier}}, \bibinfo {author} {\bibfnamefont {P.}~\bibnamefont {Marquard}}, \ and\ \bibinfo {author} {\bibfnamefont {G.}~\bibnamefont {Sch{\"a}fer}},\ }\bibfield  {title} {\enquote {\bibinfo {title} {{The fifth-order post-Newtonian Hamiltonian dynamics of two-body systems from an effective field theory approach}},}\ }\href {\doibase 10.1016/j.nuclphysb.2022.115900} {\bibfield  {journal} {\bibinfo  {journal} {Nucl. Phys. B}\ }\textbf {\bibinfo {volume} {983}},\ \bibinfo {pages} {115900} (\bibinfo {year} {2022})},\ \bibinfo {note} {[Erratum: Nucl.Phys.B 985, 115991 (2022)]},\ \Eprint {http://arxiv.org/abs/2110.13822} {arXiv:2110.13822 [gr-qc]} \BibitemShut {NoStop}%
\bibitem [{\citenamefont {Porto}\ \emph {et~al.}(2025)\citenamefont {Porto}, \citenamefont {Riva},\ and\ \citenamefont {Yang}}]{Porto:2024cwd}%
  \BibitemOpen
  \bibfield  {author} {\bibinfo {author} {\bibfnamefont {Rafael~A.}\ \bibnamefont {Porto}}, \bibinfo {author} {\bibfnamefont {Massimiliano~M.}\ \bibnamefont {Riva}}, \ and\ \bibinfo {author} {\bibfnamefont {Zixin}\ \bibnamefont {Yang}},\ }\bibfield  {title} {\enquote {\bibinfo {title} {{Nonlinear gravitational radiation reaction: failed tail, memories {\&} squares}},}\ }\href {\doibase 10.1007/JHEP04(2025)050} {\bibfield  {journal} {\bibinfo  {journal} {JHEP}\ }\textbf {\bibinfo {volume} {04}},\ \bibinfo {pages} {050} (\bibinfo {year} {2025})},\ \Eprint {http://arxiv.org/abs/2409.05860} {arXiv:2409.05860 [gr-qc]} \BibitemShut {NoStop}%
\bibitem [{\citenamefont {Bl{\"u}mlein}\ \emph {et~al.}(2021{\natexlab{b}})\citenamefont {Bl{\"u}mlein}, \citenamefont {Maier}, \citenamefont {Marquard},\ and\ \citenamefont {Sch{\"a}fer}}]{Blumlein:2021txj}%
  \BibitemOpen
  \bibfield  {author} {\bibinfo {author} {\bibfnamefont {J.}~\bibnamefont {Bl{\"u}mlein}}, \bibinfo {author} {\bibfnamefont {A.}~\bibnamefont {Maier}}, \bibinfo {author} {\bibfnamefont {P.}~\bibnamefont {Marquard}}, \ and\ \bibinfo {author} {\bibfnamefont {G.}~\bibnamefont {Sch{\"a}fer}},\ }\bibfield  {title} {\enquote {\bibinfo {title} {{The 6th post-Newtonian potential terms at $O(G_N^4)$}},}\ }\href {\doibase 10.1016/j.physletb.2021.136260} {\bibfield  {journal} {\bibinfo  {journal} {Phys. Lett. B}\ }\textbf {\bibinfo {volume} {816}},\ \bibinfo {pages} {136260} (\bibinfo {year} {2021}{\natexlab{b}})},\ \Eprint {http://arxiv.org/abs/2101.08630} {arXiv:2101.08630 [gr-qc]} \BibitemShut {NoStop}%
\bibitem [{\citenamefont {Porto}(2006)}]{Porto:2005ac}%
  \BibitemOpen
  \bibfield  {author} {\bibinfo {author} {\bibfnamefont {Rafael~A.}\ \bibnamefont {Porto}},\ }\bibfield  {title} {\enquote {\bibinfo {title} {{Post-Newtonian corrections to the motion of spinning bodies in NRGR}},}\ }\href {\doibase 10.1103/PhysRevD.73.104031} {\bibfield  {journal} {\bibinfo  {journal} {Phys. Rev. D}\ }\textbf {\bibinfo {volume} {73}},\ \bibinfo {pages} {104031} (\bibinfo {year} {2006})},\ \Eprint {http://arxiv.org/abs/gr-qc/0511061} {arXiv:gr-qc/0511061} \BibitemShut {NoStop}%
\bibitem [{\citenamefont {Levi}\ and\ \citenamefont {Steinhoff}(2015)}]{Levi:2015msa}%
  \BibitemOpen
  \bibfield  {author} {\bibinfo {author} {\bibfnamefont {Michele}\ \bibnamefont {Levi}}\ and\ \bibinfo {author} {\bibfnamefont {Jan}\ \bibnamefont {Steinhoff}},\ }\bibfield  {title} {\enquote {\bibinfo {title} {{Spinning gravitating objects in the effective field theory in the post-Newtonian scheme}},}\ }\href {\doibase 10.1007/JHEP09(2015)219} {\bibfield  {journal} {\bibinfo  {journal} {JHEP}\ }\textbf {\bibinfo {volume} {09}},\ \bibinfo {pages} {219} (\bibinfo {year} {2015})},\ \Eprint {http://arxiv.org/abs/1501.04956} {arXiv:1501.04956 [gr-qc]} \BibitemShut {NoStop}%
\bibitem [{\citenamefont {Kim}\ \emph {et~al.}(2022)\citenamefont {Kim}, \citenamefont {Levi},\ and\ \citenamefont {Yin}}]{Kim:2021rfj}%
  \BibitemOpen
  \bibfield  {author} {\bibinfo {author} {\bibfnamefont {Jung-Wook}\ \bibnamefont {Kim}}, \bibinfo {author} {\bibfnamefont {Mich{\`e}le}\ \bibnamefont {Levi}}, \ and\ \bibinfo {author} {\bibfnamefont {Zhewei}\ \bibnamefont {Yin}},\ }\bibfield  {title} {\enquote {\bibinfo {title} {{Quadratic-in-spin interactions at fifth post-Newtonian order probe new physics}},}\ }\href {\doibase 10.1016/j.physletb.2022.137410} {\bibfield  {journal} {\bibinfo  {journal} {Phys. Lett. B}\ }\textbf {\bibinfo {volume} {834}},\ \bibinfo {pages} {137410} (\bibinfo {year} {2022})},\ \Eprint {http://arxiv.org/abs/2112.01509} {arXiv:2112.01509 [hep-th]} \BibitemShut {NoStop}%
\bibitem [{\citenamefont {Levi}\ \emph {et~al.}(2021{\natexlab{a}})\citenamefont {Levi}, \citenamefont {Mcleod},\ and\ \citenamefont {Von~Hippel}}]{Levi:2020uwu}%
  \BibitemOpen
  \bibfield  {author} {\bibinfo {author} {\bibfnamefont {Mich{\`e}le}\ \bibnamefont {Levi}}, \bibinfo {author} {\bibfnamefont {Andrew~J.}\ \bibnamefont {Mcleod}}, \ and\ \bibinfo {author} {\bibfnamefont {Matthew}\ \bibnamefont {Von~Hippel}},\ }\bibfield  {title} {\enquote {\bibinfo {title} {{N$^{3}$LO gravitational quadratic-in-spin interactions at G$^{4}$}},}\ }\href {\doibase 10.1007/JHEP07(2021)116} {\bibfield  {journal} {\bibinfo  {journal} {JHEP}\ }\textbf {\bibinfo {volume} {07}},\ \bibinfo {pages} {116} (\bibinfo {year} {2021}{\natexlab{a}})},\ \Eprint {http://arxiv.org/abs/2003.07890} {arXiv:2003.07890 [hep-th]} \BibitemShut {NoStop}%
\bibitem [{\citenamefont {Levi}\ \emph {et~al.}(2021{\natexlab{b}})\citenamefont {Levi}, \citenamefont {Mougiakakos},\ and\ \citenamefont {Vieira}}]{Levi:2019kgk}%
  \BibitemOpen
  \bibfield  {author} {\bibinfo {author} {\bibfnamefont {Mich{\`e}le}\ \bibnamefont {Levi}}, \bibinfo {author} {\bibfnamefont {Stavros}\ \bibnamefont {Mougiakakos}}, \ and\ \bibinfo {author} {\bibfnamefont {Mariana}\ \bibnamefont {Vieira}},\ }\bibfield  {title} {\enquote {\bibinfo {title} {{Gravitational cubic-in-spin interaction at the next-to-leading post-Newtonian order}},}\ }\href {\doibase 10.1007/JHEP01(2021)036} {\bibfield  {journal} {\bibinfo  {journal} {JHEP}\ }\textbf {\bibinfo {volume} {01}},\ \bibinfo {pages} {036} (\bibinfo {year} {2021}{\natexlab{b}})},\ \Eprint {http://arxiv.org/abs/1912.06276} {arXiv:1912.06276 [hep-th]} \BibitemShut {NoStop}%
\bibitem [{\citenamefont {Kim}\ \emph {et~al.}(2023{\natexlab{a}})\citenamefont {Kim}, \citenamefont {Levi},\ and\ \citenamefont {Yin}}]{Kim:2022pou}%
  \BibitemOpen
  \bibfield  {author} {\bibinfo {author} {\bibfnamefont {Jung-Wook}\ \bibnamefont {Kim}}, \bibinfo {author} {\bibfnamefont {Mich{\`e}le}\ \bibnamefont {Levi}}, \ and\ \bibinfo {author} {\bibfnamefont {Zhewei}\ \bibnamefont {Yin}},\ }\bibfield  {title} {\enquote {\bibinfo {title} {{N$^{3}$LO spin-orbit interaction via the EFT of spinning gravitating objects}},}\ }\href {\doibase 10.1007/JHEP05(2023)184} {\bibfield  {journal} {\bibinfo  {journal} {JHEP}\ }\textbf {\bibinfo {volume} {05}},\ \bibinfo {pages} {184} (\bibinfo {year} {2023}{\natexlab{a}})},\ \Eprint {http://arxiv.org/abs/2208.14949} {arXiv:2208.14949 [hep-th]} \BibitemShut {NoStop}%
\bibitem [{\citenamefont {Kim}\ \emph {et~al.}(2023{\natexlab{b}})\citenamefont {Kim}, \citenamefont {Levi},\ and\ \citenamefont {Yin}}]{Kim:2022bwv}%
  \BibitemOpen
  \bibfield  {author} {\bibinfo {author} {\bibfnamefont {Jung-Wook}\ \bibnamefont {Kim}}, \bibinfo {author} {\bibfnamefont {Mich{\`e}le}\ \bibnamefont {Levi}}, \ and\ \bibinfo {author} {\bibfnamefont {Zhewei}\ \bibnamefont {Yin}},\ }\bibfield  {title} {\enquote {\bibinfo {title} {{N$^{3}$LO quadratic-in-spin interactions for generic compact binaries}},}\ }\href {\doibase 10.1007/JHEP03(2023)098} {\bibfield  {journal} {\bibinfo  {journal} {JHEP}\ }\textbf {\bibinfo {volume} {03}},\ \bibinfo {pages} {098} (\bibinfo {year} {2023}{\natexlab{b}})},\ \Eprint {http://arxiv.org/abs/2209.09235} {arXiv:2209.09235 [hep-th]} \BibitemShut {NoStop}%
\bibitem [{\citenamefont {Levi}\ \emph {et~al.}(2023)\citenamefont {Levi}, \citenamefont {Morales},\ and\ \citenamefont {Yin}}]{Levi:2022dqm}%
  \BibitemOpen
  \bibfield  {author} {\bibinfo {author} {\bibfnamefont {Mich{\`e}le}\ \bibnamefont {Levi}}, \bibinfo {author} {\bibfnamefont {Roger}\ \bibnamefont {Morales}}, \ and\ \bibinfo {author} {\bibfnamefont {Zhewei}\ \bibnamefont {Yin}},\ }\bibfield  {title} {\enquote {\bibinfo {title} {{From the EFT of spinning gravitating objects to Poincar{\'e} and gauge invariance at the 4.5PN precision frontier}},}\ }\href {\doibase 10.1007/JHEP09(2023)090} {\bibfield  {journal} {\bibinfo  {journal} {JHEP}\ }\textbf {\bibinfo {volume} {09}},\ \bibinfo {pages} {090} (\bibinfo {year} {2023})},\ \Eprint {http://arxiv.org/abs/2210.17538} {arXiv:2210.17538 [hep-th]} \BibitemShut {NoStop}%
\bibitem [{\citenamefont {Levi}\ and\ \citenamefont {Yin}(2023)}]{Levi:2022rrq}%
  \BibitemOpen
  \bibfield  {author} {\bibinfo {author} {\bibfnamefont {Mich{\`e}le}\ \bibnamefont {Levi}}\ and\ \bibinfo {author} {\bibfnamefont {Zhewei}\ \bibnamefont {Yin}},\ }\bibfield  {title} {\enquote {\bibinfo {title} {{Completing the fifth PN precision frontier via the EFT of spinning gravitating objects}},}\ }\href {\doibase 10.1007/JHEP04(2023)079} {\bibfield  {journal} {\bibinfo  {journal} {JHEP}\ }\textbf {\bibinfo {volume} {04}},\ \bibinfo {pages} {079} (\bibinfo {year} {2023})},\ \Eprint {http://arxiv.org/abs/2211.14018} {arXiv:2211.14018 [hep-th]} \BibitemShut {NoStop}%
\bibitem [{\citenamefont {Mandal}\ \emph {et~al.}(2023{\natexlab{a}})\citenamefont {Mandal}, \citenamefont {Mastrolia}, \citenamefont {Patil},\ and\ \citenamefont {Steinhoff}}]{Mandal:2022nty}%
  \BibitemOpen
  \bibfield  {author} {\bibinfo {author} {\bibfnamefont {Manoj~K.}\ \bibnamefont {Mandal}}, \bibinfo {author} {\bibfnamefont {Pierpaolo}\ \bibnamefont {Mastrolia}}, \bibinfo {author} {\bibfnamefont {Raj}\ \bibnamefont {Patil}}, \ and\ \bibinfo {author} {\bibfnamefont {Jan}\ \bibnamefont {Steinhoff}},\ }\bibfield  {title} {\enquote {\bibinfo {title} {{Gravitational spin-orbit Hamiltonian at NNNLO in the post-Newtonian framework}},}\ }\href {\doibase 10.1007/JHEP03(2023)130} {\bibfield  {journal} {\bibinfo  {journal} {JHEP}\ }\textbf {\bibinfo {volume} {03}},\ \bibinfo {pages} {130} (\bibinfo {year} {2023}{\natexlab{a}})},\ \Eprint {http://arxiv.org/abs/2209.00611} {arXiv:2209.00611 [hep-th]} \BibitemShut {NoStop}%
\bibitem [{\citenamefont {Mandal}\ \emph {et~al.}(2023{\natexlab{b}})\citenamefont {Mandal}, \citenamefont {Mastrolia}, \citenamefont {Patil},\ and\ \citenamefont {Steinhoff}}]{Mandal:2022ufb}%
  \BibitemOpen
  \bibfield  {author} {\bibinfo {author} {\bibfnamefont {Manoj~K.}\ \bibnamefont {Mandal}}, \bibinfo {author} {\bibfnamefont {Pierpaolo}\ \bibnamefont {Mastrolia}}, \bibinfo {author} {\bibfnamefont {Raj}\ \bibnamefont {Patil}}, \ and\ \bibinfo {author} {\bibfnamefont {Jan}\ \bibnamefont {Steinhoff}},\ }\bibfield  {title} {\enquote {\bibinfo {title} {{Gravitational quadratic-in-spin Hamiltonian at NNNLO in the post-Newtonian framework}},}\ }\href {\doibase 10.1007/JHEP07(2023)128} {\bibfield  {journal} {\bibinfo  {journal} {JHEP}\ }\textbf {\bibinfo {volume} {07}},\ \bibinfo {pages} {128} (\bibinfo {year} {2023}{\natexlab{b}})},\ \Eprint {http://arxiv.org/abs/2210.09176} {arXiv:2210.09176 [hep-th]} \BibitemShut {NoStop}%
\bibitem [{\citenamefont {Goldberger}\ and\ \citenamefont {Ross}(2010)}]{Goldberger:2009qd}%
  \BibitemOpen
  \bibfield  {author} {\bibinfo {author} {\bibfnamefont {Walter~D.}\ \bibnamefont {Goldberger}}\ and\ \bibinfo {author} {\bibfnamefont {Andreas}\ \bibnamefont {Ross}},\ }\bibfield  {title} {\enquote {\bibinfo {title} {{Gravitational radiative corrections from effective field theory}},}\ }\href {\doibase 10.1103/PhysRevD.81.124015} {\bibfield  {journal} {\bibinfo  {journal} {Phys. Rev. D}\ }\textbf {\bibinfo {volume} {81}},\ \bibinfo {pages} {124015} (\bibinfo {year} {2010})},\ \Eprint {http://arxiv.org/abs/0912.4254} {arXiv:0912.4254 [gr-qc]} \BibitemShut {NoStop}%
\bibitem [{\citenamefont {Ross}(2012)}]{Ross:2012fc}%
  \BibitemOpen
  \bibfield  {author} {\bibinfo {author} {\bibfnamefont {Andreas}\ \bibnamefont {Ross}},\ }\bibfield  {title} {\enquote {\bibinfo {title} {{Multipole expansion at the level of the action}},}\ }\href {\doibase 10.1103/PhysRevD.85.125033} {\bibfield  {journal} {\bibinfo  {journal} {Phys. Rev. D}\ }\textbf {\bibinfo {volume} {85}},\ \bibinfo {pages} {125033} (\bibinfo {year} {2012})},\ \Eprint {http://arxiv.org/abs/1202.4750} {arXiv:1202.4750 [gr-qc]} \BibitemShut {NoStop}%
\bibitem [{\citenamefont {Cho}\ \emph {et~al.}(2021)\citenamefont {Cho}, \citenamefont {Pardo},\ and\ \citenamefont {Porto}}]{Cho:2021mqw}%
  \BibitemOpen
  \bibfield  {author} {\bibinfo {author} {\bibfnamefont {Gihyuk}\ \bibnamefont {Cho}}, \bibinfo {author} {\bibfnamefont {Brian}\ \bibnamefont {Pardo}}, \ and\ \bibinfo {author} {\bibfnamefont {Rafael~A.}\ \bibnamefont {Porto}},\ }\bibfield  {title} {\enquote {\bibinfo {title} {{Gravitational radiation from inspiralling compact objects: Spin-spin effects completed at the next-to-leading post-Newtonian order}},}\ }\href {\doibase 10.1103/PhysRevD.104.024037} {\bibfield  {journal} {\bibinfo  {journal} {Phys. Rev. D}\ }\textbf {\bibinfo {volume} {104}},\ \bibinfo {pages} {024037} (\bibinfo {year} {2021})},\ \Eprint {http://arxiv.org/abs/2103.14612} {arXiv:2103.14612 [gr-qc]} \BibitemShut {NoStop}%
\bibitem [{\citenamefont {Cho}\ \emph {et~al.}(2022)\citenamefont {Cho}, \citenamefont {Porto},\ and\ \citenamefont {Yang}}]{Cho:2022syn}%
  \BibitemOpen
  \bibfield  {author} {\bibinfo {author} {\bibfnamefont {Gihyuk}\ \bibnamefont {Cho}}, \bibinfo {author} {\bibfnamefont {Rafael~A.}\ \bibnamefont {Porto}}, \ and\ \bibinfo {author} {\bibfnamefont {Zixin}\ \bibnamefont {Yang}},\ }\bibfield  {title} {\enquote {\bibinfo {title} {{Gravitational radiation from inspiralling compact objects: Spin effects to the fourth post-Newtonian order}},}\ }\href {\doibase 10.1103/PhysRevD.106.L101501} {\bibfield  {journal} {\bibinfo  {journal} {Phys. Rev. D}\ }\textbf {\bibinfo {volume} {106}},\ \bibinfo {pages} {L101501} (\bibinfo {year} {2022})},\ \Eprint {http://arxiv.org/abs/2201.05138} {arXiv:2201.05138 [gr-qc]} \BibitemShut {NoStop}%
\bibitem [{\citenamefont {Amalberti}\ \emph {et~al.}(2024{\natexlab{a}})\citenamefont {Amalberti}, \citenamefont {Larrouturou},\ and\ \citenamefont {Yang}}]{Amalberti:2023ohj}%
  \BibitemOpen
  \bibfield  {author} {\bibinfo {author} {\bibfnamefont {Loris}\ \bibnamefont {Amalberti}}, \bibinfo {author} {\bibfnamefont {Fran{\c{c}}ois}\ \bibnamefont {Larrouturou}}, \ and\ \bibinfo {author} {\bibfnamefont {Zixin}\ \bibnamefont {Yang}},\ }\bibfield  {title} {\enquote {\bibinfo {title} {{Multipole expansion at the level of the action in d-dimensions}},}\ }\href {\doibase 10.1103/PhysRevD.109.104027} {\bibfield  {journal} {\bibinfo  {journal} {Phys. Rev. D}\ }\textbf {\bibinfo {volume} {109}},\ \bibinfo {pages} {104027} (\bibinfo {year} {2024}{\natexlab{a}})},\ \Eprint {http://arxiv.org/abs/2312.02868} {arXiv:2312.02868 [gr-qc]} \BibitemShut {NoStop}%
\bibitem [{\citenamefont {Amalberti}\ \emph {et~al.}(2024{\natexlab{b}})\citenamefont {Amalberti}, \citenamefont {Yang},\ and\ \citenamefont {Porto}}]{Amalberti:2024jaa}%
  \BibitemOpen
  \bibfield  {author} {\bibinfo {author} {\bibfnamefont {Loris}\ \bibnamefont {Amalberti}}, \bibinfo {author} {\bibfnamefont {Zixin}\ \bibnamefont {Yang}}, \ and\ \bibinfo {author} {\bibfnamefont {Rafael~A.}\ \bibnamefont {Porto}},\ }\bibfield  {title} {\enquote {\bibinfo {title} {{Gravitational radiation from inspiralling compact binaries to N3LO in the effective field theory approach}},}\ }\href {\doibase 10.1103/PhysRevD.110.044046} {\bibfield  {journal} {\bibinfo  {journal} {Phys. Rev. D}\ }\textbf {\bibinfo {volume} {110}},\ \bibinfo {pages} {044046} (\bibinfo {year} {2024}{\natexlab{b}})},\ \Eprint {http://arxiv.org/abs/2406.03457} {arXiv:2406.03457 [gr-qc]} \BibitemShut {NoStop}%
\bibitem [{\citenamefont {Mandal}\ \emph {et~al.}(2025)\citenamefont {Mandal}, \citenamefont {Mastrolia}, \citenamefont {Patil},\ and\ \citenamefont {Steinhoff}}]{Mandal:2024iug}%
  \BibitemOpen
  \bibfield  {author} {\bibinfo {author} {\bibfnamefont {Manoj~K.}\ \bibnamefont {Mandal}}, \bibinfo {author} {\bibfnamefont {Pierpaolo}\ \bibnamefont {Mastrolia}}, \bibinfo {author} {\bibfnamefont {Raj}\ \bibnamefont {Patil}}, \ and\ \bibinfo {author} {\bibfnamefont {Jan}\ \bibnamefont {Steinhoff}},\ }\bibfield  {title} {\enquote {\bibinfo {title} {{Radiating Love: adiabatic tidal fluxes and modes up to next-to-next-to-leading post-Newtonian order}},}\ }\href {\doibase 10.1007/JHEP05(2025)008} {\bibfield  {journal} {\bibinfo  {journal} {JHEP}\ }\textbf {\bibinfo {volume} {05}},\ \bibinfo {pages} {008} (\bibinfo {year} {2025})},\ \Eprint {http://arxiv.org/abs/2412.01706} {arXiv:2412.01706 [gr-qc]} \BibitemShut {NoStop}%
\bibitem [{\citenamefont {Westpfahl}\ and\ \citenamefont {Goller}(1979)}]{Westpfahl:1979gu}%
  \BibitemOpen
  \bibfield  {author} {\bibinfo {author} {\bibfnamefont {K.}~\bibnamefont {Westpfahl}}\ and\ \bibinfo {author} {\bibfnamefont {M.}~\bibnamefont {Goller}},\ }\bibfield  {title} {\enquote {\bibinfo {title} {{GRAVITATIONAL SCATTERING OF TWO RELATIVISTIC PARTICLES IN POSTLINEAR APPROXIMATION}},}\ }\href {\doibase 10.1007/BF02817047} {\bibfield  {journal} {\bibinfo  {journal} {Lett. Nuovo Cim.}\ }\textbf {\bibinfo {volume} {26}},\ \bibinfo {pages} {573--576} (\bibinfo {year} {1979})}\BibitemShut {NoStop}%
\bibitem [{\citenamefont {Westpfahl}\ and\ \citenamefont {Hoyler}(1980)}]{Westpfahl:1980mk}%
  \BibitemOpen
  \bibfield  {author} {\bibinfo {author} {\bibfnamefont {K.}~\bibnamefont {Westpfahl}}\ and\ \bibinfo {author} {\bibfnamefont {H.}~\bibnamefont {Hoyler}},\ }\bibfield  {title} {\enquote {\bibinfo {title} {{GRAVITATIONAL BREMSSTRAHLUNG IN POSTLINEAR FAST MOTION APPROXIMATION}},}\ }\href {\doibase 10.1007/BF02750304} {\bibfield  {journal} {\bibinfo  {journal} {Lett. Nuovo Cim.}\ }\textbf {\bibinfo {volume} {27}},\ \bibinfo {pages} {581--585} (\bibinfo {year} {1980})}\BibitemShut {NoStop}%
\bibitem [{\citenamefont {Bel}\ \emph {et~al.}(1981)\citenamefont {Bel}, \citenamefont {Damour}, \citenamefont {Deruelle}, \citenamefont {Ibanez},\ and\ \citenamefont {Martin}}]{Bel:1981be}%
  \BibitemOpen
  \bibfield  {author} {\bibinfo {author} {\bibfnamefont {LLuis}\ \bibnamefont {Bel}}, \bibinfo {author} {\bibfnamefont {T.}~\bibnamefont {Damour}}, \bibinfo {author} {\bibfnamefont {N.}~\bibnamefont {Deruelle}}, \bibinfo {author} {\bibfnamefont {J.}~\bibnamefont {Ibanez}}, \ and\ \bibinfo {author} {\bibfnamefont {J.}~\bibnamefont {Martin}},\ }\bibfield  {title} {\enquote {\bibinfo {title} {{Poincar{\'e}-invariant gravitational field and equations of motion of two pointlike objects: The postlinear approximation of general relativity}},}\ }\href {\doibase 10.1007/BF00756073} {\bibfield  {journal} {\bibinfo  {journal} {Gen. Rel. Grav.}\ }\textbf {\bibinfo {volume} {13}},\ \bibinfo {pages} {963--1004} (\bibinfo {year} {1981})}\BibitemShut {NoStop}%
\bibitem [{\citenamefont {Westpfahl}(1985)}]{Westpfahl:1985tsl}%
  \BibitemOpen
  \bibfield  {author} {\bibinfo {author} {\bibfnamefont {Konradin}\ \bibnamefont {Westpfahl}},\ }\bibfield  {title} {\enquote {\bibinfo {title} {{High-Speed Scattering of Charged and Uncharged Particles in General Relativity}},}\ }\href {\doibase 10.1002/prop.2190330802} {\bibfield  {journal} {\bibinfo  {journal} {Fortsch. Phys.}\ }\textbf {\bibinfo {volume} {33}},\ \bibinfo {pages} {417--493} (\bibinfo {year} {1985})}\BibitemShut {NoStop}%
\bibitem [{\citenamefont {Sch{\"a}fer}(1986)}]{schafer1986adm}%
  \BibitemOpen
  \bibfield  {author} {\bibinfo {author} {\bibfnamefont {Gerhard}\ \bibnamefont {Sch{\"a}fer}},\ }\bibfield  {title} {\enquote {\bibinfo {title} {The adm hamiltonian at the postlinear approximation},}\ }\href@noop {} {\bibfield  {journal} {\bibinfo  {journal} {General relativity and gravitation}\ }\textbf {\bibinfo {volume} {18}},\ \bibinfo {pages} {255--270} (\bibinfo {year} {1986})}\BibitemShut {NoStop}%
\bibitem [{\citenamefont {Ledvinka}\ \emph {et~al.}(2008)\citenamefont {Ledvinka}, \citenamefont {Schaefer},\ and\ \citenamefont {Bicak}}]{Ledvinka:2008tk}%
  \BibitemOpen
  \bibfield  {author} {\bibinfo {author} {\bibfnamefont {Tomas}\ \bibnamefont {Ledvinka}}, \bibinfo {author} {\bibfnamefont {Gerhard}\ \bibnamefont {Schaefer}}, \ and\ \bibinfo {author} {\bibfnamefont {Jiri}\ \bibnamefont {Bicak}},\ }\bibfield  {title} {\enquote {\bibinfo {title} {{Relativistic Closed-Form Hamiltonian for Many-Body Gravitating Systems in the Post-Minkowskian Approximation}},}\ }\href {\doibase 10.1103/PhysRevLett.100.251101} {\bibfield  {journal} {\bibinfo  {journal} {Phys. Rev. Lett.}\ }\textbf {\bibinfo {volume} {100}},\ \bibinfo {pages} {251101} (\bibinfo {year} {2008})},\ \Eprint {http://arxiv.org/abs/0807.0214} {arXiv:0807.0214 [gr-qc]} \BibitemShut {NoStop}%
\bibitem [{\citenamefont {Buonanno}\ \emph {et~al.}(2022)\citenamefont {Buonanno}, \citenamefont {Khalil}, \citenamefont {O'Connell}, \citenamefont {Roiban}, \citenamefont {Solon},\ and\ \citenamefont {Zeng}}]{Buonanno:2022pgc}%
  \BibitemOpen
  \bibfield  {author} {\bibinfo {author} {\bibfnamefont {Alessandra}\ \bibnamefont {Buonanno}}, \bibinfo {author} {\bibfnamefont {Mohammed}\ \bibnamefont {Khalil}}, \bibinfo {author} {\bibfnamefont {Donal}\ \bibnamefont {O'Connell}}, \bibinfo {author} {\bibfnamefont {Radu}\ \bibnamefont {Roiban}}, \bibinfo {author} {\bibfnamefont {Mikhail~P.}\ \bibnamefont {Solon}}, \ and\ \bibinfo {author} {\bibfnamefont {Mao}\ \bibnamefont {Zeng}},\ }\bibfield  {title} {\enquote {\bibinfo {title} {{Snowmass White Paper: Gravitational Waves and Scattering Amplitudes}},}\ }in\ \href@noop {} {\emph {\bibinfo {booktitle} {{Snowmass 2021}}}}\ (\bibinfo {year} {2022})\ \Eprint {http://arxiv.org/abs/2204.05194} {arXiv:2204.05194 [hep-th]} \BibitemShut {NoStop}%
\bibitem [{\citenamefont {Travaglini}\ \emph {et~al.}(2022)\citenamefont {Travaglini} \emph {et~al.}}]{Travaglini:2022uwo}%
  \BibitemOpen
  \bibfield  {author} {\bibinfo {author} {\bibfnamefont {Gabriele}\ \bibnamefont {Travaglini}} \emph {et~al.},\ }\bibfield  {title} {\enquote {\bibinfo {title} {{The SAGEX review on scattering amplitudes}},}\ }\href {\doibase 10.1088/1751-8121/ac8380} {\bibfield  {journal} {\bibinfo  {journal} {J. Phys. A}\ }\textbf {\bibinfo {volume} {55}},\ \bibinfo {pages} {443001} (\bibinfo {year} {2022})},\ \Eprint {http://arxiv.org/abs/2203.13011} {arXiv:2203.13011 [hep-th]} \BibitemShut {NoStop}%
\bibitem [{\citenamefont {Bjerrum-Bohr}\ \emph {et~al.}(2022)\citenamefont {Bjerrum-Bohr}, \citenamefont {Damgaard}, \citenamefont {Plante},\ and\ \citenamefont {Vanhove}}]{Bjerrum-Bohr:2022blt}%
  \BibitemOpen
  \bibfield  {author} {\bibinfo {author} {\bibfnamefont {N.~E.~J.}\ \bibnamefont {Bjerrum-Bohr}}, \bibinfo {author} {\bibfnamefont {P.~H.}\ \bibnamefont {Damgaard}}, \bibinfo {author} {\bibfnamefont {L.}~\bibnamefont {Plante}}, \ and\ \bibinfo {author} {\bibfnamefont {P.}~\bibnamefont {Vanhove}},\ }\bibfield  {title} {\enquote {\bibinfo {title} {{The SAGEX review on scattering amplitudes Chapter 13: Post-Minkowskian expansion from scattering amplitudes}},}\ }\href {\doibase 10.1088/1751-8121/ac7a78} {\bibfield  {journal} {\bibinfo  {journal} {J. Phys. A}\ }\textbf {\bibinfo {volume} {55}},\ \bibinfo {pages} {443014} (\bibinfo {year} {2022})},\ \Eprint {http://arxiv.org/abs/2203.13024} {arXiv:2203.13024 [hep-th]} \BibitemShut {NoStop}%
\bibitem [{\citenamefont {Kosower}\ \emph {et~al.}(2022)\citenamefont {Kosower}, \citenamefont {Monteiro},\ and\ \citenamefont {O'Connell}}]{Kosower:2022yvp}%
  \BibitemOpen
  \bibfield  {author} {\bibinfo {author} {\bibfnamefont {David~A.}\ \bibnamefont {Kosower}}, \bibinfo {author} {\bibfnamefont {Ricardo}\ \bibnamefont {Monteiro}}, \ and\ \bibinfo {author} {\bibfnamefont {Donal}\ \bibnamefont {O'Connell}},\ }\bibfield  {title} {\enquote {\bibinfo {title} {{The SAGEX review on scattering amplitudes Chapter 14: Classical gravity from scattering amplitudes}},}\ }\href {\doibase 10.1088/1751-8121/ac8846} {\bibfield  {journal} {\bibinfo  {journal} {J. Phys. A}\ }\textbf {\bibinfo {volume} {55}},\ \bibinfo {pages} {443015} (\bibinfo {year} {2022})},\ \Eprint {http://arxiv.org/abs/2203.13025} {arXiv:2203.13025 [hep-th]} \BibitemShut {NoStop}%
\bibitem [{\citenamefont {Damour}(2016)}]{Damour:2016gwp}%
  \BibitemOpen
  \bibfield  {author} {\bibinfo {author} {\bibfnamefont {Thibault}\ \bibnamefont {Damour}},\ }\bibfield  {title} {\enquote {\bibinfo {title} {{Gravitational scattering, post-Minkowskian approximation and Effective One-Body theory}},}\ }\href {\doibase 10.1103/PhysRevD.94.104015} {\bibfield  {journal} {\bibinfo  {journal} {Phys. Rev. D}\ }\textbf {\bibinfo {volume} {94}},\ \bibinfo {pages} {104015} (\bibinfo {year} {2016})},\ \Eprint {http://arxiv.org/abs/1609.00354} {arXiv:1609.00354 [gr-qc]} \BibitemShut {NoStop}%
\bibitem [{\citenamefont {Cheung}\ \emph {et~al.}(2018)\citenamefont {Cheung}, \citenamefont {Rothstein},\ and\ \citenamefont {Solon}}]{Cheung:2018wkq}%
  \BibitemOpen
  \bibfield  {author} {\bibinfo {author} {\bibfnamefont {Clifford}\ \bibnamefont {Cheung}}, \bibinfo {author} {\bibfnamefont {Ira~Z.}\ \bibnamefont {Rothstein}}, \ and\ \bibinfo {author} {\bibfnamefont {Mikhail~P.}\ \bibnamefont {Solon}},\ }\bibfield  {title} {\enquote {\bibinfo {title} {{From Scattering Amplitudes to Classical Potentials in the Post-Minkowskian Expansion}},}\ }\href {\doibase 10.1103/PhysRevLett.121.251101} {\bibfield  {journal} {\bibinfo  {journal} {Phys. Rev. Lett.}\ }\textbf {\bibinfo {volume} {121}},\ \bibinfo {pages} {251101} (\bibinfo {year} {2018})},\ \Eprint {http://arxiv.org/abs/1808.02489} {arXiv:1808.02489 [hep-th]} \BibitemShut {NoStop}%
\bibitem [{\citenamefont {Bern}\ \emph {et~al.}(2019)\citenamefont {Bern}, \citenamefont {Cheung}, \citenamefont {Roiban}, \citenamefont {Shen}, \citenamefont {Solon},\ and\ \citenamefont {Zeng}}]{Bern:2019nnu}%
  \BibitemOpen
  \bibfield  {author} {\bibinfo {author} {\bibfnamefont {Zvi}\ \bibnamefont {Bern}}, \bibinfo {author} {\bibfnamefont {Clifford}\ \bibnamefont {Cheung}}, \bibinfo {author} {\bibfnamefont {Radu}\ \bibnamefont {Roiban}}, \bibinfo {author} {\bibfnamefont {Chia-Hsien}\ \bibnamefont {Shen}}, \bibinfo {author} {\bibfnamefont {Mikhail~P.}\ \bibnamefont {Solon}}, \ and\ \bibinfo {author} {\bibfnamefont {Mao}\ \bibnamefont {Zeng}},\ }\bibfield  {title} {\enquote {\bibinfo {title} {{Scattering Amplitudes and the Conservative Hamiltonian for Binary Systems at Third Post-Minkowskian Order}},}\ }\href {\doibase 10.1103/PhysRevLett.122.201603} {\bibfield  {journal} {\bibinfo  {journal} {Phys. Rev. Lett.}\ }\textbf {\bibinfo {volume} {122}},\ \bibinfo {pages} {201603} (\bibinfo {year} {2019})},\ \Eprint {http://arxiv.org/abs/1901.04424} {arXiv:1901.04424 [hep-th]} \BibitemShut {NoStop}%
\bibitem [{\citenamefont {Bern}\ \emph {et~al.}(2023)\citenamefont {Bern}, \citenamefont {Kosmopoulos}, \citenamefont {Luna}, \citenamefont {Roiban},\ and\ \citenamefont {Teng}}]{Bern:2022kto}%
  \BibitemOpen
  \bibfield  {author} {\bibinfo {author} {\bibfnamefont {Zvi}\ \bibnamefont {Bern}}, \bibinfo {author} {\bibfnamefont {Dimitrios}\ \bibnamefont {Kosmopoulos}}, \bibinfo {author} {\bibfnamefont {Andr{\'e}s}\ \bibnamefont {Luna}}, \bibinfo {author} {\bibfnamefont {Radu}\ \bibnamefont {Roiban}}, \ and\ \bibinfo {author} {\bibfnamefont {Fei}\ \bibnamefont {Teng}},\ }\bibfield  {title} {\enquote {\bibinfo {title} {{Binary Dynamics through the Fifth Power of Spin at O(G2)}},}\ }\href {\doibase 10.1103/PhysRevLett.130.201402} {\bibfield  {journal} {\bibinfo  {journal} {Phys. Rev. Lett.}\ }\textbf {\bibinfo {volume} {130}},\ \bibinfo {pages} {201402} (\bibinfo {year} {2023})},\ \Eprint {http://arxiv.org/abs/2203.06202} {arXiv:2203.06202 [hep-th]} \BibitemShut {NoStop}%
\bibitem [{\citenamefont {Bern}\ \emph {et~al.}(2024)\citenamefont {Bern}, \citenamefont {Herrmann}, \citenamefont {Roiban}, \citenamefont {Ruf}, \citenamefont {Smirnov}, \citenamefont {Smirnov},\ and\ \citenamefont {Zeng}}]{Bern:2024adl}%
  \BibitemOpen
  \bibfield  {author} {\bibinfo {author} {\bibfnamefont {Zvi}\ \bibnamefont {Bern}}, \bibinfo {author} {\bibfnamefont {Enrico}\ \bibnamefont {Herrmann}}, \bibinfo {author} {\bibfnamefont {Radu}\ \bibnamefont {Roiban}}, \bibinfo {author} {\bibfnamefont {Michael~S.}\ \bibnamefont {Ruf}}, \bibinfo {author} {\bibfnamefont {Alexander~V.}\ \bibnamefont {Smirnov}}, \bibinfo {author} {\bibfnamefont {Vladimir~A.}\ \bibnamefont {Smirnov}}, \ and\ \bibinfo {author} {\bibfnamefont {Mao}\ \bibnamefont {Zeng}},\ }\bibfield  {title} {\enquote {\bibinfo {title} {{Amplitudes, supersymmetric black hole scattering at $ \mathcal{O}\left({G}^5\right) $, and loop integration}},}\ }\href {\doibase 10.1007/JHEP10(2024)023} {\bibfield  {journal} {\bibinfo  {journal} {JHEP}\ }\textbf {\bibinfo {volume} {10}},\ \bibinfo {pages} {023} (\bibinfo {year} {2024})},\ \Eprint {http://arxiv.org/abs/2406.01554} {arXiv:2406.01554 [hep-th]} \BibitemShut {NoStop}%
\bibitem [{\citenamefont {Kosower}\ \emph {et~al.}(2019)\citenamefont {Kosower}, \citenamefont {Maybee},\ and\ \citenamefont {O'Connell}}]{Kosower:2018adc}%
  \BibitemOpen
  \bibfield  {author} {\bibinfo {author} {\bibfnamefont {David~A.}\ \bibnamefont {Kosower}}, \bibinfo {author} {\bibfnamefont {Ben}\ \bibnamefont {Maybee}}, \ and\ \bibinfo {author} {\bibfnamefont {Donal}\ \bibnamefont {O'Connell}},\ }\bibfield  {title} {\enquote {\bibinfo {title} {{Amplitudes, Observables, and Classical Scattering}},}\ }\href {\doibase 10.1007/JHEP02(2019)137} {\bibfield  {journal} {\bibinfo  {journal} {JHEP}\ }\textbf {\bibinfo {volume} {02}},\ \bibinfo {pages} {137} (\bibinfo {year} {2019})},\ \Eprint {http://arxiv.org/abs/1811.10950} {arXiv:1811.10950 [hep-th]} \BibitemShut {NoStop}%
\bibitem [{\citenamefont {Bjerrum-Bohr}\ \emph {et~al.}(2018)\citenamefont {Bjerrum-Bohr}, \citenamefont {Damgaard}, \citenamefont {Festuccia}, \citenamefont {Plant{\'e}},\ and\ \citenamefont {Vanhove}}]{Bjerrum-Bohr:2018xdl}%
  \BibitemOpen
  \bibfield  {author} {\bibinfo {author} {\bibfnamefont {N.~E.~J.}\ \bibnamefont {Bjerrum-Bohr}}, \bibinfo {author} {\bibfnamefont {Poul~H.}\ \bibnamefont {Damgaard}}, \bibinfo {author} {\bibfnamefont {Guido}\ \bibnamefont {Festuccia}}, \bibinfo {author} {\bibfnamefont {Ludovic}\ \bibnamefont {Plant{\'e}}}, \ and\ \bibinfo {author} {\bibfnamefont {Pierre}\ \bibnamefont {Vanhove}},\ }\bibfield  {title} {\enquote {\bibinfo {title} {{General Relativity from Scattering Amplitudes}},}\ }\href {\doibase 10.1103/PhysRevLett.121.171601} {\bibfield  {journal} {\bibinfo  {journal} {Phys. Rev. Lett.}\ }\textbf {\bibinfo {volume} {121}},\ \bibinfo {pages} {171601} (\bibinfo {year} {2018})},\ \Eprint {http://arxiv.org/abs/1806.04920} {arXiv:1806.04920 [hep-th]} \BibitemShut {NoStop}%
\bibitem [{\citenamefont {Cristofoli}\ \emph {et~al.}(2019)\citenamefont {Cristofoli}, \citenamefont {Bjerrum-Bohr}, \citenamefont {Damgaard},\ and\ \citenamefont {Vanhove}}]{Cristofoli:2019neg}%
  \BibitemOpen
  \bibfield  {author} {\bibinfo {author} {\bibfnamefont {Andrea}\ \bibnamefont {Cristofoli}}, \bibinfo {author} {\bibfnamefont {N.~E.~J.}\ \bibnamefont {Bjerrum-Bohr}}, \bibinfo {author} {\bibfnamefont {Poul~H.}\ \bibnamefont {Damgaard}}, \ and\ \bibinfo {author} {\bibfnamefont {Pierre}\ \bibnamefont {Vanhove}},\ }\bibfield  {title} {\enquote {\bibinfo {title} {{Post-Minkowskian Hamiltonians in general relativity}},}\ }\href {\doibase 10.1103/PhysRevD.100.084040} {\bibfield  {journal} {\bibinfo  {journal} {Phys. Rev. D}\ }\textbf {\bibinfo {volume} {100}},\ \bibinfo {pages} {084040} (\bibinfo {year} {2019})},\ \Eprint {http://arxiv.org/abs/1906.01579} {arXiv:1906.01579 [hep-th]} \BibitemShut {NoStop}%
\bibitem [{\citenamefont {Damgaard}\ \emph {et~al.}(2019)\citenamefont {Damgaard}, \citenamefont {Haddad},\ and\ \citenamefont {Helset}}]{Damgaard:2019lfh}%
  \BibitemOpen
  \bibfield  {author} {\bibinfo {author} {\bibfnamefont {Poul~H.}\ \bibnamefont {Damgaard}}, \bibinfo {author} {\bibfnamefont {Kays}\ \bibnamefont {Haddad}}, \ and\ \bibinfo {author} {\bibfnamefont {Andreas}\ \bibnamefont {Helset}},\ }\bibfield  {title} {\enquote {\bibinfo {title} {{Heavy Black Hole Effective Theory}},}\ }\href {\doibase 10.1007/JHEP11(2019)070} {\bibfield  {journal} {\bibinfo  {journal} {JHEP}\ }\textbf {\bibinfo {volume} {11}},\ \bibinfo {pages} {070} (\bibinfo {year} {2019})},\ \Eprint {http://arxiv.org/abs/1908.10308} {arXiv:1908.10308 [hep-ph]} \BibitemShut {NoStop}%
\bibitem [{\citenamefont {Brandhuber}\ \emph {et~al.}(2021)\citenamefont {Brandhuber}, \citenamefont {Chen}, \citenamefont {Travaglini},\ and\ \citenamefont {Wen}}]{Brandhuber:2021eyq}%
  \BibitemOpen
  \bibfield  {author} {\bibinfo {author} {\bibfnamefont {Andreas}\ \bibnamefont {Brandhuber}}, \bibinfo {author} {\bibfnamefont {Gang}\ \bibnamefont {Chen}}, \bibinfo {author} {\bibfnamefont {Gabriele}\ \bibnamefont {Travaglini}}, \ and\ \bibinfo {author} {\bibfnamefont {Congkao}\ \bibnamefont {Wen}},\ }\bibfield  {title} {\enquote {\bibinfo {title} {{Classical gravitational scattering from a gauge-invariant double copy}},}\ }\href {\doibase 10.1007/JHEP10(2021)118} {\bibfield  {journal} {\bibinfo  {journal} {JHEP}\ }\textbf {\bibinfo {volume} {10}},\ \bibinfo {pages} {118} (\bibinfo {year} {2021})},\ \Eprint {http://arxiv.org/abs/2108.04216} {arXiv:2108.04216 [hep-th]} \BibitemShut {NoStop}%
\bibitem [{\citenamefont {Vines}(2018)}]{Vines:2017hyw}%
  \BibitemOpen
  \bibfield  {author} {\bibinfo {author} {\bibfnamefont {Justin}\ \bibnamefont {Vines}},\ }\bibfield  {title} {\enquote {\bibinfo {title} {{Scattering of two spinning black holes in post-Minkowskian gravity, to all orders in spin, and effective-one-body mappings}},}\ }\href {\doibase 10.1088/1361-6382/aaa3a8} {\bibfield  {journal} {\bibinfo  {journal} {Class. Quant. Grav.}\ }\textbf {\bibinfo {volume} {35}},\ \bibinfo {pages} {084002} (\bibinfo {year} {2018})},\ \Eprint {http://arxiv.org/abs/1709.06016} {arXiv:1709.06016 [gr-qc]} \BibitemShut {NoStop}%
\bibitem [{\citenamefont {K{\"a}lin}\ and\ \citenamefont {Porto}(2020{\natexlab{a}})}]{Kalin:2020mvi}%
  \BibitemOpen
  \bibfield  {author} {\bibinfo {author} {\bibfnamefont {Gregor}\ \bibnamefont {K{\"a}lin}}\ and\ \bibinfo {author} {\bibfnamefont {Rafael~A.}\ \bibnamefont {Porto}},\ }\bibfield  {title} {\enquote {\bibinfo {title} {{Post-Minkowskian Effective Field Theory for Conservative Binary Dynamics}},}\ }\href {\doibase 10.1007/JHEP11(2020)106} {\bibfield  {journal} {\bibinfo  {journal} {JHEP}\ }\textbf {\bibinfo {volume} {11}},\ \bibinfo {pages} {106} (\bibinfo {year} {2020}{\natexlab{a}})},\ \Eprint {http://arxiv.org/abs/2006.01184} {arXiv:2006.01184 [hep-th]} \BibitemShut {NoStop}%
\bibitem [{\citenamefont {K{\"a}lin}\ \emph {et~al.}(2020)\citenamefont {K{\"a}lin}, \citenamefont {Liu},\ and\ \citenamefont {Porto}}]{Kalin:2020fhe}%
  \BibitemOpen
  \bibfield  {author} {\bibinfo {author} {\bibfnamefont {Gregor}\ \bibnamefont {K{\"a}lin}}, \bibinfo {author} {\bibfnamefont {Zhengwen}\ \bibnamefont {Liu}}, \ and\ \bibinfo {author} {\bibfnamefont {Rafael~A.}\ \bibnamefont {Porto}},\ }\bibfield  {title} {\enquote {\bibinfo {title} {{Conservative Dynamics of Binary Systems to Third Post-Minkowskian Order from the Effective Field Theory Approach}},}\ }\href {\doibase 10.1103/PhysRevLett.125.261103} {\bibfield  {journal} {\bibinfo  {journal} {Phys. Rev. Lett.}\ }\textbf {\bibinfo {volume} {125}},\ \bibinfo {pages} {261103} (\bibinfo {year} {2020})},\ \Eprint {http://arxiv.org/abs/2007.04977} {arXiv:2007.04977 [hep-th]} \BibitemShut {NoStop}%
\bibitem [{\citenamefont {Mogull}\ \emph {et~al.}(2021)\citenamefont {Mogull}, \citenamefont {Plefka},\ and\ \citenamefont {Steinhoff}}]{Mogull:2020sak}%
  \BibitemOpen
  \bibfield  {author} {\bibinfo {author} {\bibfnamefont {Gustav}\ \bibnamefont {Mogull}}, \bibinfo {author} {\bibfnamefont {Jan}\ \bibnamefont {Plefka}}, \ and\ \bibinfo {author} {\bibfnamefont {Jan}\ \bibnamefont {Steinhoff}},\ }\bibfield  {title} {\enquote {\bibinfo {title} {{Classical black hole scattering from a worldline quantum field theory}},}\ }\href {\doibase 10.1007/JHEP02(2021)048} {\bibfield  {journal} {\bibinfo  {journal} {JHEP}\ }\textbf {\bibinfo {volume} {02}},\ \bibinfo {pages} {048} (\bibinfo {year} {2021})},\ \Eprint {http://arxiv.org/abs/2010.02865} {arXiv:2010.02865 [hep-th]} \BibitemShut {NoStop}%
\bibitem [{\citenamefont {Jakobsen}\ \emph {et~al.}(2022{\natexlab{a}})\citenamefont {Jakobsen}, \citenamefont {Mogull}, \citenamefont {Plefka},\ and\ \citenamefont {Steinhoff}}]{Jakobsen:2021zvh}%
  \BibitemOpen
  \bibfield  {author} {\bibinfo {author} {\bibfnamefont {Gustav~Uhre}\ \bibnamefont {Jakobsen}}, \bibinfo {author} {\bibfnamefont {Gustav}\ \bibnamefont {Mogull}}, \bibinfo {author} {\bibfnamefont {Jan}\ \bibnamefont {Plefka}}, \ and\ \bibinfo {author} {\bibfnamefont {Jan}\ \bibnamefont {Steinhoff}},\ }\bibfield  {title} {\enquote {\bibinfo {title} {{SUSY in the sky with gravitons}},}\ }\href {\doibase 10.1007/JHEP01(2022)027} {\bibfield  {journal} {\bibinfo  {journal} {JHEP}\ }\textbf {\bibinfo {volume} {01}},\ \bibinfo {pages} {027} (\bibinfo {year} {2022}{\natexlab{a}})},\ \Eprint {http://arxiv.org/abs/2109.04465} {arXiv:2109.04465 [hep-th]} \BibitemShut {NoStop}%
\bibitem [{\citenamefont {Jakobsen}\ \emph {et~al.}(2022{\natexlab{b}})\citenamefont {Jakobsen}, \citenamefont {Mogull}, \citenamefont {Plefka},\ and\ \citenamefont {Sauer}}]{Jakobsen:2022psy}%
  \BibitemOpen
  \bibfield  {author} {\bibinfo {author} {\bibfnamefont {Gustav~Uhre}\ \bibnamefont {Jakobsen}}, \bibinfo {author} {\bibfnamefont {Gustav}\ \bibnamefont {Mogull}}, \bibinfo {author} {\bibfnamefont {Jan}\ \bibnamefont {Plefka}}, \ and\ \bibinfo {author} {\bibfnamefont {Benjamin}\ \bibnamefont {Sauer}},\ }\bibfield  {title} {\enquote {\bibinfo {title} {{All things retarded: radiation-reaction in worldline quantum field theory}},}\ }\href {\doibase 10.1007/JHEP10(2022)128} {\bibfield  {journal} {\bibinfo  {journal} {JHEP}\ }\textbf {\bibinfo {volume} {10}},\ \bibinfo {pages} {128} (\bibinfo {year} {2022}{\natexlab{b}})},\ \Eprint {http://arxiv.org/abs/2207.00569} {arXiv:2207.00569 [hep-th]} \BibitemShut {NoStop}%
\bibitem [{\citenamefont {Driesse}\ \emph {et~al.}(2024)\citenamefont {Driesse}, \citenamefont {Jakobsen}, \citenamefont {Mogull}, \citenamefont {Plefka}, \citenamefont {Sauer},\ and\ \citenamefont {Usovitsch}}]{Driesse:2024xad}%
  \BibitemOpen
  \bibfield  {author} {\bibinfo {author} {\bibfnamefont {Mathias}\ \bibnamefont {Driesse}}, \bibinfo {author} {\bibfnamefont {Gustav~Uhre}\ \bibnamefont {Jakobsen}}, \bibinfo {author} {\bibfnamefont {Gustav}\ \bibnamefont {Mogull}}, \bibinfo {author} {\bibfnamefont {Jan}\ \bibnamefont {Plefka}}, \bibinfo {author} {\bibfnamefont {Benjamin}\ \bibnamefont {Sauer}}, \ and\ \bibinfo {author} {\bibfnamefont {Johann}\ \bibnamefont {Usovitsch}},\ }\bibfield  {title} {\enquote {\bibinfo {title} {{Conservative Black Hole Scattering at Fifth Post-Minkowskian and First Self-Force Order}},}\ }\href {\doibase 10.1103/PhysRevLett.132.241402} {\bibfield  {journal} {\bibinfo  {journal} {Phys. Rev. Lett.}\ }\textbf {\bibinfo {volume} {132}},\ \bibinfo {pages} {241402} (\bibinfo {year} {2024})},\ \Eprint {http://arxiv.org/abs/2403.07781} {arXiv:2403.07781 [hep-th]} \BibitemShut {NoStop}%
\bibitem [{\citenamefont {K{\"a}lin}\ \emph {et~al.}(2023)\citenamefont {K{\"a}lin}, \citenamefont {Neef},\ and\ \citenamefont {Porto}}]{Kalin:2022hph}%
  \BibitemOpen
  \bibfield  {author} {\bibinfo {author} {\bibfnamefont {Gregor}\ \bibnamefont {K{\"a}lin}}, \bibinfo {author} {\bibfnamefont {Jakob}\ \bibnamefont {Neef}}, \ and\ \bibinfo {author} {\bibfnamefont {Rafael~A.}\ \bibnamefont {Porto}},\ }\bibfield  {title} {\enquote {\bibinfo {title} {{Radiation-reaction in the Effective Field Theory approach to Post-Minkowskian dynamics}},}\ }\href {\doibase 10.1007/JHEP01(2023)140} {\bibfield  {journal} {\bibinfo  {journal} {JHEP}\ }\textbf {\bibinfo {volume} {01}},\ \bibinfo {pages} {140} (\bibinfo {year} {2023})},\ \Eprint {http://arxiv.org/abs/2207.00580} {arXiv:2207.00580 [hep-th]} \BibitemShut {NoStop}%
\bibitem [{\citenamefont {Driesse}\ \emph {et~al.}(2025)\citenamefont {Driesse}, \citenamefont {Jakobsen}, \citenamefont {Klemm}, \citenamefont {Mogull}, \citenamefont {Nega}, \citenamefont {Plefka}, \citenamefont {Sauer},\ and\ \citenamefont {Usovitsch}}]{Driesse:2024feo}%
  \BibitemOpen
  \bibfield  {author} {\bibinfo {author} {\bibfnamefont {Mathias}\ \bibnamefont {Driesse}}, \bibinfo {author} {\bibfnamefont {Gustav~Uhre}\ \bibnamefont {Jakobsen}}, \bibinfo {author} {\bibfnamefont {Albrecht}\ \bibnamefont {Klemm}}, \bibinfo {author} {\bibfnamefont {Gustav}\ \bibnamefont {Mogull}}, \bibinfo {author} {\bibfnamefont {Christoph}\ \bibnamefont {Nega}}, \bibinfo {author} {\bibfnamefont {Jan}\ \bibnamefont {Plefka}}, \bibinfo {author} {\bibfnamefont {Benjamin}\ \bibnamefont {Sauer}}, \ and\ \bibinfo {author} {\bibfnamefont {Johann}\ \bibnamefont {Usovitsch}},\ }\bibfield  {title} {\enquote {\bibinfo {title} {{Emergence of Calabi{\textendash}Yau manifolds in high-precision black-hole scattering}},}\ }\href {\doibase 10.1038/s41586-025-08984-2} {\bibfield  {journal} {\bibinfo  {journal} {Nature}\ }\textbf {\bibinfo {volume} {641}},\ \bibinfo {pages} {603--607} (\bibinfo {year} {2025})},\ \Eprint {http://arxiv.org/abs/2411.11846} {arXiv:2411.11846 [hep-th]} \BibitemShut {NoStop}%
\bibitem [{\citenamefont {Bern}\ \emph {et~al.}(2026)\citenamefont {Bern}, \citenamefont {Herrmann}, \citenamefont {Roiban}, \citenamefont {Ruf}, \citenamefont {Smirnov}, \citenamefont {Smirnov},\ and\ \citenamefont {Zeng}}]{Bern:2025zno}%
  \BibitemOpen
  \bibfield  {author} {\bibinfo {author} {\bibfnamefont {Zvi}\ \bibnamefont {Bern}}, \bibinfo {author} {\bibfnamefont {Enrico}\ \bibnamefont {Herrmann}}, \bibinfo {author} {\bibfnamefont {Radu}\ \bibnamefont {Roiban}}, \bibinfo {author} {\bibfnamefont {Michael~S.}\ \bibnamefont {Ruf}}, \bibinfo {author} {\bibfnamefont {Alexander~V.}\ \bibnamefont {Smirnov}}, \bibinfo {author} {\bibfnamefont {Vladimir~A.}\ \bibnamefont {Smirnov}}, \ and\ \bibinfo {author} {\bibfnamefont {Mao}\ \bibnamefont {Zeng}},\ }\bibfield  {title} {\enquote {\bibinfo {title} {{Second-Order Self-Force Potential-Region Binary Dynamics at O(G5) in Supergravity}},}\ }\href {\doibase 10.1103/jmby-htz9} {\bibfield  {journal} {\bibinfo  {journal} {Phys. Rev. Lett.}\ }\textbf {\bibinfo {volume} {136}},\ \bibinfo {pages} {081401} (\bibinfo {year} {2026})},\ \Eprint {http://arxiv.org/abs/2509.17412} {arXiv:2509.17412 [hep-th]} \BibitemShut {NoStop}%
\bibitem [{\citenamefont {Dlapa}\ \emph {et~al.}(2025)\citenamefont {Dlapa}, \citenamefont {K{\"a}lin}, \citenamefont {Liu},\ and\ \citenamefont {Porto}}]{Dlapa:2025biy}%
  \BibitemOpen
  \bibfield  {author} {\bibinfo {author} {\bibfnamefont {Christoph}\ \bibnamefont {Dlapa}}, \bibinfo {author} {\bibfnamefont {Gregor}\ \bibnamefont {K{\"a}lin}}, \bibinfo {author} {\bibfnamefont {Zhengwen}\ \bibnamefont {Liu}}, \ and\ \bibinfo {author} {\bibfnamefont {Rafael~A.}\ \bibnamefont {Porto}},\ }\bibfield  {title} {\enquote {\bibinfo {title} {{Local-in-Time Conservative Binary Dynamics at Fifth Post-Minkowskian and First Self-Force Orders}},}\ }\href@noop {} {\  (\bibinfo {year} {2025})},\ \Eprint {http://arxiv.org/abs/2506.20665} {arXiv:2506.20665 [hep-th]} \BibitemShut {NoStop}%
\bibitem [{\citenamefont {K{\"a}lin}\ and\ \citenamefont {Porto}(2020{\natexlab{b}})}]{Kalin:2019rwq}%
  \BibitemOpen
  \bibfield  {author} {\bibinfo {author} {\bibfnamefont {Gregor}\ \bibnamefont {K{\"a}lin}}\ and\ \bibinfo {author} {\bibfnamefont {Rafael~A.}\ \bibnamefont {Porto}},\ }\bibfield  {title} {\enquote {\bibinfo {title} {{From Boundary Data to Bound States}},}\ }\href {\doibase 10.1007/JHEP01(2020)072} {\bibfield  {journal} {\bibinfo  {journal} {JHEP}\ }\textbf {\bibinfo {volume} {01}},\ \bibinfo {pages} {072} (\bibinfo {year} {2020}{\natexlab{b}})},\ \Eprint {http://arxiv.org/abs/1910.03008} {arXiv:1910.03008 [hep-th]} \BibitemShut {NoStop}%
\bibitem [{\citenamefont {Dlapa}\ \emph {et~al.}(2024)\citenamefont {Dlapa}, \citenamefont {K{\"a}lin}, \citenamefont {Liu},\ and\ \citenamefont {Porto}}]{Dlapa:2024cje}%
  \BibitemOpen
  \bibfield  {author} {\bibinfo {author} {\bibfnamefont {Christoph}\ \bibnamefont {Dlapa}}, \bibinfo {author} {\bibfnamefont {Gregor}\ \bibnamefont {K{\"a}lin}}, \bibinfo {author} {\bibfnamefont {Zhengwen}\ \bibnamefont {Liu}}, \ and\ \bibinfo {author} {\bibfnamefont {Rafael~A.}\ \bibnamefont {Porto}},\ }\bibfield  {title} {\enquote {\bibinfo {title} {{Local in Time Conservative Binary Dynamics at Fourth Post-Minkowskian Order}},}\ }\href {\doibase 10.1103/PhysRevLett.132.221401} {\bibfield  {journal} {\bibinfo  {journal} {Phys. Rev. Lett.}\ }\textbf {\bibinfo {volume} {132}},\ \bibinfo {pages} {221401} (\bibinfo {year} {2024})},\ \Eprint {http://arxiv.org/abs/2403.04853} {arXiv:2403.04853 [hep-th]} \BibitemShut {NoStop}%
\bibitem [{\citenamefont {Bern}\ \emph {et~al.}(2025)\citenamefont {Bern}, \citenamefont {Herrmann}, \citenamefont {Roiban}, \citenamefont {Ruf}, \citenamefont {Smirnov}, \citenamefont {Smith},\ and\ \citenamefont {Zeng}}]{Bern:2025wyd}%
  \BibitemOpen
  \bibfield  {author} {\bibinfo {author} {\bibfnamefont {Zvi}\ \bibnamefont {Bern}}, \bibinfo {author} {\bibfnamefont {Enrico}\ \bibnamefont {Herrmann}}, \bibinfo {author} {\bibfnamefont {Radu}\ \bibnamefont {Roiban}}, \bibinfo {author} {\bibfnamefont {Michael~S.}\ \bibnamefont {Ruf}}, \bibinfo {author} {\bibfnamefont {Alexander~V.}\ \bibnamefont {Smirnov}}, \bibinfo {author} {\bibfnamefont {Sid}\ \bibnamefont {Smith}}, \ and\ \bibinfo {author} {\bibfnamefont {Mao}\ \bibnamefont {Zeng}},\ }\bibfield  {title} {\enquote {\bibinfo {title} {{Scattering Amplitudes and Conservative Binary Dynamics at $O(G^5)$ without Self-Force Truncation}},}\ }\href@noop {} {\  (\bibinfo {year} {2025})},\ \Eprint {http://arxiv.org/abs/2512.23654} {arXiv:2512.23654 [hep-th]} \BibitemShut {NoStop}%
\bibitem [{\citenamefont {Mino}\ \emph {et~al.}(1997)\citenamefont {Mino}, \citenamefont {Sasaki},\ and\ \citenamefont {Tanaka}}]{Mino:1996nk}%
  \BibitemOpen
  \bibfield  {author} {\bibinfo {author} {\bibfnamefont {Yasushi}\ \bibnamefont {Mino}}, \bibinfo {author} {\bibfnamefont {Misao}\ \bibnamefont {Sasaki}}, \ and\ \bibinfo {author} {\bibfnamefont {Takahiro}\ \bibnamefont {Tanaka}},\ }\bibfield  {title} {\enquote {\bibinfo {title} {{Gravitational radiation reaction to a particle motion}},}\ }\href {\doibase 10.1103/PhysRevD.55.3457} {\bibfield  {journal} {\bibinfo  {journal} {Phys. Rev. D}\ }\textbf {\bibinfo {volume} {55}},\ \bibinfo {pages} {3457--3476} (\bibinfo {year} {1997})},\ \Eprint {http://arxiv.org/abs/gr-qc/9606018} {arXiv:gr-qc/9606018} \BibitemShut {NoStop}%
\bibitem [{\citenamefont {Quinn}\ and\ \citenamefont {Wald}(1997)}]{Quinn:1996am}%
  \BibitemOpen
  \bibfield  {author} {\bibinfo {author} {\bibfnamefont {Theodore~C.}\ \bibnamefont {Quinn}}\ and\ \bibinfo {author} {\bibfnamefont {Robert~M.}\ \bibnamefont {Wald}},\ }\bibfield  {title} {\enquote {\bibinfo {title} {{An Axiomatic approach to electromagnetic and gravitational radiation reaction of particles in curved space-time}},}\ }\href {\doibase 10.1103/PhysRevD.56.3381} {\bibfield  {journal} {\bibinfo  {journal} {Phys. Rev. D}\ }\textbf {\bibinfo {volume} {56}},\ \bibinfo {pages} {3381--3394} (\bibinfo {year} {1997})},\ \Eprint {http://arxiv.org/abs/gr-qc/9610053} {arXiv:gr-qc/9610053} \BibitemShut {NoStop}%
\bibitem [{\citenamefont {Barack}\ \emph {et~al.}(2002)\citenamefont {Barack}, \citenamefont {Mino}, \citenamefont {Nakano}, \citenamefont {Ori},\ and\ \citenamefont {Sasaki}}]{Barack:2001gx}%
  \BibitemOpen
  \bibfield  {author} {\bibinfo {author} {\bibfnamefont {Leor}\ \bibnamefont {Barack}}, \bibinfo {author} {\bibfnamefont {Yasushi}\ \bibnamefont {Mino}}, \bibinfo {author} {\bibfnamefont {Hiroyuki}\ \bibnamefont {Nakano}}, \bibinfo {author} {\bibfnamefont {Amos}\ \bibnamefont {Ori}}, \ and\ \bibinfo {author} {\bibfnamefont {Misao}\ \bibnamefont {Sasaki}},\ }\bibfield  {title} {\enquote {\bibinfo {title} {{Calculating the gravitational selfforce in Schwarzschild space-time}},}\ }\href {\doibase 10.1103/PhysRevLett.88.091101} {\bibfield  {journal} {\bibinfo  {journal} {Phys. Rev. Lett.}\ }\textbf {\bibinfo {volume} {88}},\ \bibinfo {pages} {091101} (\bibinfo {year} {2002})},\ \Eprint {http://arxiv.org/abs/gr-qc/0111001} {arXiv:gr-qc/0111001} \BibitemShut {NoStop}%
\bibitem [{\citenamefont {Barack}\ and\ \citenamefont {Ori}(2003)}]{Barack:2002mh}%
  \BibitemOpen
  \bibfield  {author} {\bibinfo {author} {\bibfnamefont {Leor}\ \bibnamefont {Barack}}\ and\ \bibinfo {author} {\bibfnamefont {Amos}\ \bibnamefont {Ori}},\ }\bibfield  {title} {\enquote {\bibinfo {title} {{Gravitational selfforce on a particle orbiting a Kerr black hole}},}\ }\href {\doibase 10.1103/PhysRevLett.90.111101} {\bibfield  {journal} {\bibinfo  {journal} {Phys. Rev. Lett.}\ }\textbf {\bibinfo {volume} {90}},\ \bibinfo {pages} {111101} (\bibinfo {year} {2003})},\ \Eprint {http://arxiv.org/abs/gr-qc/0212103} {arXiv:gr-qc/0212103} \BibitemShut {NoStop}%
\bibitem [{\citenamefont {Gralla}\ and\ \citenamefont {Wald}(2008)}]{Gralla:2008fg}%
  \BibitemOpen
  \bibfield  {author} {\bibinfo {author} {\bibfnamefont {Samuel~E.}\ \bibnamefont {Gralla}}\ and\ \bibinfo {author} {\bibfnamefont {Robert~M.}\ \bibnamefont {Wald}},\ }\bibfield  {title} {\enquote {\bibinfo {title} {{A Rigorous Derivation of Gravitational Self-force}},}\ }\href {\doibase 10.1088/0264-9381/25/20/205009} {\bibfield  {journal} {\bibinfo  {journal} {Class. Quant. Grav.}\ }\textbf {\bibinfo {volume} {25}},\ \bibinfo {pages} {205009} (\bibinfo {year} {2008})},\ \bibinfo {note} {[Erratum: Class.Quant.Grav. 28, 159501 (2011)]},\ \Eprint {http://arxiv.org/abs/0806.3293} {arXiv:0806.3293 [gr-qc]} \BibitemShut {NoStop}%
\bibitem [{\citenamefont {Detweiler}(2008)}]{Detweiler:2008ft}%
  \BibitemOpen
  \bibfield  {author} {\bibinfo {author} {\bibfnamefont {Steven~L.}\ \bibnamefont {Detweiler}},\ }\bibfield  {title} {\enquote {\bibinfo {title} {{A Consequence of the gravitational self-force for circular orbits of the Schwarzschild geometry}},}\ }\href {\doibase 10.1103/PhysRevD.77.124026} {\bibfield  {journal} {\bibinfo  {journal} {Phys. Rev. D}\ }\textbf {\bibinfo {volume} {77}},\ \bibinfo {pages} {124026} (\bibinfo {year} {2008})},\ \Eprint {http://arxiv.org/abs/0804.3529} {arXiv:0804.3529 [gr-qc]} \BibitemShut {NoStop}%
\bibitem [{\citenamefont {Keidl}\ \emph {et~al.}(2010)\citenamefont {Keidl}, \citenamefont {Shah}, \citenamefont {Friedman}, \citenamefont {Kim},\ and\ \citenamefont {Price}}]{Keidl:2010pm}%
  \BibitemOpen
  \bibfield  {author} {\bibinfo {author} {\bibfnamefont {Tobias~S.}\ \bibnamefont {Keidl}}, \bibinfo {author} {\bibfnamefont {Abhay~G.}\ \bibnamefont {Shah}}, \bibinfo {author} {\bibfnamefont {John~L.}\ \bibnamefont {Friedman}}, \bibinfo {author} {\bibfnamefont {Dong-Hoon}\ \bibnamefont {Kim}}, \ and\ \bibinfo {author} {\bibfnamefont {Larry~R.}\ \bibnamefont {Price}},\ }\bibfield  {title} {\enquote {\bibinfo {title} {{Gravitational Self-force in a Radiation Gauge}},}\ }\href {\doibase 10.1103/PhysRevD.82.124012} {\bibfield  {journal} {\bibinfo  {journal} {Phys. Rev. D}\ }\textbf {\bibinfo {volume} {82}},\ \bibinfo {pages} {124012} (\bibinfo {year} {2010})},\ \bibinfo {note} {[Erratum: Phys.Rev.D 90, 109902 (2014)]},\ \Eprint {http://arxiv.org/abs/1004.2276} {arXiv:1004.2276 [gr-qc]} \BibitemShut {NoStop}%
\bibitem [{\citenamefont {van~de Meent}(2018)}]{vandeMeent:2017bcc}%
  \BibitemOpen
  \bibfield  {author} {\bibinfo {author} {\bibfnamefont {Maarten}\ \bibnamefont {van~de Meent}},\ }\bibfield  {title} {\enquote {\bibinfo {title} {{Gravitational self-force on generic bound geodesics in Kerr spacetime}},}\ }\href {\doibase 10.1103/PhysRevD.97.104033} {\bibfield  {journal} {\bibinfo  {journal} {Phys. Rev. D}\ }\textbf {\bibinfo {volume} {97}},\ \bibinfo {pages} {104033} (\bibinfo {year} {2018})},\ \Eprint {http://arxiv.org/abs/1711.09607} {arXiv:1711.09607 [gr-qc]} \BibitemShut {NoStop}%
\bibitem [{\citenamefont {Pound}(2012)}]{Pound:2012nt}%
  \BibitemOpen
  \bibfield  {author} {\bibinfo {author} {\bibfnamefont {Adam}\ \bibnamefont {Pound}},\ }\bibfield  {title} {\enquote {\bibinfo {title} {{Second-order gravitational self-force}},}\ }\href {\doibase 10.1103/PhysRevLett.109.051101} {\bibfield  {journal} {\bibinfo  {journal} {Phys. Rev. Lett.}\ }\textbf {\bibinfo {volume} {109}},\ \bibinfo {pages} {051101} (\bibinfo {year} {2012})},\ \Eprint {http://arxiv.org/abs/1201.5089} {arXiv:1201.5089 [gr-qc]} \BibitemShut {NoStop}%
\bibitem [{\citenamefont {Pound}\ \emph {et~al.}(2020)\citenamefont {Pound}, \citenamefont {Wardell}, \citenamefont {Warburton},\ and\ \citenamefont {Miller}}]{Pound:2019lzj}%
  \BibitemOpen
  \bibfield  {author} {\bibinfo {author} {\bibfnamefont {Adam}\ \bibnamefont {Pound}}, \bibinfo {author} {\bibfnamefont {Barry}\ \bibnamefont {Wardell}}, \bibinfo {author} {\bibfnamefont {Niels}\ \bibnamefont {Warburton}}, \ and\ \bibinfo {author} {\bibfnamefont {Jeremy}\ \bibnamefont {Miller}},\ }\bibfield  {title} {\enquote {\bibinfo {title} {{Second-Order Self-Force Calculation of Gravitational Binding Energy in Compact Binaries}},}\ }\href {\doibase 10.1103/PhysRevLett.124.021101} {\bibfield  {journal} {\bibinfo  {journal} {Phys. Rev. Lett.}\ }\textbf {\bibinfo {volume} {124}},\ \bibinfo {pages} {021101} (\bibinfo {year} {2020})},\ \Eprint {http://arxiv.org/abs/1908.07419} {arXiv:1908.07419 [gr-qc]} \BibitemShut {NoStop}%
\bibitem [{\citenamefont {Gralla}\ and\ \citenamefont {Lobo}(2022)}]{Gralla:2021qaf}%
  \BibitemOpen
  \bibfield  {author} {\bibinfo {author} {\bibfnamefont {Samuel~E.}\ \bibnamefont {Gralla}}\ and\ \bibinfo {author} {\bibfnamefont {Kunal}\ \bibnamefont {Lobo}},\ }\bibfield  {title} {\enquote {\bibinfo {title} {{Self-force effects in post-Minkowskian scattering}},}\ }\href {\doibase 10.1088/1361-6382/ac5d88} {\bibfield  {journal} {\bibinfo  {journal} {Class. Quant. Grav.}\ }\textbf {\bibinfo {volume} {39}},\ \bibinfo {pages} {095001} (\bibinfo {year} {2022})},\ \bibinfo {note} {[Erratum: Class.Quant.Grav. 41, 179501 (2024)]},\ \Eprint {http://arxiv.org/abs/2110.08681} {arXiv:2110.08681 [gr-qc]} \BibitemShut {NoStop}%
\bibitem [{\citenamefont {Pound}\ and\ \citenamefont {Wardell}(2021)}]{Pound:2021qin}%
  \BibitemOpen
  \bibfield  {author} {\bibinfo {author} {\bibfnamefont {Adam}\ \bibnamefont {Pound}}\ and\ \bibinfo {author} {\bibfnamefont {Barry}\ \bibnamefont {Wardell}},\ }\href {\doibase 10.1007/978-981-15-4702-7_38-1} {\enquote {\bibinfo {title} {{Black hole perturbation theory and gravitational self-force}},}\ } (\bibinfo {year} {2021}),\ \Eprint {http://arxiv.org/abs/2101.04592} {arXiv:2101.04592 [gr-qc]} \BibitemShut {NoStop}%
\bibitem [{\citenamefont {Warburton}\ \emph {et~al.}(2021)\citenamefont {Warburton}, \citenamefont {Pound}, \citenamefont {Wardell}, \citenamefont {Miller},\ and\ \citenamefont {Durkan}}]{Warburton:2021kwk}%
  \BibitemOpen
  \bibfield  {author} {\bibinfo {author} {\bibfnamefont {Niels}\ \bibnamefont {Warburton}}, \bibinfo {author} {\bibfnamefont {Adam}\ \bibnamefont {Pound}}, \bibinfo {author} {\bibfnamefont {Barry}\ \bibnamefont {Wardell}}, \bibinfo {author} {\bibfnamefont {Jeremy}\ \bibnamefont {Miller}}, \ and\ \bibinfo {author} {\bibfnamefont {Leanne}\ \bibnamefont {Durkan}},\ }\bibfield  {title} {\enquote {\bibinfo {title} {{Gravitational-Wave Energy Flux for Compact Binaries through Second Order in the Mass Ratio}},}\ }\href {\doibase 10.1103/PhysRevLett.127.151102} {\bibfield  {journal} {\bibinfo  {journal} {Phys. Rev. Lett.}\ }\textbf {\bibinfo {volume} {127}},\ \bibinfo {pages} {151102} (\bibinfo {year} {2021})},\ \Eprint {http://arxiv.org/abs/2107.01298} {arXiv:2107.01298 [gr-qc]} \BibitemShut {NoStop}%
\bibitem [{\citenamefont {Wardell}\ \emph {et~al.}(2023)\citenamefont {Wardell}, \citenamefont {Pound}, \citenamefont {Warburton}, \citenamefont {Miller}, \citenamefont {Durkan},\ and\ \citenamefont {Le~Tiec}}]{Wardell:2021fyy}%
  \BibitemOpen
  \bibfield  {author} {\bibinfo {author} {\bibfnamefont {Barry}\ \bibnamefont {Wardell}}, \bibinfo {author} {\bibfnamefont {Adam}\ \bibnamefont {Pound}}, \bibinfo {author} {\bibfnamefont {Niels}\ \bibnamefont {Warburton}}, \bibinfo {author} {\bibfnamefont {Jeremy}\ \bibnamefont {Miller}}, \bibinfo {author} {\bibfnamefont {Leanne}\ \bibnamefont {Durkan}}, \ and\ \bibinfo {author} {\bibfnamefont {Alexandre}\ \bibnamefont {Le~Tiec}},\ }\bibfield  {title} {\enquote {\bibinfo {title} {{Gravitational Waveforms for Compact Binaries from Second-Order Self-Force Theory}},}\ }\href {\doibase 10.1103/PhysRevLett.130.241402} {\bibfield  {journal} {\bibinfo  {journal} {Phys. Rev. Lett.}\ }\textbf {\bibinfo {volume} {130}},\ \bibinfo {pages} {241402} (\bibinfo {year} {2023})},\ \Eprint {http://arxiv.org/abs/2112.12265} {arXiv:2112.12265 [gr-qc]} \BibitemShut {NoStop}%
\bibitem [{\citenamefont {Ajith}\ \emph {et~al.}(2007)\citenamefont {Ajith} \emph {et~al.}}]{Ajith:2007qp}%
  \BibitemOpen
  \bibfield  {author} {\bibinfo {author} {\bibfnamefont {Parameswaran}\ \bibnamefont {Ajith}} \emph {et~al.},\ }\bibfield  {title} {\enquote {\bibinfo {title} {{Phenomenological template family for black-hole coalescence waveforms}},}\ }\href {\doibase 10.1088/0264-9381/24/19/S31} {\bibfield  {journal} {\bibinfo  {journal} {Class. Quant. Grav.}\ }\textbf {\bibinfo {volume} {24}},\ \bibinfo {pages} {S689--S700} (\bibinfo {year} {2007})},\ \Eprint {http://arxiv.org/abs/0704.3764} {arXiv:0704.3764 [gr-qc]} \BibitemShut {NoStop}%
\bibitem [{\citenamefont {Pratten}\ \emph {et~al.}(2021)\citenamefont {Pratten} \emph {et~al.}}]{Pratten:2020ceb}%
  \BibitemOpen
  \bibfield  {author} {\bibinfo {author} {\bibfnamefont {Geraint}\ \bibnamefont {Pratten}} \emph {et~al.},\ }\bibfield  {title} {\enquote {\bibinfo {title} {{Computationally efficient models for the dominant and subdominant harmonic modes of precessing binary black holes}},}\ }\href {\doibase 10.1103/PhysRevD.103.104056} {\bibfield  {journal} {\bibinfo  {journal} {Phys. Rev. D}\ }\textbf {\bibinfo {volume} {103}},\ \bibinfo {pages} {104056} (\bibinfo {year} {2021})},\ \Eprint {http://arxiv.org/abs/2004.06503} {arXiv:2004.06503 [gr-qc]} \BibitemShut {NoStop}%
\bibitem [{\citenamefont {Estell{\'e}s}\ \emph {et~al.}(2022)\citenamefont {Estell{\'e}s}, \citenamefont {Colleoni}, \citenamefont {Garc{\'\i}a-Quir{\'o}s}, \citenamefont {Husa}, \citenamefont {Keitel}, \citenamefont {Mateu-Lucena}, \citenamefont {Planas},\ and\ \citenamefont {Ramos-Buades}}]{Estelles:2021gvs}%
  \BibitemOpen
  \bibfield  {author} {\bibinfo {author} {\bibfnamefont {H{\'e}ctor}\ \bibnamefont {Estell{\'e}s}}, \bibinfo {author} {\bibfnamefont {Marta}\ \bibnamefont {Colleoni}}, \bibinfo {author} {\bibfnamefont {Cecilio}\ \bibnamefont {Garc{\'\i}a-Quir{\'o}s}}, \bibinfo {author} {\bibfnamefont {Sascha}\ \bibnamefont {Husa}}, \bibinfo {author} {\bibfnamefont {David}\ \bibnamefont {Keitel}}, \bibinfo {author} {\bibfnamefont {Maite}\ \bibnamefont {Mateu-Lucena}}, \bibinfo {author} {\bibfnamefont {Maria de~Lluc}\ \bibnamefont {Planas}}, \ and\ \bibinfo {author} {\bibfnamefont {Antoni}\ \bibnamefont {Ramos-Buades}},\ }\bibfield  {title} {\enquote {\bibinfo {title} {{New twists in compact binary waveform modeling: A fast time-domain model for precession}},}\ }\href {\doibase 10.1103/PhysRevD.105.084040} {\bibfield  {journal} {\bibinfo  {journal} {Phys. Rev. D}\ }\textbf {\bibinfo {volume} {105}},\ \bibinfo {pages} {084040} (\bibinfo {year} {2022})},\ \Eprint {http://arxiv.org/abs/2105.05872} {arXiv:2105.05872 [gr-qc]} \BibitemShut {NoStop}%
\bibitem [{\citenamefont {Buonanno}\ and\ \citenamefont {Damour}(1999)}]{Buonanno:1998gg}%
  \BibitemOpen
  \bibfield  {author} {\bibinfo {author} {\bibfnamefont {A.}~\bibnamefont {Buonanno}}\ and\ \bibinfo {author} {\bibfnamefont {T.}~\bibnamefont {Damour}},\ }\bibfield  {title} {\enquote {\bibinfo {title} {{Effective one-body approach to general relativistic two-body dynamics}},}\ }\href {\doibase 10.1103/PhysRevD.59.084006} {\bibfield  {journal} {\bibinfo  {journal} {Phys. Rev. D}\ }\textbf {\bibinfo {volume} {59}},\ \bibinfo {pages} {084006} (\bibinfo {year} {1999})},\ \Eprint {http://arxiv.org/abs/gr-qc/9811091} {arXiv:gr-qc/9811091} \BibitemShut {NoStop}%
\bibitem [{\citenamefont {Buonanno}\ and\ \citenamefont {Damour}(2000)}]{Buonanno:2000ef}%
  \BibitemOpen
  \bibfield  {author} {\bibinfo {author} {\bibfnamefont {Alessandra}\ \bibnamefont {Buonanno}}\ and\ \bibinfo {author} {\bibfnamefont {Thibault}\ \bibnamefont {Damour}},\ }\bibfield  {title} {\enquote {\bibinfo {title} {{Transition from inspiral to plunge in binary black hole coalescences}},}\ }\href {\doibase 10.1103/PhysRevD.62.064015} {\bibfield  {journal} {\bibinfo  {journal} {Phys. Rev. D}\ }\textbf {\bibinfo {volume} {62}},\ \bibinfo {pages} {064015} (\bibinfo {year} {2000})},\ \Eprint {http://arxiv.org/abs/gr-qc/0001013} {arXiv:gr-qc/0001013} \BibitemShut {NoStop}%
\bibitem [{\citenamefont {Damour}\ \emph {et~al.}(2000)\citenamefont {Damour}, \citenamefont {Jaranowski},\ and\ \citenamefont {Schaefer}}]{Damour:2000we}%
  \BibitemOpen
  \bibfield  {author} {\bibinfo {author} {\bibfnamefont {Thibault}\ \bibnamefont {Damour}}, \bibinfo {author} {\bibfnamefont {Piotr}\ \bibnamefont {Jaranowski}}, \ and\ \bibinfo {author} {\bibfnamefont {Gerhard}\ \bibnamefont {Schaefer}},\ }\bibfield  {title} {\enquote {\bibinfo {title} {{On the determination of the last stable orbit for circular general relativistic binaries at the third postNewtonian approximation}},}\ }\href {\doibase 10.1103/PhysRevD.62.084011} {\bibfield  {journal} {\bibinfo  {journal} {Phys. Rev. D}\ }\textbf {\bibinfo {volume} {62}},\ \bibinfo {pages} {084011} (\bibinfo {year} {2000})},\ \Eprint {http://arxiv.org/abs/gr-qc/0005034} {arXiv:gr-qc/0005034} \BibitemShut {NoStop}%
\bibitem [{\citenamefont {Damour}(2001)}]{Damour:2001tu}%
  \BibitemOpen
  \bibfield  {author} {\bibinfo {author} {\bibfnamefont {Thibault}\ \bibnamefont {Damour}},\ }\bibfield  {title} {\enquote {\bibinfo {title} {{Coalescence of two spinning black holes: an effective one-body approach}},}\ }\href {\doibase 10.1103/PhysRevD.64.124013} {\bibfield  {journal} {\bibinfo  {journal} {Phys. Rev. D}\ }\textbf {\bibinfo {volume} {64}},\ \bibinfo {pages} {124013} (\bibinfo {year} {2001})},\ \Eprint {http://arxiv.org/abs/gr-qc/0103018} {arXiv:gr-qc/0103018} \BibitemShut {NoStop}%
\bibitem [{\citenamefont {Buonanno}\ \emph {et~al.}(2006)\citenamefont {Buonanno}, \citenamefont {Chen},\ and\ \citenamefont {Damour}}]{Buonanno:2005xu}%
  \BibitemOpen
  \bibfield  {author} {\bibinfo {author} {\bibfnamefont {Alessandra}\ \bibnamefont {Buonanno}}, \bibinfo {author} {\bibfnamefont {Yanbei}\ \bibnamefont {Chen}}, \ and\ \bibinfo {author} {\bibfnamefont {Thibault}\ \bibnamefont {Damour}},\ }\bibfield  {title} {\enquote {\bibinfo {title} {{Transition from inspiral to plunge in precessing binaries of spinning black holes}},}\ }\href {\doibase 10.1103/PhysRevD.74.104005} {\bibfield  {journal} {\bibinfo  {journal} {Phys. Rev. D}\ }\textbf {\bibinfo {volume} {74}},\ \bibinfo {pages} {104005} (\bibinfo {year} {2006})},\ \Eprint {http://arxiv.org/abs/gr-qc/0508067} {arXiv:gr-qc/0508067} \BibitemShut {NoStop}%
\bibitem [{\citenamefont {Damour}\ \emph {et~al.}(2009)\citenamefont {Damour}, \citenamefont {Iyer},\ and\ \citenamefont {Nagar}}]{Damour:2008gu}%
  \BibitemOpen
  \bibfield  {author} {\bibinfo {author} {\bibfnamefont {Thibault}\ \bibnamefont {Damour}}, \bibinfo {author} {\bibfnamefont {Bala~R.}\ \bibnamefont {Iyer}}, \ and\ \bibinfo {author} {\bibfnamefont {Alessandro}\ \bibnamefont {Nagar}},\ }\bibfield  {title} {\enquote {\bibinfo {title} {{Improved resummation of post-Newtonian multipolar waveforms from circularized compact binaries}},}\ }\href {\doibase 10.1103/PhysRevD.79.064004} {\bibfield  {journal} {\bibinfo  {journal} {Phys. Rev. D}\ }\textbf {\bibinfo {volume} {79}},\ \bibinfo {pages} {064004} (\bibinfo {year} {2009})},\ \Eprint {http://arxiv.org/abs/0811.2069} {arXiv:0811.2069 [gr-qc]} \BibitemShut {NoStop}%
\bibitem [{\citenamefont {P{\"u}rrer}\ and\ \citenamefont {Haster}(2020)}]{Purrer:2019jcp}%
  \BibitemOpen
  \bibfield  {author} {\bibinfo {author} {\bibfnamefont {Michael}\ \bibnamefont {P{\"u}rrer}}\ and\ \bibinfo {author} {\bibfnamefont {Carl-Johan}\ \bibnamefont {Haster}},\ }\bibfield  {title} {\enquote {\bibinfo {title} {{Gravitational waveform accuracy requirements for future ground-based detectors}},}\ }\href {\doibase 10.1103/PhysRevResearch.2.023151} {\bibfield  {journal} {\bibinfo  {journal} {Phys. Rev. Res.}\ }\textbf {\bibinfo {volume} {2}},\ \bibinfo {pages} {023151} (\bibinfo {year} {2020})},\ \Eprint {http://arxiv.org/abs/1912.10055} {arXiv:1912.10055 [gr-qc]} \BibitemShut {NoStop}%
\bibitem [{\citenamefont {Hu}\ and\ \citenamefont {Veitch}(2022)}]{Hu:2022rjq}%
  \BibitemOpen
  \bibfield  {author} {\bibinfo {author} {\bibfnamefont {Qian}\ \bibnamefont {Hu}}\ and\ \bibinfo {author} {\bibfnamefont {John}\ \bibnamefont {Veitch}},\ }\bibfield  {title} {\enquote {\bibinfo {title} {{Assessing the model waveform accuracy of gravitational waves}},}\ }\href {\doibase 10.1103/PhysRevD.106.044042} {\bibfield  {journal} {\bibinfo  {journal} {Phys. Rev. D}\ }\textbf {\bibinfo {volume} {106}},\ \bibinfo {pages} {044042} (\bibinfo {year} {2022})},\ \Eprint {http://arxiv.org/abs/2205.08448} {arXiv:2205.08448 [gr-qc]} \BibitemShut {NoStop}%
\bibitem [{\citenamefont {Mandal}\ \emph {et~al.}()\citenamefont {Mandal}, \citenamefont {Mastrolia}, \citenamefont {Patil},\ and\ \citenamefont {Steinhoff}}]{panther:inprep}%
  \BibitemOpen
  \bibfield  {author} {\bibinfo {author} {\bibfnamefont {Manoj~K.}\ \bibnamefont {Mandal}}, \bibinfo {author} {\bibfnamefont {Pierpaolo}\ \bibnamefont {Mastrolia}}, \bibinfo {author} {\bibfnamefont {Raj}\ \bibnamefont {Patil}}, \ and\ \bibinfo {author} {\bibfnamefont {Jan}\ \bibnamefont {Steinhoff}},\ }\bibfield  {title} {\enquote {\bibinfo {title} {{\texttt{PNTHR}: Post-Newtonian Toolkit for Hamiltonian and Radiation}},}\ }\href@noop {} {\ }\bibinfo {note} {(unpublished)}\BibitemShut {NoStop}%
\bibitem [{\citenamefont {Tkachov}(1981)}]{Tkachov:1981wb}%
  \BibitemOpen
  \bibfield  {author} {\bibinfo {author} {\bibfnamefont {F.~V.}\ \bibnamefont {Tkachov}},\ }\bibfield  {title} {\enquote {\bibinfo {title} {{A theorem on analytical calculability of 4-loop renormalization group functions}},}\ }\href {\doibase 10.1016/0370-2693(81)90288-4} {\bibfield  {journal} {\bibinfo  {journal} {Phys. Lett. B}\ }\textbf {\bibinfo {volume} {100}},\ \bibinfo {pages} {65--68} (\bibinfo {year} {1981})}\BibitemShut {NoStop}%
\bibitem [{\citenamefont {Chetyrkin}\ and\ \citenamefont {Tkachov}(1981)}]{Chetyrkin:1981qh}%
  \BibitemOpen
  \bibfield  {author} {\bibinfo {author} {\bibfnamefont {K.~G.}\ \bibnamefont {Chetyrkin}}\ and\ \bibinfo {author} {\bibfnamefont {F.~V.}\ \bibnamefont {Tkachov}},\ }\bibfield  {title} {\enquote {\bibinfo {title} {{Integration by parts: The algorithm to calculate $\beta$-functions in 4 loops}},}\ }\href {\doibase 10.1016/0550-3213(81)90199-1} {\bibfield  {journal} {\bibinfo  {journal} {Nucl. Phys. B}\ }\textbf {\bibinfo {volume} {192}},\ \bibinfo {pages} {159--204} (\bibinfo {year} {1981})}\BibitemShut {NoStop}%
\bibitem [{\citenamefont {Laporta}(2000)}]{Laporta:2000dsw}%
  \BibitemOpen
  \bibfield  {author} {\bibinfo {author} {\bibfnamefont {S.}~\bibnamefont {Laporta}},\ }\bibfield  {title} {\enquote {\bibinfo {title} {{High-precision calculation of multiloop Feynman integrals by difference equations}},}\ }\href {\doibase 10.1142/S0217751X00002159} {\bibfield  {journal} {\bibinfo  {journal} {Int. J. Mod. Phys. A}\ }\textbf {\bibinfo {volume} {15}},\ \bibinfo {pages} {5087--5159} (\bibinfo {year} {2000})},\ \Eprint {http://arxiv.org/abs/hep-ph/0102033} {arXiv:hep-ph/0102033} \BibitemShut {NoStop}%
\bibitem [{\citenamefont {Larsen}\ and\ \citenamefont {Zhang}(2016)}]{Larsen:2015ped}%
  \BibitemOpen
  \bibfield  {author} {\bibinfo {author} {\bibfnamefont {Kasper~J.}\ \bibnamefont {Larsen}}\ and\ \bibinfo {author} {\bibfnamefont {Yang}\ \bibnamefont {Zhang}},\ }\bibfield  {title} {\enquote {\bibinfo {title} {{Integration-by-parts reductions from unitarity cuts and algebraic geometry}},}\ }\href {\doibase 10.1103/PhysRevD.93.041701} {\bibfield  {journal} {\bibinfo  {journal} {Phys. Rev. D}\ }\textbf {\bibinfo {volume} {93}},\ \bibinfo {pages} {041701} (\bibinfo {year} {2016})},\ \Eprint {http://arxiv.org/abs/1511.01071} {arXiv:1511.01071 [hep-th]} \BibitemShut {NoStop}%
\bibitem [{\citenamefont {Gluza}\ \emph {et~al.}(2011)\citenamefont {Gluza}, \citenamefont {Kajda},\ and\ \citenamefont {Kosower}}]{Gluza:2010ws}%
  \BibitemOpen
  \bibfield  {author} {\bibinfo {author} {\bibfnamefont {Janusz}\ \bibnamefont {Gluza}}, \bibinfo {author} {\bibfnamefont {Krzysztof}\ \bibnamefont {Kajda}}, \ and\ \bibinfo {author} {\bibfnamefont {David~A.}\ \bibnamefont {Kosower}},\ }\bibfield  {title} {\enquote {\bibinfo {title} {{Towards a Basis for Planar Two-Loop Integrals}},}\ }\href {\doibase 10.1103/PhysRevD.83.045012} {\bibfield  {journal} {\bibinfo  {journal} {Phys. Rev. D}\ }\textbf {\bibinfo {volume} {83}},\ \bibinfo {pages} {045012} (\bibinfo {year} {2011})},\ \Eprint {http://arxiv.org/abs/1009.0472} {arXiv:1009.0472 [hep-th]} \BibitemShut {NoStop}%
\bibitem [{\citenamefont {Wu}\ \emph {et~al.}(2024)\citenamefont {Wu}, \citenamefont {Boehm}, \citenamefont {Ma}, \citenamefont {Xu},\ and\ \citenamefont {Zhang}}]{Wu:2023upw}%
  \BibitemOpen
  \bibfield  {author} {\bibinfo {author} {\bibfnamefont {Zihao}\ \bibnamefont {Wu}}, \bibinfo {author} {\bibfnamefont {Janko}\ \bibnamefont {Boehm}}, \bibinfo {author} {\bibfnamefont {Rourou}\ \bibnamefont {Ma}}, \bibinfo {author} {\bibfnamefont {Hefeng}\ \bibnamefont {Xu}}, \ and\ \bibinfo {author} {\bibfnamefont {Yang}\ \bibnamefont {Zhang}},\ }\bibfield  {title} {\enquote {\bibinfo {title} {{NeatIBP 1.0, a package generating small-size integration-by-parts relations for Feynman integrals}},}\ }\href {\doibase 10.1016/j.cpc.2023.108999} {\bibfield  {journal} {\bibinfo  {journal} {Comput. Phys. Commun.}\ }\textbf {\bibinfo {volume} {295}},\ \bibinfo {pages} {108999} (\bibinfo {year} {2024})},\ \Eprint {http://arxiv.org/abs/2305.08783} {arXiv:2305.08783 [hep-ph]} \BibitemShut {NoStop}%
\bibitem [{\citenamefont {Wu}\ \emph {et~al.}(2025)\citenamefont {Wu}, \citenamefont {B{\"o}hm}, \citenamefont {Ma}, \citenamefont {Usovitsch}, \citenamefont {Xu},\ and\ \citenamefont {Zhang}}]{Wu:2025aeg}%
  \BibitemOpen
  \bibfield  {author} {\bibinfo {author} {\bibfnamefont {Zihao}\ \bibnamefont {Wu}}, \bibinfo {author} {\bibfnamefont {Janko}\ \bibnamefont {B{\"o}hm}}, \bibinfo {author} {\bibfnamefont {Rourou}\ \bibnamefont {Ma}}, \bibinfo {author} {\bibfnamefont {Johann}\ \bibnamefont {Usovitsch}}, \bibinfo {author} {\bibfnamefont {Yingxuan}\ \bibnamefont {Xu}}, \ and\ \bibinfo {author} {\bibfnamefont {Yang}\ \bibnamefont {Zhang}},\ }\bibfield  {title} {\enquote {\bibinfo {title} {{Performing integration-by-parts reductions using NeatIBP 1.1 + Kira}},}\ }\href {\doibase 10.1016/j.cpc.2025.109798} {\bibfield  {journal} {\bibinfo  {journal} {Comput. Phys. Commun.}\ }\textbf {\bibinfo {volume} {316}},\ \bibinfo {pages} {109798} (\bibinfo {year} {2025})},\ \Eprint {http://arxiv.org/abs/2502.20778} {arXiv:2502.20778 [hep-ph]} \BibitemShut {NoStop}%
\bibitem [{\citenamefont {Smith}\ and\ \citenamefont {Zeng}(2025)}]{Smith:2025xes}%
  \BibitemOpen
  \bibfield  {author} {\bibinfo {author} {\bibfnamefont {Sid}\ \bibnamefont {Smith}}\ and\ \bibinfo {author} {\bibfnamefont {Mao}\ \bibnamefont {Zeng}},\ }\bibfield  {title} {\enquote {\bibinfo {title} {{Feynman Integral Reduction using Syzygy-Constrained Symbolic Reduction Rules}},}\ }\href@noop {} {\  (\bibinfo {year} {2025})},\ \Eprint {http://arxiv.org/abs/2507.11140} {arXiv:2507.11140 [hep-th]} \BibitemShut {NoStop}%
\bibitem [{\citenamefont {Lange}\ \emph {et~al.}(2025)\citenamefont {Lange}, \citenamefont {Usovitsch},\ and\ \citenamefont {Wu}}]{Lange:2025fba}%
  \BibitemOpen
  \bibfield  {author} {\bibinfo {author} {\bibfnamefont {Fabian}\ \bibnamefont {Lange}}, \bibinfo {author} {\bibfnamefont {Johann}\ \bibnamefont {Usovitsch}}, \ and\ \bibinfo {author} {\bibfnamefont {Zihao}\ \bibnamefont {Wu}},\ }\bibfield  {title} {\enquote {\bibinfo {title} {{Kira 3: integral reduction with efficient seeding and optimized equation selection}},}\ }\href@noop {} {\  (\bibinfo {year} {2025})},\ \Eprint {http://arxiv.org/abs/2505.20197} {arXiv:2505.20197 [hep-ph]} \BibitemShut {NoStop}%
\bibitem [{\citenamefont {von Hippel}\ and\ \citenamefont {Wilhelm}(2025)}]{vonHippel:2025okr}%
  \BibitemOpen
  \bibfield  {author} {\bibinfo {author} {\bibfnamefont {Matt}\ \bibnamefont {von Hippel}}\ and\ \bibinfo {author} {\bibfnamefont {Matthias}\ \bibnamefont {Wilhelm}},\ }\bibfield  {title} {\enquote {\bibinfo {title} {{Refining Integration-by-Parts Reduction of Feynman Integrals with Machine Learning}},}\ }\href {\doibase 10.1007/JHEP05(2025)185} {\bibfield  {journal} {\bibinfo  {journal} {JHEP}\ }\textbf {\bibinfo {volume} {05}},\ \bibinfo {pages} {185} (\bibinfo {year} {2025})},\ \Eprint {http://arxiv.org/abs/2502.05121} {arXiv:2502.05121 [hep-th]} \BibitemShut {NoStop}%
\bibitem [{\citenamefont {Song}\ \emph {et~al.}(2025)\citenamefont {Song}, \citenamefont {Yang}, \citenamefont {Cao}, \citenamefont {Luo},\ and\ \citenamefont {Zhu}}]{Song:2025pwy}%
  \BibitemOpen
  \bibfield  {author} {\bibinfo {author} {\bibfnamefont {Zhuo-Yang}\ \bibnamefont {Song}}, \bibinfo {author} {\bibfnamefont {Tong-Zhi}\ \bibnamefont {Yang}}, \bibinfo {author} {\bibfnamefont {Qing-Hong}\ \bibnamefont {Cao}}, \bibinfo {author} {\bibfnamefont {Ming-xing}\ \bibnamefont {Luo}}, \ and\ \bibinfo {author} {\bibfnamefont {Hua~Xing}\ \bibnamefont {Zhu}},\ }\bibfield  {title} {\enquote {\bibinfo {title} {{Explainable AI-assisted Optimization for Feynman Integral Reduction}},}\ }\href@noop {} {\  (\bibinfo {year} {2025})},\ \Eprint {http://arxiv.org/abs/2502.09544} {arXiv:2502.09544 [hep-ph]} \BibitemShut {NoStop}%
\bibitem [{\citenamefont {Zeng}(2025)}]{Zeng:2025xbh}%
  \BibitemOpen
  \bibfield  {author} {\bibinfo {author} {\bibfnamefont {Mao}\ \bibnamefont {Zeng}},\ }\bibfield  {title} {\enquote {\bibinfo {title} {{Reinforcement Learning and Metaheuristics for Feynman Integral Reduction}},}\ }\href@noop {} {\  (\bibinfo {year} {2025})},\ \Eprint {http://arxiv.org/abs/2504.16045} {arXiv:2504.16045 [hep-ph]} \BibitemShut {NoStop}%
\bibitem [{\citenamefont {Liu}\ \emph {et~al.}(2018)\citenamefont {Liu}, \citenamefont {Ma},\ and\ \citenamefont {Wang}}]{Liu:2017jxz}%
  \BibitemOpen
  \bibfield  {author} {\bibinfo {author} {\bibfnamefont {Xiao}\ \bibnamefont {Liu}}, \bibinfo {author} {\bibfnamefont {Yan-Qing}\ \bibnamefont {Ma}}, \ and\ \bibinfo {author} {\bibfnamefont {Chen-Yu}\ \bibnamefont {Wang}},\ }\bibfield  {title} {\enquote {\bibinfo {title} {{A Systematic and Efficient Method to Compute Multi-loop Master Integrals}},}\ }\href {\doibase 10.1016/j.physletb.2018.02.026} {\bibfield  {journal} {\bibinfo  {journal} {Phys. Lett. B}\ }\textbf {\bibinfo {volume} {779}},\ \bibinfo {pages} {353--357} (\bibinfo {year} {2018})},\ \Eprint {http://arxiv.org/abs/1711.09572} {arXiv:1711.09572 [hep-ph]} \BibitemShut {NoStop}%
\bibitem [{\citenamefont {Liu}\ and\ \citenamefont {Ma}(2023)}]{Liu:2022chg}%
  \BibitemOpen
  \bibfield  {author} {\bibinfo {author} {\bibfnamefont {Xiao}\ \bibnamefont {Liu}}\ and\ \bibinfo {author} {\bibfnamefont {Yan-Qing}\ \bibnamefont {Ma}},\ }\bibfield  {title} {\enquote {\bibinfo {title} {{AMFlow: A Mathematica package for Feynman integrals computation via auxiliary mass flow}},}\ }\href {\doibase 10.1016/j.cpc.2022.108565} {\bibfield  {journal} {\bibinfo  {journal} {Comput. Phys. Commun.}\ }\textbf {\bibinfo {volume} {283}},\ \bibinfo {pages} {108565} (\bibinfo {year} {2023})},\ \Eprint {http://arxiv.org/abs/2201.11669} {arXiv:2201.11669 [hep-ph]} \BibitemShut {NoStop}%
\bibitem [{\citenamefont {Ferguson}\ \emph {et~al.}(1999)\citenamefont {Ferguson}, \citenamefont {Bailey},\ and\ \citenamefont {Arno}}]{PSLQ}%
  \BibitemOpen
  \bibfield  {author} {\bibinfo {author} {\bibfnamefont {Helaman~R.P.}\ \bibnamefont {Ferguson}}, \bibinfo {author} {\bibfnamefont {David~H.}\ \bibnamefont {Bailey}}, \ and\ \bibinfo {author} {\bibfnamefont {Steve}\ \bibnamefont {Arno}},\ }\bibfield  {title} {\enquote {\bibinfo {title} {Analysis of pslq, an integer relation finding algorithm},}\ }\href {\doibase 10.1090/S0025-5718-99-00995-3} {\bibfield  {journal} {\bibinfo  {journal} {Mathematics of Computation}\ }\textbf {\bibinfo {volume} {68}},\ \bibinfo {pages} {351--369} (\bibinfo {year} {1999})}\BibitemShut {NoStop}%
\bibitem [{\citenamefont {Kol}\ and\ \citenamefont {Smolkin}(2008{\natexlab{a}})}]{Kol:2007bc}%
  \BibitemOpen
  \bibfield  {author} {\bibinfo {author} {\bibfnamefont {Barak}\ \bibnamefont {Kol}}\ and\ \bibinfo {author} {\bibfnamefont {Michael}\ \bibnamefont {Smolkin}},\ }\bibfield  {title} {\enquote {\bibinfo {title} {{Non-Relativistic Gravitation: From Newton to Einstein and Back}},}\ }\href {\doibase 10.1088/0264-9381/25/14/145011} {\bibfield  {journal} {\bibinfo  {journal} {Class. Quant. Grav.}\ }\textbf {\bibinfo {volume} {25}},\ \bibinfo {pages} {145011} (\bibinfo {year} {2008}{\natexlab{a}})},\ \Eprint {http://arxiv.org/abs/0712.4116} {arXiv:0712.4116 [hep-th]} \BibitemShut {NoStop}%
\bibitem [{\citenamefont {Kol}\ and\ \citenamefont {Smolkin}(2008{\natexlab{b}})}]{Kol:2007rx}%
  \BibitemOpen
  \bibfield  {author} {\bibinfo {author} {\bibfnamefont {Barak}\ \bibnamefont {Kol}}\ and\ \bibinfo {author} {\bibfnamefont {Michael}\ \bibnamefont {Smolkin}},\ }\bibfield  {title} {\enquote {\bibinfo {title} {{Classical Effective Field Theory and Caged Black Holes}},}\ }\href {\doibase 10.1103/PhysRevD.77.064033} {\bibfield  {journal} {\bibinfo  {journal} {Phys. Rev. D}\ }\textbf {\bibinfo {volume} {77}},\ \bibinfo {pages} {064033} (\bibinfo {year} {2008}{\natexlab{b}})},\ \Eprint {http://arxiv.org/abs/0712.2822} {arXiv:0712.2822 [hep-th]} \BibitemShut {NoStop}%
\bibitem [{\citenamefont {Mandal}\ \emph {et~al.}(2023{\natexlab{c}})\citenamefont {Mandal}, \citenamefont {Mastrolia}, \citenamefont {Silva}, \citenamefont {Patil},\ and\ \citenamefont {Steinhoff}}]{Mandal:2023lgy}%
  \BibitemOpen
  \bibfield  {author} {\bibinfo {author} {\bibfnamefont {Manoj~K.}\ \bibnamefont {Mandal}}, \bibinfo {author} {\bibfnamefont {Pierpaolo}\ \bibnamefont {Mastrolia}}, \bibinfo {author} {\bibfnamefont {Hector~O.}\ \bibnamefont {Silva}}, \bibinfo {author} {\bibfnamefont {Raj}\ \bibnamefont {Patil}}, \ and\ \bibinfo {author} {\bibfnamefont {Jan}\ \bibnamefont {Steinhoff}},\ }\bibfield  {title} {\enquote {\bibinfo {title} {{Gravitoelectric dynamical tides at second post-Newtonian order}},}\ }\href {\doibase 10.1007/JHEP11(2023)067} {\bibfield  {journal} {\bibinfo  {journal} {JHEP}\ }\textbf {\bibinfo {volume} {11}},\ \bibinfo {pages} {067} (\bibinfo {year} {2023}{\natexlab{c}})},\ \Eprint {http://arxiv.org/abs/2304.02030} {arXiv:2304.02030 [hep-th]} \BibitemShut {NoStop}%
\bibitem [{\citenamefont {Mandal}\ \emph {et~al.}(2024)\citenamefont {Mandal}, \citenamefont {Mastrolia}, \citenamefont {Silva}, \citenamefont {Patil},\ and\ \citenamefont {Steinhoff}}]{Mandal:2023hqa}%
  \BibitemOpen
  \bibfield  {author} {\bibinfo {author} {\bibfnamefont {Manoj~K.}\ \bibnamefont {Mandal}}, \bibinfo {author} {\bibfnamefont {Pierpaolo}\ \bibnamefont {Mastrolia}}, \bibinfo {author} {\bibfnamefont {Hector~O.}\ \bibnamefont {Silva}}, \bibinfo {author} {\bibfnamefont {Raj}\ \bibnamefont {Patil}}, \ and\ \bibinfo {author} {\bibfnamefont {Jan}\ \bibnamefont {Steinhoff}},\ }\bibfield  {title} {\enquote {\bibinfo {title} {{Renormalizing Love: tidal effects at the third post-Newtonian order}},}\ }\href {\doibase 10.1007/JHEP02(2024)188} {\bibfield  {journal} {\bibinfo  {journal} {JHEP}\ }\textbf {\bibinfo {volume} {02}},\ \bibinfo {pages} {188} (\bibinfo {year} {2024})},\ \Eprint {http://arxiv.org/abs/2308.01865} {arXiv:2308.01865 [hep-th]} \BibitemShut {NoStop}%
\bibitem [{\citenamefont {Nogueira}(1993)}]{Nogueira:1991ex}%
  \BibitemOpen
  \bibfield  {author} {\bibinfo {author} {\bibfnamefont {Paulo}\ \bibnamefont {Nogueira}},\ }\bibfield  {title} {\enquote {\bibinfo {title} {{Automatic Feynman Graph Generation}},}\ }\href {\doibase 10.1006/jcph.1993.1074} {\bibfield  {journal} {\bibinfo  {journal} {J. Comput. Phys.}\ }\textbf {\bibinfo {volume} {105}},\ \bibinfo {pages} {279--289} (\bibinfo {year} {1993})}\BibitemShut {NoStop}%
\bibitem [{\citenamefont {Garc\'ia}()}]{xAct}%
  \BibitemOpen
  \bibfield  {author} {\bibinfo {author} {\bibfnamefont {J.~M.~Mart\'in}\ \bibnamefont {Garc\'ia}},\ }\href {http://www.xact.es} {\enquote {\bibinfo {title} {xact: Efficient tensor computer algebra for mathematica},}\ }\BibitemShut {NoStop}%
\bibitem [{\citenamefont {Brunello}\ \emph {et~al.}()\citenamefont {Brunello}, \citenamefont {Mastrolia}, \citenamefont {Ronca}, \citenamefont {Smith},\ and\ \citenamefont {Zeng}}]{PRISM}%
  \BibitemOpen
  \bibfield  {author} {\bibinfo {author} {\bibfnamefont {Giacomo}\ \bibnamefont {Brunello}}, \bibinfo {author} {\bibfnamefont {Pierpaolo}\ \bibnamefont {Mastrolia}}, \bibinfo {author} {\bibfnamefont {Jonathan}\ \bibnamefont {Ronca}}, \bibinfo {author} {\bibfnamefont {Sid}\ \bibnamefont {Smith}}, \ and\ \bibinfo {author} {\bibfnamefont {Mao}\ \bibnamefont {Zeng}},\ }\bibfield  {title} {\enquote {\bibinfo {title} {{PRISM: Prime-field Reduction of Integrals using Syzygy Methods}},}\ }\href@noop {} {\bibinfo  {journal} {in progress}\ }\BibitemShut {NoStop}%
\bibitem [{\citenamefont {Lee}(2014)}]{Lee:2013mka}%
  \BibitemOpen
\bibfield  {journal} {  }\bibfield  {author} {\bibinfo {author} {\bibfnamefont {Roman~N.}\ \bibnamefont {Lee}},\ }\bibfield  {title} {\enquote {\bibinfo {title} {{LiteRed 1.4: a powerful tool for reduction of multiloop integrals}},}\ }\href {\doibase 10.1088/1742-6596/523/1/012059} {\bibfield  {journal} {\bibinfo  {journal} {J. Phys. Conf. Ser.}\ }\textbf {\bibinfo {volume} {523}},\ \bibinfo {pages} {012059} (\bibinfo {year} {2014})},\ \Eprint {http://arxiv.org/abs/1310.1145} {arXiv:1310.1145 [hep-ph]} \BibitemShut {NoStop}%
\bibitem [{\citenamefont {Maierh{\"o}fer}\ \emph {et~al.}(2018)\citenamefont {Maierh{\"o}fer}, \citenamefont {Usovitsch},\ and\ \citenamefont {Uwer}}]{Maierhofer:2017gsa}%
  \BibitemOpen
  \bibfield  {author} {\bibinfo {author} {\bibfnamefont {Philipp}\ \bibnamefont {Maierh{\"o}fer}}, \bibinfo {author} {\bibfnamefont {Johann}\ \bibnamefont {Usovitsch}}, \ and\ \bibinfo {author} {\bibfnamefont {Peter}\ \bibnamefont {Uwer}},\ }\bibfield  {title} {\enquote {\bibinfo {title} {{Kira{\textemdash}A Feynman integral reduction program}},}\ }\href {\doibase 10.1016/j.cpc.2018.04.012} {\bibfield  {journal} {\bibinfo  {journal} {Comput. Phys. Commun.}\ }\textbf {\bibinfo {volume} {230}},\ \bibinfo {pages} {99--112} (\bibinfo {year} {2018})},\ \Eprint {http://arxiv.org/abs/1705.05610} {arXiv:1705.05610 [hep-ph]} \BibitemShut {NoStop}%
\bibitem [{\citenamefont {Decker}\ \emph {et~al.}(2024)\citenamefont {Decker}, \citenamefont {Greuel}, \citenamefont {Pfister},\ and\ \citenamefont {Sch\"onemann}}]{DGPS}%
  \BibitemOpen
  \bibfield  {author} {\bibinfo {author} {\bibfnamefont {Wolfram}\ \bibnamefont {Decker}}, \bibinfo {author} {\bibfnamefont {Gert-Martin}\ \bibnamefont {Greuel}}, \bibinfo {author} {\bibfnamefont {Gerhard}\ \bibnamefont {Pfister}}, \ and\ \bibinfo {author} {\bibfnamefont {Hans}\ \bibnamefont {Sch\"onemann}},\ }\href@noop {} {\enquote {\bibinfo {title} {{\sc Singular} {4-4-0} --- {A} computer algebra system for polynomial computations},}\ }\bibinfo {howpublished} {\url{http://www.singular.uni-kl.de}} (\bibinfo {year} {2024})\BibitemShut {NoStop}%
\bibitem [{\citenamefont {Peraro}(2019)}]{Peraro:2019svx}%
  \BibitemOpen
  \bibfield  {author} {\bibinfo {author} {\bibfnamefont {Tiziano}\ \bibnamefont {Peraro}},\ }\bibfield  {title} {\enquote {\bibinfo {title} {{$\text{FiniteFlow}$: multivariate functional reconstruction using finite fields and dataflow graphs}},}\ }\href {\doibase 10.1007/JHEP07(2019)031} {\bibfield  {journal} {\bibinfo  {journal} {JHEP}\ }\textbf {\bibinfo {volume} {07}},\ \bibinfo {pages} {031} (\bibinfo {year} {2019})},\ \Eprint {http://arxiv.org/abs/1905.08019} {arXiv:1905.08019 [hep-ph]} \BibitemShut {NoStop}%
\bibitem [{\citenamefont {Damgaard}\ and\ \citenamefont {Lee}(2024)}]{Damgaard:2024fqj}%
  \BibitemOpen
  \bibfield  {author} {\bibinfo {author} {\bibfnamefont {Poul~H.}\ \bibnamefont {Damgaard}}\ and\ \bibinfo {author} {\bibfnamefont {Kanghoon}\ \bibnamefont {Lee}},\ }\bibfield  {title} {\enquote {\bibinfo {title} {{Schwarzschild Black Hole from Perturbation Theory to All Orders}},}\ }\href {\doibase 10.1103/PhysRevLett.132.251603} {\bibfield  {journal} {\bibinfo  {journal} {Phys. Rev. Lett.}\ }\textbf {\bibinfo {volume} {132}},\ \bibinfo {pages} {251603} (\bibinfo {year} {2024})},\ \Eprint {http://arxiv.org/abs/2403.13216} {arXiv:2403.13216 [hep-th]} \BibitemShut {NoStop}%
\bibitem [{\citenamefont {Mougiakakos}\ and\ \citenamefont {Vanhove}(2024)}]{Mougiakakos:2024lif}%
  \BibitemOpen
  \bibfield  {author} {\bibinfo {author} {\bibfnamefont {Stavros}\ \bibnamefont {Mougiakakos}}\ and\ \bibinfo {author} {\bibfnamefont {Pierre}\ \bibnamefont {Vanhove}},\ }\bibfield  {title} {\enquote {\bibinfo {title} {{Schwarzschild geodesics from scattering amplitudes to all orders in G$_{N}$}},}\ }\href {\doibase 10.1007/JHEP10(2024)152} {\bibfield  {journal} {\bibinfo  {journal} {JHEP}\ }\textbf {\bibinfo {volume} {10}},\ \bibinfo {pages} {152} (\bibinfo {year} {2024})},\ \Eprint {http://arxiv.org/abs/2407.09448} {arXiv:2407.09448 [hep-th]} \BibitemShut {NoStop}%
\bibitem [{\citenamefont {Chetyrkin}\ \emph {et~al.}(1980)\citenamefont {Chetyrkin}, \citenamefont {Kataev},\ and\ \citenamefont {Tkachov}}]{Chetyrkin:1980pr}%
  \BibitemOpen
  \bibfield  {author} {\bibinfo {author} {\bibfnamefont {K.~G.}\ \bibnamefont {Chetyrkin}}, \bibinfo {author} {\bibfnamefont {A.~L.}\ \bibnamefont {Kataev}}, \ and\ \bibinfo {author} {\bibfnamefont {F.~V.}\ \bibnamefont {Tkachov}},\ }\bibfield  {title} {\enquote {\bibinfo {title} {{New Approach to Evaluation of Multiloop Feynman Integrals: The Gegenbauer Polynomial x Space Technique}},}\ }\href {\doibase 10.1016/0550-3213(80)90289-8} {\bibfield  {journal} {\bibinfo  {journal} {Nucl. Phys. B}\ }\textbf {\bibinfo {volume} {174}},\ \bibinfo {pages} {345--377} (\bibinfo {year} {1980})}\BibitemShut {NoStop}%
\bibitem [{\citenamefont {Kotikov}(1996)}]{Kotikov:1995cw}%
  \BibitemOpen
  \bibfield  {author} {\bibinfo {author} {\bibfnamefont {A.~V.}\ \bibnamefont {Kotikov}},\ }\bibfield  {title} {\enquote {\bibinfo {title} {{The Gegenbauer polynomial technique: The Evaluation of a class of Feynman diagrams}},}\ }\href {\doibase 10.1016/0370-2693(96)00226-2} {\bibfield  {journal} {\bibinfo  {journal} {Phys. Lett. B}\ }\textbf {\bibinfo {volume} {375}},\ \bibinfo {pages} {240--248} (\bibinfo {year} {1996})},\ \Eprint {http://arxiv.org/abs/hep-ph/9512270} {arXiv:hep-ph/9512270} \BibitemShut {NoStop}%
\bibitem [{\citenamefont {Moch}\ \emph {et~al.}(2002)\citenamefont {Moch}, \citenamefont {Uwer},\ and\ \citenamefont {Weinzierl}}]{Moch:2001zr}%
  \BibitemOpen
  \bibfield  {author} {\bibinfo {author} {\bibfnamefont {Sven}\ \bibnamefont {Moch}}, \bibinfo {author} {\bibfnamefont {Peter}\ \bibnamefont {Uwer}}, \ and\ \bibinfo {author} {\bibfnamefont {Stefan}\ \bibnamefont {Weinzierl}},\ }\bibfield  {title} {\enquote {\bibinfo {title} {{Nested sums, expansion of transcendental functions and multiscale multiloop integrals}},}\ }\href {\doibase 10.1063/1.1471366} {\bibfield  {journal} {\bibinfo  {journal} {J. Math. Phys.}\ }\textbf {\bibinfo {volume} {43}},\ \bibinfo {pages} {3363--3386} (\bibinfo {year} {2002})},\ \Eprint {http://arxiv.org/abs/hep-ph/0110083} {arXiv:hep-ph/0110083} \BibitemShut {NoStop}%
\bibitem [{\citenamefont {Bierenbaum}\ and\ \citenamefont {Weinzierl}(2003)}]{Bierenbaum:2003ud}%
  \BibitemOpen
  \bibfield  {author} {\bibinfo {author} {\bibfnamefont {Isabella}\ \bibnamefont {Bierenbaum}}\ and\ \bibinfo {author} {\bibfnamefont {Stefan}\ \bibnamefont {Weinzierl}},\ }\bibfield  {title} {\enquote {\bibinfo {title} {{The Massless two loop two point function}},}\ }\href {\doibase 10.1140/epjc/s2003-01389-7} {\bibfield  {journal} {\bibinfo  {journal} {Eur. Phys. J. C}\ }\textbf {\bibinfo {volume} {32}},\ \bibinfo {pages} {67--78} (\bibinfo {year} {2003})},\ \Eprint {http://arxiv.org/abs/hep-ph/0308311} {arXiv:hep-ph/0308311} \BibitemShut {NoStop}%
\bibitem [{\citenamefont {Grozin}(2012)}]{Grozin:2012xi}%
  \BibitemOpen
  \bibfield  {author} {\bibinfo {author} {\bibfnamefont {A.~G.}\ \bibnamefont {Grozin}},\ }\bibfield  {title} {\enquote {\bibinfo {title} {{Massless two-loop self-energy diagram: Historical review}},}\ }\href {\doibase 10.1142/S0217751X12300189} {\bibfield  {journal} {\bibinfo  {journal} {Int. J. Mod. Phys. A}\ }\textbf {\bibinfo {volume} {27}},\ \bibinfo {pages} {1230018} (\bibinfo {year} {2012})},\ \Eprint {http://arxiv.org/abs/1206.2572} {arXiv:1206.2572 [hep-ph]} \BibitemShut {NoStop}%
\bibitem [{\citenamefont {Remiddi}\ and\ \citenamefont {Vermaseren}(2000)}]{Remiddi:1999ew}%
  \BibitemOpen
  \bibfield  {author} {\bibinfo {author} {\bibfnamefont {E.}~\bibnamefont {Remiddi}}\ and\ \bibinfo {author} {\bibfnamefont {J.~A.~M.}\ \bibnamefont {Vermaseren}},\ }\bibfield  {title} {\enquote {\bibinfo {title} {{Harmonic polylogarithms}},}\ }\href {\doibase 10.1142/S0217751X00000367} {\bibfield  {journal} {\bibinfo  {journal} {Int. J. Mod. Phys. A}\ }\textbf {\bibinfo {volume} {15}},\ \bibinfo {pages} {725--754} (\bibinfo {year} {2000})},\ \Eprint {http://arxiv.org/abs/hep-ph/9905237} {arXiv:hep-ph/9905237} \BibitemShut {NoStop}%
\bibitem [{\citenamefont {Lee}(2010{\natexlab{a}})}]{Lee:2009dh}%
  \BibitemOpen
  \bibfield  {author} {\bibinfo {author} {\bibfnamefont {R.~N.}\ \bibnamefont {Lee}},\ }\bibfield  {title} {\enquote {\bibinfo {title} {{Space-time dimensionality D as complex variable: Calculating loop integrals using dimensional recurrence relation and analytical properties with respect to D}},}\ }\href {\doibase 10.1016/j.nuclphysb.2009.12.025} {\bibfield  {journal} {\bibinfo  {journal} {Nucl. Phys. B}\ }\textbf {\bibinfo {volume} {830}},\ \bibinfo {pages} {474--492} (\bibinfo {year} {2010}{\natexlab{a}})},\ \Eprint {http://arxiv.org/abs/0911.0252} {arXiv:0911.0252 [hep-ph]} \BibitemShut {NoStop}%
\bibitem [{\citenamefont {Lee}(2010{\natexlab{b}})}]{Lee:2010wea}%
  \BibitemOpen
  \bibfield  {author} {\bibinfo {author} {\bibfnamefont {R.~N.}\ \bibnamefont {Lee}},\ }\bibfield  {title} {\enquote {\bibinfo {title} {{Calculating multiloop integrals using dimensional recurrence relation and $D$-analyticity}},}\ }\href {\doibase 10.1016/j.nuclphysbps.2010.08.032} {\bibfield  {journal} {\bibinfo  {journal} {Nucl. Phys. B Proc. Suppl.}\ }\textbf {\bibinfo {volume} {205-206}},\ \bibinfo {pages} {135--140} (\bibinfo {year} {2010}{\natexlab{b}})},\ \Eprint {http://arxiv.org/abs/1007.2256} {arXiv:1007.2256 [hep-ph]} \BibitemShut {NoStop}%
\bibitem [{\citenamefont {Lee}\ and\ \citenamefont {Mingulov}(2016)}]{Lee:2015eva}%
  \BibitemOpen
  \bibfield  {author} {\bibinfo {author} {\bibfnamefont {Roman~N.}\ \bibnamefont {Lee}}\ and\ \bibinfo {author} {\bibfnamefont {Kirill~T.}\ \bibnamefont {Mingulov}},\ }\bibfield  {title} {\enquote {\bibinfo {title} {{Introducing SummerTime: a package for high-precision computation of sums appearing in DRA method}},}\ }\href {\doibase 10.1016/j.cpc.2016.02.018} {\bibfield  {journal} {\bibinfo  {journal} {Comput. Phys. Commun.}\ }\textbf {\bibinfo {volume} {203}},\ \bibinfo {pages} {255--267} (\bibinfo {year} {2016})},\ \Eprint {http://arxiv.org/abs/1507.04256} {arXiv:1507.04256 [hep-ph]} \BibitemShut {NoStop}%
\bibitem [{\citenamefont {Lee}\ and\ \citenamefont {Mingulov}(2017)}]{Lee:2017ftw}%
  \BibitemOpen
  \bibfield  {author} {\bibinfo {author} {\bibfnamefont {Roman~N.}\ \bibnamefont {Lee}}\ and\ \bibinfo {author} {\bibfnamefont {Kirill~T.}\ \bibnamefont {Mingulov}},\ }\bibfield  {title} {\enquote {\bibinfo {title} {{DREAM, a program for arbitrary-precision computation of dimensional recurrence relations solutions, and its applications}},}\ }\href@noop {} {\  (\bibinfo {year} {2017})},\ \Eprint {http://arxiv.org/abs/1712.05173} {arXiv:1712.05173 [hep-ph]} \BibitemShut {NoStop}%
\bibitem [{\citenamefont {Borinsky}\ \emph {et~al.}(2023)\citenamefont {Borinsky}, \citenamefont {Munch},\ and\ \citenamefont {Tellander}}]{Borinsky:2023jdv}%
  \BibitemOpen
  \bibfield  {author} {\bibinfo {author} {\bibfnamefont {Michael}\ \bibnamefont {Borinsky}}, \bibinfo {author} {\bibfnamefont {Henrik~J.}\ \bibnamefont {Munch}}, \ and\ \bibinfo {author} {\bibfnamefont {Felix}\ \bibnamefont {Tellander}},\ }\bibfield  {title} {\enquote {\bibinfo {title} {{Tropical Feynman integration in the Minkowski regime}},}\ }\href {\doibase 10.1016/j.cpc.2023.108874} {\bibfield  {journal} {\bibinfo  {journal} {Comput. Phys. Commun.}\ }\textbf {\bibinfo {volume} {292}},\ \bibinfo {pages} {108874} (\bibinfo {year} {2023})},\ \Eprint {http://arxiv.org/abs/2302.08955} {arXiv:2302.08955 [hep-ph]} \BibitemShut {NoStop}%
\bibitem [{\citenamefont {Bini}\ \emph {et~al.}(2020{\natexlab{a}})\citenamefont {Bini}, \citenamefont {Damour},\ and\ \citenamefont {Geralico}}]{Bini:2020nsb}%
  \BibitemOpen
  \bibfield  {author} {\bibinfo {author} {\bibfnamefont {Donato}\ \bibnamefont {Bini}}, \bibinfo {author} {\bibfnamefont {Thibault}\ \bibnamefont {Damour}}, \ and\ \bibinfo {author} {\bibfnamefont {Andrea}\ \bibnamefont {Geralico}},\ }\bibfield  {title} {\enquote {\bibinfo {title} {{Sixth post-Newtonian local-in-time dynamics of binary systems}},}\ }\href {\doibase 10.1103/PhysRevD.102.024061} {\bibfield  {journal} {\bibinfo  {journal} {Phys. Rev. D}\ }\textbf {\bibinfo {volume} {102}},\ \bibinfo {pages} {024061} (\bibinfo {year} {2020}{\natexlab{a}})},\ \Eprint {http://arxiv.org/abs/2004.05407} {arXiv:2004.05407 [gr-qc]} \BibitemShut {NoStop}%
\bibitem [{\citenamefont {Bini}\ \emph {et~al.}(2021)\citenamefont {Bini}, \citenamefont {Damour}, \citenamefont {Geralico}, \citenamefont {Laporta},\ and\ \citenamefont {Mastrolia}}]{Bini:2020rzn}%
  \BibitemOpen
  \bibfield  {author} {\bibinfo {author} {\bibfnamefont {Donato}\ \bibnamefont {Bini}}, \bibinfo {author} {\bibfnamefont {Thibault}\ \bibnamefont {Damour}}, \bibinfo {author} {\bibfnamefont {Andrea}\ \bibnamefont {Geralico}}, \bibinfo {author} {\bibfnamefont {Stefano}\ \bibnamefont {Laporta}}, \ and\ \bibinfo {author} {\bibfnamefont {Pierpaolo}\ \bibnamefont {Mastrolia}},\ }\bibfield  {title} {\enquote {\bibinfo {title} {{Gravitational scattering at the seventh order in $G$: nonlocal contribution at the sixth post-Newtonian accuracy}},}\ }\href {\doibase 10.1103/PhysRevD.103.044038} {\bibfield  {journal} {\bibinfo  {journal} {Phys. Rev. D}\ }\textbf {\bibinfo {volume} {103}},\ \bibinfo {pages} {044038} (\bibinfo {year} {2021})},\ \Eprint {http://arxiv.org/abs/2012.12918} {arXiv:2012.12918 [gr-qc]} \BibitemShut {NoStop}%
\bibitem [{\citenamefont {Bini}\ \emph {et~al.}(2020{\natexlab{b}})\citenamefont {Bini}, \citenamefont {Damour}, \citenamefont {Geralico}, \citenamefont {Laporta},\ and\ \citenamefont {Mastrolia}}]{Bini:2020uiq}%
  \BibitemOpen
  \bibfield  {author} {\bibinfo {author} {\bibfnamefont {Donato}\ \bibnamefont {Bini}}, \bibinfo {author} {\bibfnamefont {Thibault}\ \bibnamefont {Damour}}, \bibinfo {author} {\bibfnamefont {Andrea}\ \bibnamefont {Geralico}}, \bibinfo {author} {\bibfnamefont {Stefano}\ \bibnamefont {Laporta}}, \ and\ \bibinfo {author} {\bibfnamefont {Pierpaolo}\ \bibnamefont {Mastrolia}},\ }\bibfield  {title} {\enquote {\bibinfo {title} {{Gravitational dynamics at $O(G^6)$: perturbative gravitational scattering meets experimental mathematics}},}\ }\href@noop {} {\  (\bibinfo {year} {2020}{\natexlab{b}})},\ \Eprint {http://arxiv.org/abs/2008.09389} {arXiv:2008.09389 [gr-qc]} \BibitemShut {NoStop}%
\bibitem [{\citenamefont {Almeida}\ \emph {et~al.}(2021)\citenamefont {Almeida}, \citenamefont {Foffa},\ and\ \citenamefont {Sturani}}]{Almeida:2021xwn}%
  \BibitemOpen
  \bibfield  {author} {\bibinfo {author} {\bibfnamefont {Gabriel~Luz}\ \bibnamefont {Almeida}}, \bibinfo {author} {\bibfnamefont {Stefano}\ \bibnamefont {Foffa}}, \ and\ \bibinfo {author} {\bibfnamefont {Riccardo}\ \bibnamefont {Sturani}},\ }\bibfield  {title} {\enquote {\bibinfo {title} {{Tail contributions to gravitational conservative dynamics}},}\ }\href {\doibase 10.1103/PhysRevD.104.124075} {\bibfield  {journal} {\bibinfo  {journal} {Phys. Rev. D}\ }\textbf {\bibinfo {volume} {104}},\ \bibinfo {pages} {124075} (\bibinfo {year} {2021})},\ \Eprint {http://arxiv.org/abs/2110.14146} {arXiv:2110.14146 [gr-qc]} \BibitemShut {NoStop}%
\bibitem [{\citenamefont {Bini}\ \emph {et~al.}(2020{\natexlab{c}})\citenamefont {Bini}, \citenamefont {Damour},\ and\ \citenamefont {Geralico}}]{Bini:2020hmy}%
  \BibitemOpen
  \bibfield  {author} {\bibinfo {author} {\bibfnamefont {Donato}\ \bibnamefont {Bini}}, \bibinfo {author} {\bibfnamefont {Thibault}\ \bibnamefont {Damour}}, \ and\ \bibinfo {author} {\bibfnamefont {Andrea}\ \bibnamefont {Geralico}},\ }\bibfield  {title} {\enquote {\bibinfo {title} {{Sixth post-Newtonian nonlocal-in-time dynamics of binary systems}},}\ }\href {\doibase 10.1103/PhysRevD.102.084047} {\bibfield  {journal} {\bibinfo  {journal} {Phys. Rev. D}\ }\textbf {\bibinfo {volume} {102}},\ \bibinfo {pages} {084047} (\bibinfo {year} {2020}{\natexlab{c}})},\ \Eprint {http://arxiv.org/abs/2007.11239} {arXiv:2007.11239 [gr-qc]} \BibitemShut {NoStop}%
\bibitem [{\citenamefont {Edison}\ and\ \citenamefont {Levi}(2024)}]{Edison:2023qvg}%
  \BibitemOpen
  \bibfield  {author} {\bibinfo {author} {\bibfnamefont {Alex}\ \bibnamefont {Edison}}\ and\ \bibinfo {author} {\bibfnamefont {Mich{\`e}le}\ \bibnamefont {Levi}},\ }\bibfield  {title} {\enquote {\bibinfo {title} {{Higher-order tails and RG flows due to scattering of gravitational radiation from binary inspirals}},}\ }\href {\doibase 10.1007/JHEP08(2024)161} {\bibfield  {journal} {\bibinfo  {journal} {JHEP}\ }\textbf {\bibinfo {volume} {08}},\ \bibinfo {pages} {161} (\bibinfo {year} {2024})},\ \Eprint {http://arxiv.org/abs/2310.20066} {arXiv:2310.20066 [hep-th]} \BibitemShut {NoStop}%
\bibitem [{\citenamefont {Remiddi}\ and\ \citenamefont {Tancredi}(2014)}]{Remiddi:2013joa}%
  \BibitemOpen
  \bibfield  {author} {\bibinfo {author} {\bibfnamefont {Ettore}\ \bibnamefont {Remiddi}}\ and\ \bibinfo {author} {\bibfnamefont {Lorenzo}\ \bibnamefont {Tancredi}},\ }\bibfield  {title} {\enquote {\bibinfo {title} {{Schouten identities for Feynman graph amplitudes; The Master Integrals for the two-loop massive sunrise graph}},}\ }\href {\doibase 10.1016/j.nuclphysb.2014.01.009} {\bibfield  {journal} {\bibinfo  {journal} {Nucl. Phys. B}\ }\textbf {\bibinfo {volume} {880}},\ \bibinfo {pages} {343--377} (\bibinfo {year} {2014})},\ \Eprint {http://arxiv.org/abs/1311.3342} {arXiv:1311.3342 [hep-ph]} \BibitemShut {NoStop}%
\end{thebibliography}%

\end{document}